\documentclass[aps,prx,reprint,showpacs,showkeys,superscriptaddress,floatfix]{revtex4-2}
\usepackage{float}

\usepackage{multirow}
\usepackage{graphicx}
\usepackage{bm}
\usepackage{dcolumn}
\usepackage[colorlinks=true,linkcolor=blue,urlcolor=blue,citecolor=blue]{hyperref}
\usepackage{natbib}
\usepackage{amsmath}
\usepackage{times}
\usepackage{booktabs}  % for better horizontal lines
\usepackage{color}
\usepackage{xcolor}
\usepackage{soul}
\usepackage[normalem]{ulem}
\usepackage{textcomp,mathcomp}

\usepackage{amsmath,amssymb}  % Standard math packages
\usepackage{graphicx}  % Use graphicx.sty for figures
\usepackage{bm}  % For bold math symbols
\usepackage{dcolumn}  % Align table columns on decimal point
\usepackage[colorlinks=true,linkcolor=blue,urlcolor=blue,citecolor=blue]{hyperref}  % Hyperlinks
\usepackage{mhchem}
\usepackage{booktabs}
\usepackage{xcolor}
\usepackage[normalem]{ulem}
\usepackage{physics}
\usepackage{siunitx}
\usepackage{tablefootnote}
\usepackage{dsfont}

\usepackage{multirow} % Include this in your preamble

\usepackage{gensymb}  % For degree sign
\usepackage{chemformula}
\usepackage{xcolor}
\definecolor{mypurple}{RGB}{128,0,128}
\definecolor{myorange}{RGB}{237,119,0}

\newcommand{\ErSSL}{\ch{Er_2Be_2GeO_7}}
\newcommand{\SpaceGroup}{$P\bar 42_1m$}
\newcommand{\LuSSL}{\ch{Lu_2Be_2GeO_7}}

\newcommand{\SCBO}{\ch{SrCu2(BO3)2}}
\newcommand{\RNum}[1]{\uppercase\expandafter{\romannumeral #1\relax}}

\begin{document}

%%%%%%%%%%%

\title{Observation of unprecedented fractional magnetization plateaus in a new Shastry-Sutherland Ising compound}

\author{Lalit Yadav}
\thanks{equal contribution}
\affiliation{Department of Physics, Duke University, Durham, NC 27708, USA}

\author{Afonso Rufino}
\thanks{equal contribution}
\affiliation{Institute of Physics, Ecole Polytechnique Fédérale de Lausanne (EPFL), CH-1015 Lausanne, Switzerland}

\author{Rabindranath Bag}
\affiliation{Department of Physics, Duke University, Durham, NC 27708, USA}

\author{Matthew Ennis}
\affiliation{Department of Physics, Duke University, Durham, NC 27708, USA}

\author{Jan Alexander Koziol}
\affiliation{Department of Physics, Friedrich-Alexander-Universität Erlangen-Nürnberg, D-91058 Erlangen, Germany}

\author{Clarina dela Cruz}
\affiliation{Neutron Scattering Division, Oak Ridge National Laboratory, Oak Ridge, TN 37831, USA}

\author{Alexander I. Kolesnikov}
\affiliation{Neutron Scattering Division, Oak Ridge National Laboratory, Oak Ridge, TN 37831, USA}

\author{V. Ovidiu Garlea}
\affiliation{Neutron Scattering Division, Oak Ridge National Laboratory, Oak Ridge, TN 37831, USA}

\author{Keith M. Taddei}
\affiliation{Neutron Scattering Division, Oak Ridge National Laboratory, Oak Ridge, TN 37831, USA}

\author{David Graf}
\affiliation{National High Magnetic Field Laboratory and Department of Physics, Florida State University, Tallahassee, FL 32310, USA}

\author{Kai Phillip Schmidt}
\affiliation{Department of Physics, Friedrich-Alexander-Universität Erlangen-Nürnberg, D-91058 Erlangen, Germany}

\author{Frédéric Mila}
\affiliation{Institute of Physics, Ecole Polytechnique Fédérale de Lausanne (EPFL), CH-1015 Lausanne, Switzerland}

\author{Sara Haravifard}
\email[email:]{sara.haravifard@duke.edu}
\affiliation{Department of Physics, Duke University, Durham, NC 27708, USA}
\affiliation{Department of Mechanical Engineering and Materials Science, Duke University, Durham, NC 27708, USA}
\affiliation{Department of Electrical and Computer Engineering, Duke University, Durham, NC 27708, USA}

%\date{\today}
%%%%%%%%%%%%%%%%%%%%%%%%%%%%% ABSTRACT %%%%%%%%%%%%%%%%%%%%%%
\begin{abstract}
Geometrically frustrated magnetic systems, such as those based on the Shastry-Sutherland lattice (SSL), offer a rich playground for exploring unconventional magnetic states. The delicate balance between competing interactions in these systems leads to the emergence of novel phases. We present the characterization of \ErSSL{}, an SSL compound with \ch{Er^{3+}} ions forming orthogonal dimers separated by non-magnetic layers whose structure is invariant under the \SpaceGroup{} space group. Neutron scattering reveals an antiferromagnetic dimer structure at zero field, typical of Ising spins on that lattice and consistent with the anisotropic magnetization observed. However, magnetization measurements exhibit fractional plateaus at 1/4 and 1/2 of saturation, in contrast to the expected 1/3 plateau of the SSL Ising model. By comparing the energy of candidate states with ground-state lower bounds we show that this behavior requires spatially anisotropic interactions, leading to an anisotropic Shastry-Sutherland Ising Model (ASSLIM) symmetric under the $Cmm2$ space group. This anisotropy is consistent with the small orthorhombic distortion observed with single-crystal neutron diffraction. The other properties, including thermodynamics, which have been investigated theoretically using tensor networks, point to small residual interactions, potentially due to further couplings and quantum fluctuations. This study highlights \ErSSL{} as a promising platform for investigating exotic magnetic phenomena.
\end{abstract}
%%%%%%%%%%%%%%%%%%%%%%

\maketitle

%%%%%%%%%%%%%%%%%%%%%%%%%%%%%% INTRODUCTION %%%%%%%%%%%%%%%%%%%%%%%%
\section{Introduction}

Geometrically frustrated magnetic systems offer a platform to explore magnetic states with suppressed long-range ordering and unconventional excitations \cite{Balents2010SpinMagnets, Knolle2019ALiquids}. Fractional magnetization plateaus are one of the most remarkable consequences of geometrical frustration \cite{TakigawaMila2011}, but  in spite of over three decades of intense research, the discovery of plateaus in a new compound often poses a challenge to explain. The most famous example is the Shastry-Sutherland Heisenberg compound SrCu$_2$(BO$_3$)$_2$ and its improbable sequence of plateaus at 1/8, 2/15, 1/6, 1/4, 1/3, 2/5 and 1/2 \cite{Kageyama1999,Onizuka2000,Kodama2002,Jaime2012,Takigawa2013,Matsuda2013} that resisted for 15 years until a plausible explanation was put forward \cite{Corboz2014}. These magnetization plateaus have been associated with a distinct set of exotic super-lattice spin structures originating from the bosonic crystallization of the triplets or of boundstates of triplets. However, due to the high magnetic fields required for these transitions, such as 27.2 T for the first plateau, the direct evidence for the actual spin structure remains to be measured using probes such as neutron scattering. The difficulty in reaching the magnetic fields required for the higher magnetization plateaus in SrCu$_2$(BO$_3$)$_2$ and the desire to study the implications of the geometric frustration  in a different setting than that of the spin-1/2 Heisenberg model motivates the search for other magnetic compounds hosting the Shastry-Sutherland lattice (SSL).

The SSL \cite{SriramShastry1981} can be viewed as an orthogonal arrangement of dimers, and this structure has already been realized in several rare-earth compounds where the spins are Ising-like, but a clean realization of the Ising model with just inter- and intra-dimer couplings is still missing. The interest in that model lies in the very solid prediction of a unique 1/3 magnetization plateau \cite{mengPhasesMagnetizationProcess2008,dublenychGroundStatesIsing2012}, a prediction that has not been verified so far. In the \ch{ReB4} family, long-range Ruderman–Kittel–Kasuya–Yosida (RKKY) interactions give rise to various fractional magnetization plateaus: a 1/5 plateau in \ch{NdB4}; a 1/3 plateau with Up-Up-Down (UUD) ferrimagnetic ordering, as well as narrow 1/2 and 3/5 plateaus in \ch{HoB4} \cite{bruntMagnetisationProcessRare2018}; and multiple plateaus, including a 1/2 plateau, in \ch{TmB4} \cite{suzukiFinitetemperaturePhaseTransition2010, orendacGroundStateStability2021}.  Conversely,  the insulating compounds \ch{BaNd2ZnO5} \cite{Ishii2021} and \ch{BaNd2ZnS5} \cite{Marshall2023} exhibit ferromagnetic intra-dimer interactions stabilizing a double-Q magnetic structure without fractional plateaus.

\begin{figure*}[]
\centering
\includegraphics[width=\textwidth]{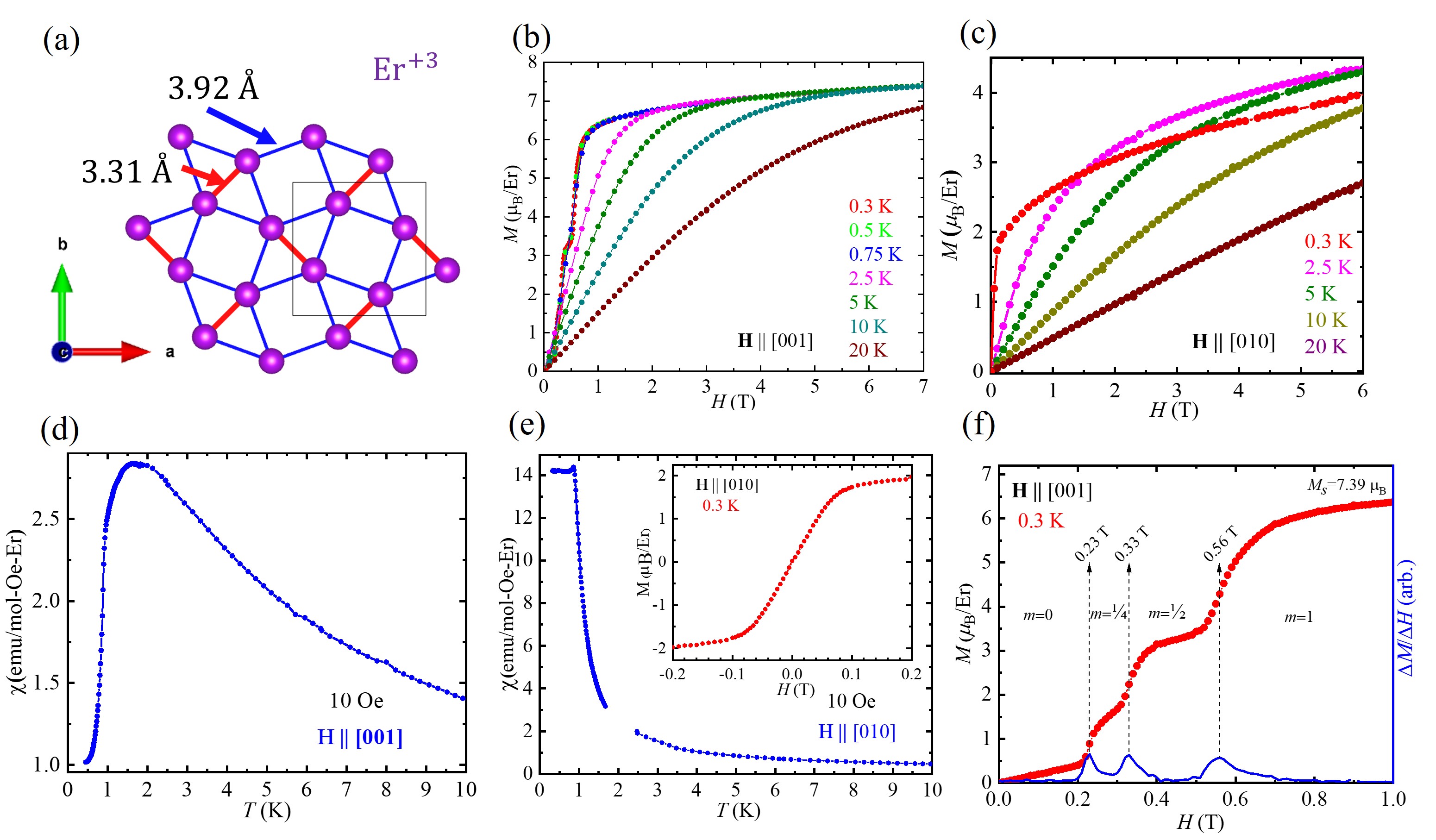}
  \caption{Crystal structure and magnetic properties of \ErSSL{}.
(a) Schematic representation of \ErSSL{} crystal structure, highlighting the 2D \ch{Er^{3+}} network with nearest and next nearest neighbor Er-Er bond lengths of \SI{3.31}{\angstrom} and \SI{3.92}{\angstrom}, respectively, and interlayer separation of \SI{4.72}{\angstrom}.
(b) Isothermal magnetization \(M_{[001]}\) measurements at various temperatures, exhibiting saturation to \(7.39~\mu_B\) at 7 T.
(c) Isothermal magnetization \(M_{[010]}\) at various temperatures, showing a tendency to saturate around \(4.5~\mu_B\).
(d) Low-temperature magnetic susceptibility (\(\chi_{[001]}\)) showcasing an antiferromagnetic transition near 0.85 K with $H$=10 Oe applied in [001] direction.
(e) Low-temperature magnetic susceptibility (\(\chi_{[010]}\)) revealing a ferromagnetic transition around 0.85 K when $H$=10 Oe applied in [010] direction. The missing data points from 1.5 K and 2.5 K are due to the use of different Helium-3 and Helium-4 setups. The inset presents a zoomed-in view of \(M_{[010]}(H)\) at 0.3 K. Within the -0.2 T to +0.2 T range, a sigmoid-like curve is observed, altering \(M_{[010]}\) from -1.5 \(\mu_B\) to +1.5 \(\mu_B\) over a mere \(\pm0.07\) T field range. No hysteresis is observed during field sweeps.
(f) Zoomed-in isothermal magnetization \(M_{[001]}(H)\) at 0.3 K, illustrating the emergence of fractional magnetization plateaus in presence of the applied magnetic field, with prominent 1/4 and 1/2 plateaus. The right axis displays the derivative of the magnetization, $\Delta M/\Delta H$. The peaks of the $\Delta M/\Delta H$ curve are used to identify the critical fields associated with the beginning and end of each plateau.}
 \label{magnetization}
\end{figure*}

Recently, a novel family of insulating rare-earth-based SSL compounds, the melilites \ch{Re2Be2GeO7}, has been reported, crystallizing in the tetragonal $P\bar 42_1m$ space group \cite{Ashtar2021}. In these compounds, Re magnetic ions form SSL orthogonal dimer planes (see FIG.~\ref{magnetization}(a)). Recently, this family has garnered significant interest, sparking active research efforts. For instance, work performed on \ch{Nd2Be2GeO7} has revealed both short-range spin correlations and long-range magnetic ordering, while studies on \ch{Pr2Be2GeO7} have identified dynamic spin-freezing behavior \cite{NdSSL_2024}. Similarly, investigations of \ch{Yb2Be2GeO7} indicate the absence of long-range order, pointing to a possible quantum spin liquid state \cite{pula2024}. Despite these advances, comprehensive magnetic property measurements across this family remain scarce, leaving many fundamental aspects yet to be explored.

In this paper, we present \ch{Er2Be2GeO7} as the closest realization of the Ising model on the SSL to date. As an insulating system, it lacks long-range RKKY interactions, and its dimers adopt an antiferromagnetic configuration in zero field, consistent with the expectations for an Ising model with strong intra-dimer interactions. However, when a magnetic field is applied along the [001] direction—parallel to the Ising spins—the magnetization curve reveals two well-defined plateaus at 1/4 and 1/2 of the saturation magnetization, with no sign of the conventionally expected 1/3 plateau. This finding is entirely unexpected, as all known variations of the Ising model on the SSL predict either a single 1/3 plateau for short-range interactions or multiple plateaus for long-range interactions, but never exclusively the 1/4 and 1/2 plateaus. Notably, these unexpected plateaus emerge at experimentally accessible magnetic fields, making their detailed investigation both feasible and compelling. 

Intrigued by this unexpected result, we conducted a theoretical exploration of several new model variants and revisited the sample’s structure using single-crystal neutron scattering. This combined investigation revealed a clear and elegant solution: a subtle orthorhombic distortion in the lattice. This distortion induces a slight asymmetry in intra-dimer interactions along the two orthogonal directions, destabilizing the expected 1/3 magnetization plateau and giving rise to the observed 1/4 and 1/2 plateaus.

%%%%%%%%%%%%%%%%%%%%%%%%FIG 1 %%%%%%%%%%%%%%%%%%%%%%%%

%TC:ignore
%TC:endignore
%%%%%%%%%%%%%%%%%%%%%%%%%%%%%%%%%%%%%%%%%%%%%%%%

%%%%%%%%%%%%%%%%%%%%%%%%%%% EXPERIMENTAL RESULTS %%%%%%%%%%%%%%%%%%%%%%%%%%%
\section{Experimental Results} 

%\subsection{Crystal structure and phase purity}

The crystal structure and phase purity of \ErSSL{} were thoroughly analyzed using powder X-ray diffraction and Rietveld refinement (see Appendix \ref{App:Crystal Structure and Phase Purity}). No sign of an impurity phase was detected. Detailed results, including the single crystal synthesis, can be found in the Appendix \ref{App:Crystal Structure and Phase Purity}. Temperature dependent magnetic susceptibility and isothermal magnetization measurements were conducted on high-quality single crystals of \ErSSL{} for both \( H \parallel [001] \) and \( H \parallel [010] \) directions. 

In FIG.~\ref{magnetization} (b,c), we present isothermal magnetization measurements along both crystallographic directions, which confirm the presence of strong directional magnetic anisotropy. A pronounced Ising-like anisotropy is evident, with a dominant spin component aligned along the \([001]\) direction. This significant anisotropy highlights the restricted spin dynamics characteristic of an Ising system. At 2.5 K, \(M_{[001]}\) approaches \(7.39~\mu_B\) at 7 T, while \(M_{[010]}\) reaches \(4.5~\mu_B\), confirming that the direction [001] has the largest projection of the easy axis. A more quantitative analysis will be performed in the theory section, after the Van Vleck contribution to the magnetization has been calculated and subtracted. We take the saturation for the [001] direction as \(7.39~\mu_B\), which can be used to label the field-emergent magnetization plateaus as fractions of \(M_S\). Additionally, Tunnel Diode Oscillator (TDO) measurements, detailed in Appendix \ref{App_TDO}, showed no evidence of any further magnetic transitions up to 35 T.

%%%%%%%%%%%%%%%%%%%%%%%% FIG 2 %%%%%%%%%%%%%%%%%%%%%%%%
\begin{figure*}[]
\centering
\includegraphics[width=\textwidth]{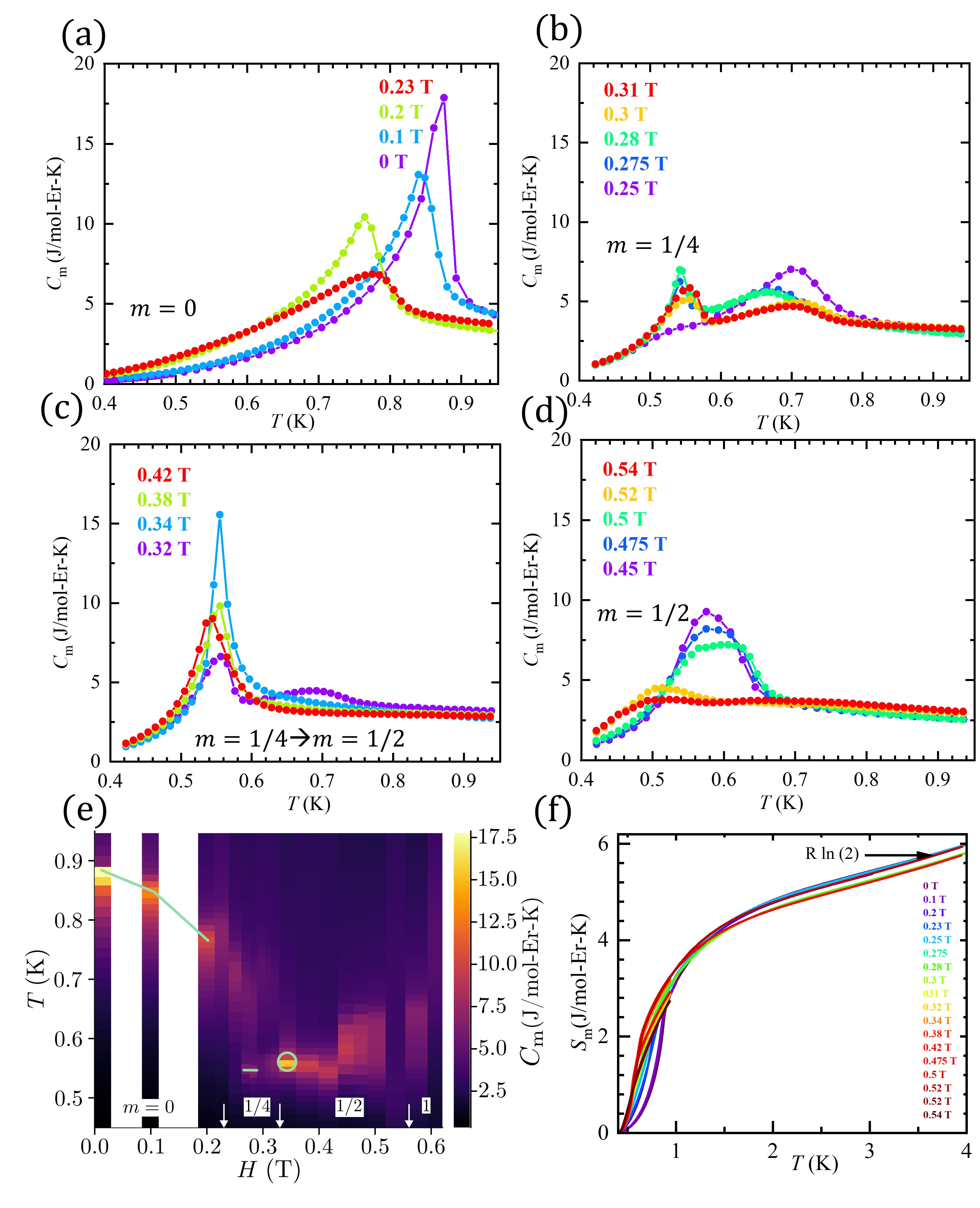}
\caption{Field-dependent $C_m(T)$ of \ErSSL{} single crystal for $\mathbf{H} \parallel$ [001]. (a--d) $C_m(T)$ grouped by fields corresponding to plateau phases, illustrating transitions and peak evolution. (e) Phase diagram in the $(\mathbf{H},T)$ plane as revealed by a color plot of $C_m$. The green lines connect the phase transitions detected in the $m=0$ and $m=1/4$ plateaus, while the green circle emphasizes an isolated critical point at the boundary between the $m=1/4$ and $m=1/2$ plateaus. The arrows mark the positions of the first-order transitions between the plateaus as deduced from magnetization at low temperature.  (f) $S_m$ calculated by $\int_{0.4~\mathrm{K}}^{T_{\mathrm{max}}} (C_m/T) \, dT$, showing similar entropy loss across plateau phases and exceeding $R\ln(2)$ above 3.5 K, indicating contributions from a low-lying CEF level.}
\label{hc}
\end{figure*}
%%%%%%%%%%%%%%%%%%%%%%%%%%%%%%%%%%%%%%%%%%%%%%%%%%

In FIG.~\ref{magnetization} (d), the low-temperature magnetic susceptibility \( \chi_{[001]} \) initially increases to 2 K, then decreases to around 0.85 K, indicating an antiferromagnetic transition. In contrast, \( \chi_{[010]} \) increases sharply from 2 K to 0.85 K and remains constant below this transition temperature (see FIG.~\ref{magnetization}(e)), indicating ferromagnetic behavior in the [010] direction. In FIG.~\ref{magnetization} (f), we present \(M_{[001]}(H)\) below the ordering temperature at 0.3~K and the corresponding \(d M/d H\) curve. Interestingly, we observe the emergence of multiple fractional magnetization plateaus as a function of field. These plateaus correspond to distinct magnetic phases that are stable within different field ranges and appear at 1/4 and 1/2 of the saturation magnetization. The critical fields associated with these plateaus are identified by differentiating \(M_{[001]}(H)\) with respect to \(H\) and using the peaks of the \(d M/d H\) curve to determine the start and end of the plateaus. The critical fields associated with the start and end of the 1/4 plateau are 0.23~T to 0.33~T, and for the 1/2 plateau, they are at 0.33~T and 0.56~T. It should be noted that the critical fields for these plateaus are about two orders of magnitude smaller than those required for \SCBO{}, making them more accessible for investigation using standard experimental techniques. For a magnetic field applied along the \( [010] \) direction at 0.3 K, \(M_{[010]}(H)\) does not exhibit any magnetization plateaus. However, as shown in the inset of FIG.~\ref{magnetization}(e), \(M_{[010]}(H)\) undergoes a rapid increase at a rate of 21.8 $\mu_B$/T for fields between 0 T and 0.07 T, reaching approximately 1.5 $\mu_B$ and following a sigmoid-like curve. Remarkably, within the narrow field range \( H \in [-0.07, 0.07] \, \text{T} \), the moment shifts linearly from \(-1.5 \, \mu_B\) to \(+1.5 \, \mu_B\) without any observable hysteresis. Beyond 0.1 T, \(M_{[010]}(H)\) gradually approaches 4 $\mu_B$, with no further transitions detected.

%\subsection{Heat Capacity} 

The specific heat of the \ErSSL{} single crystal was measured under magnetic fields applied parallel to the \([001]\) direction. Measurements were conducted at multiple fields ranging from 0 to 1~T, with fine steps in both field and temperature to capture the behavior across the narrow plateau phases. To optimize measurement time in the sub-Kelvin range, data collection primarily focused on the 0.4~K to 0.9~K range for most fields, while a few representative fields—one from each plateau—were measured over the full 0.055~K to 4~K range to ensure no significant features were overlooked. Panels in FIG.~\ref{hc}(a--d) display the magnetic heat capacity \(C_m\), grouped by field values corresponding to different plateau phases. In the \(m=0\) phase at 0~T, a sharp peak is observed at 0.88~K, consistent with the antiferromagnetic transition identified in the magnetic susceptibility. As the field increases to 0.23~T, this peak shifts to lower temperatures (down to \(\approx\) 0.75~K), decreases in intensity, and broadens significantly. In the \(m=1/4\) regime, the broad peak continues to shift to lower temperatures while becoming even wider. Additionally, a new sharp feature emerges around \(\approx\) 0.55~K, leading to a two-peak structure: a sharp peak at \(\approx\) 0.55~K and a broader peak at \(\approx\) 0.67~K. As the system transitions from the \(m=1/4\) phase to the \(m=1/2\) phase, these two peaks merge into a single, sharper peak centered at 0.55~K. Within the \(m=1/2\) phase, the sharp peak gradually flattens into a plateau-like feature at 0.5~T. Finally, in the fully polarized state, these features are completely suppressed. To better illustrate the field and temperature dependence of the specific heat, a color plot of \(C_m(T,H)\) is presented in FIG.~\ref{hc}(e). Additionally, the magnetization \(M(H)\) at 0.3~K is shown to correlate the \(C_m/T\) features with the different plateau phases. The magnetic entropy \(S_m\) was determined by integrating the \(C_m/T\) curves up to the highest available temperature. As depicted in FIG.~\ref{hc}(f), the entropy loss remains uniform across all plateau regimes. Furthermore, the magnetic entropy \(S_m\) exceeds \(R\ln(2)\), indicating additional contributions from the population of a low-lying crystal electric field (CEF) excited level above 3~K. This result aligns with the CEF analysis discussed in the following section.
%%%%%%%%%%%%%%%%%%%%%%%%%%%%%%%%%%%%%%%%%%%%%%%%%%%%%%%

%\subsection{Crystal Electric Field (CEF) Analysis}

%%%%%%%%%%%%%%%%%%%%%%%% FIG 3 %%%%%%%%%%%%%%%%%%%%%%%%
\begin{figure*}[]
\centering
\includegraphics[width=\textwidth]{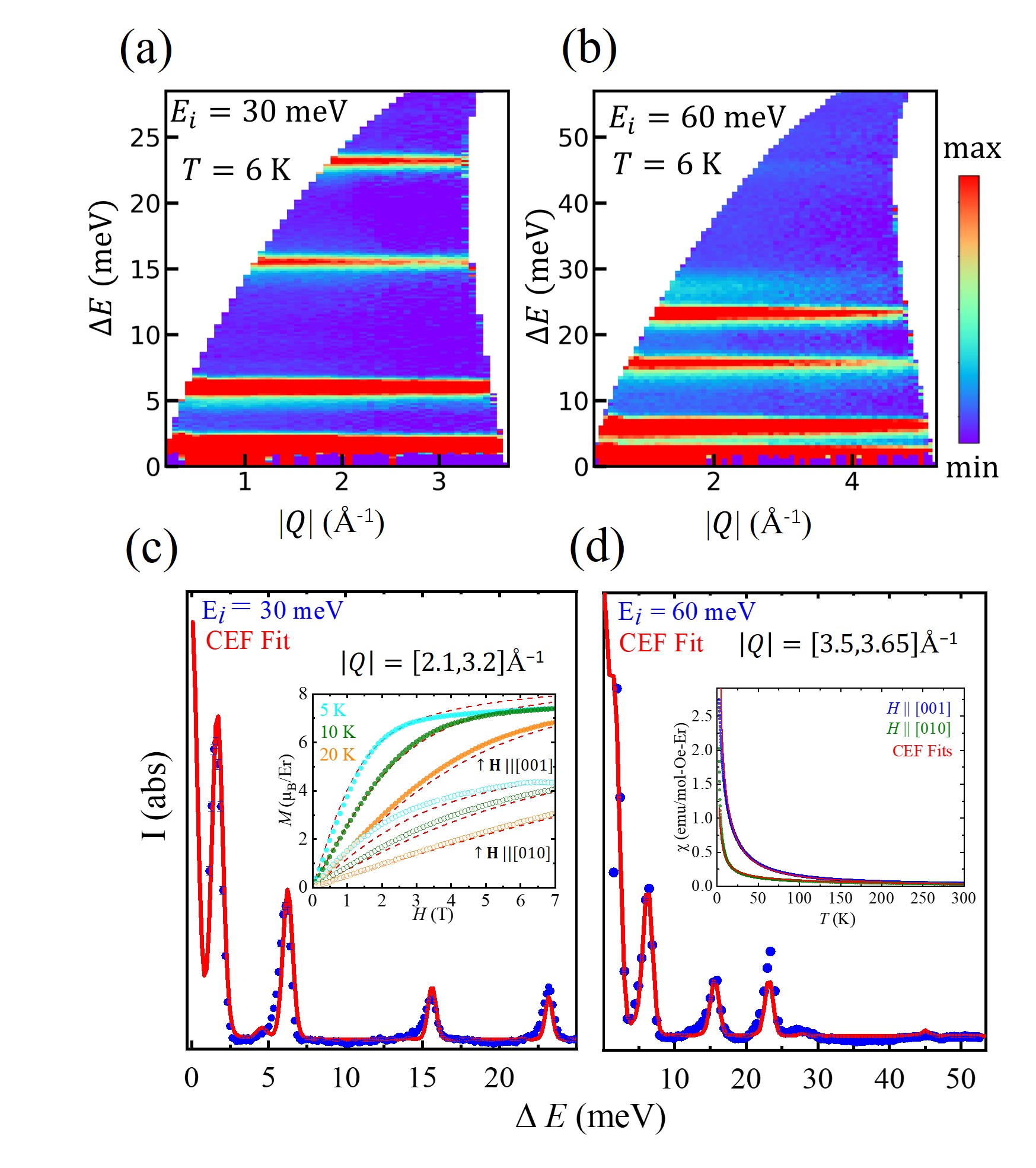}
  \caption{CEF excitations of \ErSSL{} obtained from inelastic neutron scattering. (a, b) Representative INS spectrum at 6 K using incident energies $E_i$ of 30 meV and 60 meV, respectively, collected at SEQUOIA. The non-magnetic \LuSSL{} spectrum is subtracted for background, removing phonon contributions. (c,d) Constant $Q$ cuts of the experimental data spanning $\Delta E$ from 0 to 25 meV with $E_i=30$ meV and from 0 to 55 meV with $E_i=60$ meV. The red lines represent the CEF fits to the experimental data. The inset in (d) and (e) shows the CEF fits to the isothermal magnetization and magnetic susceptibilities for $[001]$ and $[010]$ directions.
}\label{CEF}
%\endminipage\hfill
\end{figure*}
%%%%%%%%%%%%%%%%%%%%%%%%%%%%%%%%%%%%%%%%%%%%%%%%

To investigate the CEF levels of \ErSSL{}, inelastic neutron scattering (INS) measurements were performed on high-purity powder samples. According to Hund's rules, the ground-state multiplet of the \ch{Er^{3+}} ion is \(^4I_{15/2}\), with a \(2J+1 = 16\)-fold degeneracy that splits into 8 doublets under the influence of the crystal field. The INS experiments revealed energy bands corresponding to transitions between these split CEF levels. FIG.~\ref{CEF}(a,b) shows representative INS spectra at 6 K using \(E_i = 30\) meV and 60 meV. Six out of seven expected transition bands were identified, with the lowest CEF excitation observed at 1.58 meV. Measurements with higher energy coverage (\(E_i = 150\) meV) did not reveal any additional CEF levels. The CEF scheme was determined by fitting the single-ion CEF Hamiltonian as detailed in the Methods section. FIG.~\ref{CEF}(c) and (d) present constant-\(Q\) cuts of the experimental data in the energy ranges \(\Delta E = 0\)-25~meV (\(E_i = 30\)~meV) and \(\Delta E = 0\)-50~meV (\(E_i = 60\)~meV), respectively. The experimental data are overlaid with the CEF fit (red line), which closely aligns with the observed spectra. Minor discrepancies in intensity may arise from unaccounted phonon contributions due to imperfect background subtraction. The insets of FIG.~\ref{CEF}(c,d) show the experimental $M(H)$ and $\chi(T)$ along the $[001]$ and $[010]$ directions, overlaid with the CEF-calculated curves. The CEF calculations closely match the experimental data, capturing the anisotropy. Minor deviations may result from diamagnetic effects at higher temperatures and exchange interactions at lower temperatures. The fitted \(B^n_m\) parameters, energy levels, eigenvectors, and \(g\)-tensor components are provided in the Appendix \ref{CEF_app}. The analysis reveals that the easy axis of the magnetic moment lies in the plane defined by the \(c\)-direction and the dimer bond.
%%%%%%%%%%%%%%%%%%%%%%%%%%%%%%%%%%%%%%%%%%%%%%%%%%%%%%%

%\subsection{Powder neutron diffraction results}

\begin{figure*}[]
\centering
\includegraphics[width=1\textwidth]{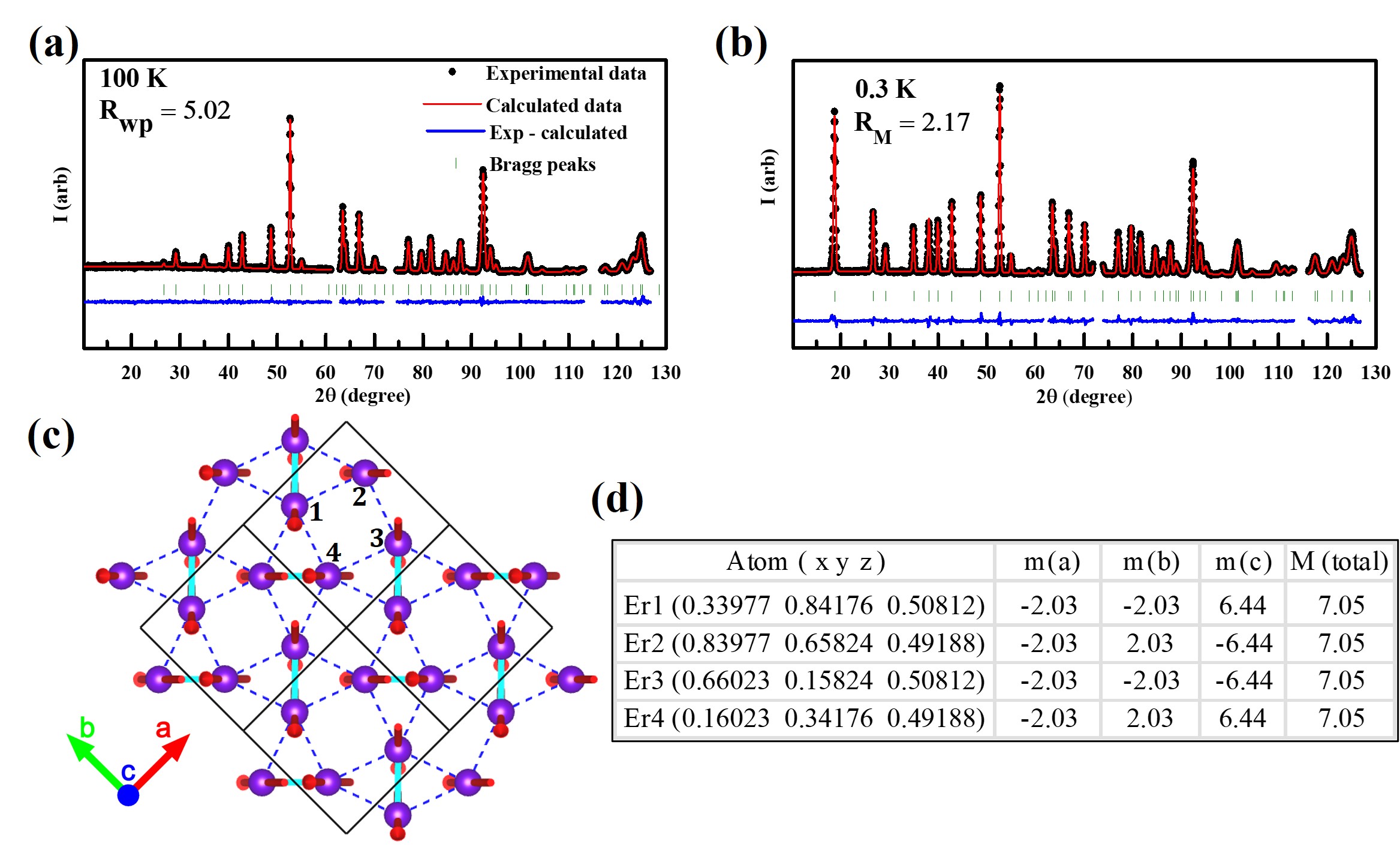}
  \caption{Magnetic structure characterization at 0 T in the ordered state of \ErSSL{}. (a) NPD spectrum at 100 K, illustrating a successful refinement within the space group \( P\overline{4}2_1m \) with a low \(R_{wp}\) of 5.02\%. Points around \( 2\theta= (61.4^\circ-62.3^\circ, 72.0^\circ-73.8^\circ, 113.1^\circ-115.9^\circ ) \) were excluded from the refinement due to prominent Al can peaks. (b) NPD spectra at 0.3 K, showcasing the emergence of magnetic Bragg peaks atop the nuclear Bragg peaks. Fit from the proposed magnetic model is shown in red and agrees well with the data achieving a low \( R_M \) value of 2.17. (c) Schematic representation of magnetic structures of \ErSSL{}, showing the magnetic unit cell outlined in black, with Er atoms labeled 1, 2, 3, and 4, and corresponding moments are listed in table in panel (d). Each dimer pair has an antiferromagnetic arrangement of spins with a slight canting along the dimer bond. (d) Magnetic moments (in $\mu_B$) and fractional atomic coordinates of Er atoms. The lattice constants are $a=b=$\SI{7.377}{\angstrom} and $c=$\SI{4.777}{\angstrom}.}
\label{npd}
%\endminipage\hfill
\end{figure*}

To determine the magnetic structure of \ErSSL{} below the transition at \( T_N = 0.9 \) K, two neutron powder diffraction (NPD) spectra were obtained: one at 0.3~K below the transition and another in the paramagnetic phase at 100~K. Initial Rietveld refinement of the Bragg peaks at 100 K successfully refined all peaks within the space group \( P\overline{4}2_1m \), yielding a low $R_{wp}$ of 5.02~\% as shown in FIG.~ \ref{npd} (a). At 0.3~K, magnetic Bragg peaks became evident on top of the nuclear Bragg peaks, as illustrated in FIG.~ \ref{npd} (b), pointing to a magnetic propagation vector \textbf{k} = (0, 0, 0). High-intensity peaks at \( 2\theta = 18.8^\circ \) and \( 26.7^\circ \), indexed as (1 0 0) and (1 1 0), hint at a significant magnetic moment outside the a-b plane in \ErSSL{}. A magnetic symmetry analysis was performed to investigate the maximal magnetic subgroups compatible with the space group \( P\overline{4}2_1m \) with propagation vector \(\mathbf{k} = (0, 0, 0)\) using the Bilbao Crystallographic Server (MAXMAGN program \cite{perez2015symmetry}).  There are six \(k\)-maximal magnetic subgroups identified. The best fitting model is given by \(P2_1 2_1 2'\) (No. 18.19), where all Er atoms and the corresponding ordered moments are described by a single Wyckoff site. In this subgroup, the magnetic moment components \(m(a)\), \(m(b)\), and \(m(c)\) are independent. Allowing these components to vary freely during the refinement process resulted in \(m(a)\) and \(m(b)\) being comparable. The CEF analysis, as discussed in the previous section, indicates that the easy-axis anisotropy lies in the plane defined by the $c$ axis and dimer bond direction, and thus we introduced the constraint \(|m(a)| = |m(b)|\) into the model. The model aligns well with the data, showing a magnetic R-factor \(R_{M}\) value of 2.17 as shown in FIG.~\ref{npd}(b). The magnetic structure is characterized by antiferromagnetic dimer pairs with a dominant moment in the crystallographic \(c\) direction as illustrated in FIG.~ \ref{npd} (c). The ordered moment tilts away from the \(c\) axis by \(24^\circ\) towards the dimer bond direction. Each Er ion exhibits an ordered moment of \(\mu \approx 7 \mu_B\). The model predicts a net ferromagnetic component in the \textit{ab} plane.  This agrees with the magnetic susceptibility behavior where \( \chi_{[001]} \) shows an antiferromagnetic transition but \( \chi_{[010]} \)  shows a ferromagnetic transition at 0.83 K. The deduced magnetic moment components from the fit for all four atoms within the unit cell are documented in FIG.~\ref{npd}(d). We compare the magnetic moment determined in this way to the one predicted in the ground-state doublet of the CEF Hamiltonian (see appendix \ref{section:zeeman_vv}):
%%%%%%%%%%%%%%%%%%% EQUATION 1 %%%%%%%%%%%%%%%%%%%%%%
\begin{equation}
    \begin{aligned}
    \boldsymbol{\mu}_{1,\text{CEF}}=g_J\bra{+}\mathbf{J}_1\ket{+}&=(-0.58,-0.58,4.96)\mu_B\\
    ||\boldsymbol{\mu}_{1,\text{CEF}}||&=5.03\mu_B\label{eq:CEF_moment}
    \end{aligned}
\end{equation}
%%%%%%%%%%%%%%%%%%%%%%%%%%%%%%%%%%%%%%%%%%%%%%%%%

for the \ce{Er^3+} site labeled as 1 in FIG. \ref{npd}, where $g_J=1.2$ is the Landé g-factor (the magnetic moment of the remaining sites of the unit cell are related to that one by symmetry). NPD and CEF both predict magnetic moments pointing in the dimer plane and with dominant component along the $z$ direction, but show a significant difference in the size of the moment (7$\mu_B$ versus 5$\mu_B$). A possible solution for this discrepancy is the enhancement of the ordered magnetic moment by exchange-induced mixing with excited CEF levels. This phenomena  has previously been reported in non-Kramers rare earth compounds such as \ce{Tb2Ti2O7} \cite{liu2024theoryrareearthkramersmagnets} and \ce{PrRu2Si2} \cite{mulders1997}, where the rare-earth ions develop a magnetic moment in spite of having a singlet CEF ground state.

%%%%%%%%%%%%%%%%%%%%%%%% FIG 4 %%%%%%%%%%%%%%%%%%%%%%%%
%%%%%%%%%%%%%%%%%%%%%%%%%%%%%%%%%%%%%%%%%%%%%%%%

%%%%%%%%%%%%%%%%%%%%%%%%%%%%%%%%%%%%%%%%%%%%%%%%%%%%%%%%%%%%%%%%%%%%%%%%%%%%%%%%
%\subsection{Single Crystal Neutron scattering results -- Elastic}

\begin{figure*}[]
\centering
\includegraphics[width=1\textwidth]{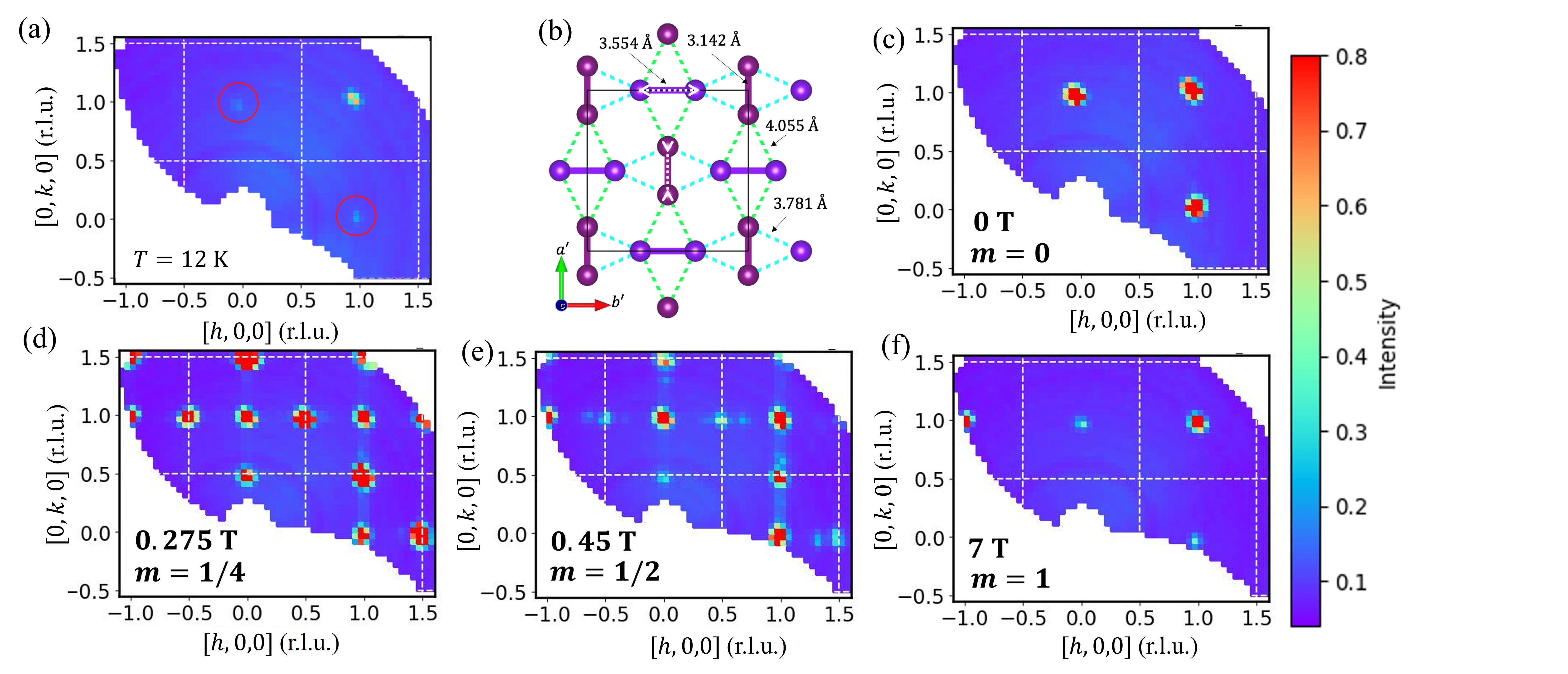}
\caption{Single-crystal neutron diffraction results for the $(h, k, 0)$ plane obtained by integrating over the elastic energy range $-0.05 < \Delta E < 0.05$ meV and along $l$ with $-0.1 < l < 0.1$. The Brillouin zones are outlined with white dashed lines for reference. (a) Diffraction data measured at $T = 12$ K in the paramagnetic phase revealing weak intensities at the forbidden $(1,0,0)$ and $(0,1,0)$ reflections. (b) Structural distortion in the \(Cmm2\) space group, described by a basis change \(\mathbf{a}' = \mathbf{a} + \mathbf{b}\), \(\mathbf{b}' = -\mathbf{a} + \mathbf{b}\), \(\mathbf{c}' = \mathbf{c}\), and an origin shift \(\left(0, \tfrac{1}{2}, 0\right)\). Here, one dimer sublattice (purple) expands while the other sublattice (magenta) contracts, as indicated by white arrows. This distortion results in anisotropic inter- and intra-dimer bond lengths. (c-f) Magnetic diffraction data collected at $T = 60$ mK, obtained by subtracting the paramagnetic dataset shown in (a). Panels depict different magnetization phases: (c) $m=0$ phase at $H=0$ T, (d) $m=1/4$ phase at $H=0.275$ T, (e) $m=1/2$ phase at $H=0.45$ T, and (f) $m=1$ phase at $H=7$ T.}
\label{sc_diff}
\end{figure*}

Single-crystal neutron scattering measurements were performed using the HYSPEC spectrometer\cite{Zaliznyak_2017} at Oak Ridge National Laboratory (ORNL) to investigate the emergence of the magnetic structure under a magnetic field. The sample was oriented in the $[h,k,0]$ scattering plane with the field along the crystallographic $c$ direction. In the paramagnetic phase at 12 K, weak $(1,0,0)$ and $(0,1,0)$ reflections were observed. For the tetragonal space group $P\overline{4}2_1m$, reflections such as $(h,0,0)$ and $(0,k,0)$ with odd $h$ and $k$ are forbidden by the $2_1$ symmetry element. The $(1,0,0)$ reflection is approximately 5\% of the intensity of the allowed $(1,1,0)$ reflection, explaining why it was undetectable in NPD due to background noise. To rule out the possibility of multiple scattering or higher-order wavelength contamination, a separate neutron single-crystal diffraction experiment was conducted (see Appendix \ref{Wand}), confirming that the forbidden peaks are intrinsic to the crystal structure. Two possible subgroups of $P\overline{4}2_1m$ that allow these reflections were identified: $P\overline{4}$ (No. 81) and $Cmm2$ (No. 35). Among these, only $Cmm2$ was found to be consistent with the observed magnetization plateaus of this compound (see below). The limited number of observed reflections made it impractical to refine the magnetic structure fully. The low intensity of these forbidden peaks indicates minimal distortion, and the lattice constants satisfy $a = b$ within the experimental resolution, supporting the validity of the high-symmetry $P\overline{4}2_1m$ approximation. To approximate the magnitude of the distortion, the NPD data at 2 K was refined using the $Cmm2$ space group under the condition that it reproduces the observed intensity ratio $I(100)/I(110) = 0.05$, thus constraining the distortion within the experimental resolution. In the refined structure (see FIG.\,\ref{sc_diff}), Er\(^{3+}\) ions were found to displace along the dimer bonds, splitting the average intra-dimer bond length of 3.3468 \AA{} into 3.142 \AA{} and 3.554 \AA{}, leading to anisotropy in the intra-dimer bonds. 

In FIG.~\ref{sc_diff}, we present the diffraction channel obtained by integrating the energy around the elastic line. The paramagnetic background at 12 K has been subtracted from the 60 mK data to isolate the magnetic contribution. In zero field, the diffraction pattern in the $[h, k, 0]$ scattering plane displays magnetic peaks indexed by the magnetic propagation vector $\mathbf{k} = (0, 0, 0)$, in agreement with NPD data. Upon applying a field of $H=0.275$ T, corresponding to $1/4$ plateau, additional magnetic Bragg peaks emerge, indexable by $\mathbf{k} = (0.5, 0, 0)$ and $(0, 0.5, 0)$. This indicates a cell doubling in the $a$ and $b$ directions with possibility of domain formation. Increasing the field to $0.45$ T, corresponding to $1/2$ plateau, results in a loss of intensity of $\mathbf{k} = (0.5, 0, 0)$ and $(0, 0.5, 0)$ peaks with the emergence of faint $(1/3, 0, 0)$ and $(2/3, 0, 0)$ satellite peaks. Finally, in the polarized state ($m=1$) at $7$ T, only peaks at $\mathbf{k} = (0, 0, 0)$ remain. Although a full magnetic refinement could not be performed due to limited access to the peaks in the $[h,k,0]$ plane, theoretical calculations discussed below propose plausible magnetic structures for the $m=1/4$ and $m=1/2$ phases, which agree with the observed peaks.

%%%%%%%%%%%%%%%%%%%%%%%%%% FIG 5 %%%%%%%%%%%%%%%%%%%%%%%%%%%%%%%%%%%%%
%%%%%%%%%%%%%%%%%%%%%%%%%%%%%%%%%%%%%%%%%%%%%%%%%%%%%%%%%%%%%%%%%%%

%%%%%%%%%%%%%%%%%%%%%%%%%%%%%%FIG 6 %%%%%%%%%%%%%%%%%%%%%%%%%%%%%%
\begin{figure*} \centering \includegraphics[width=1\linewidth]{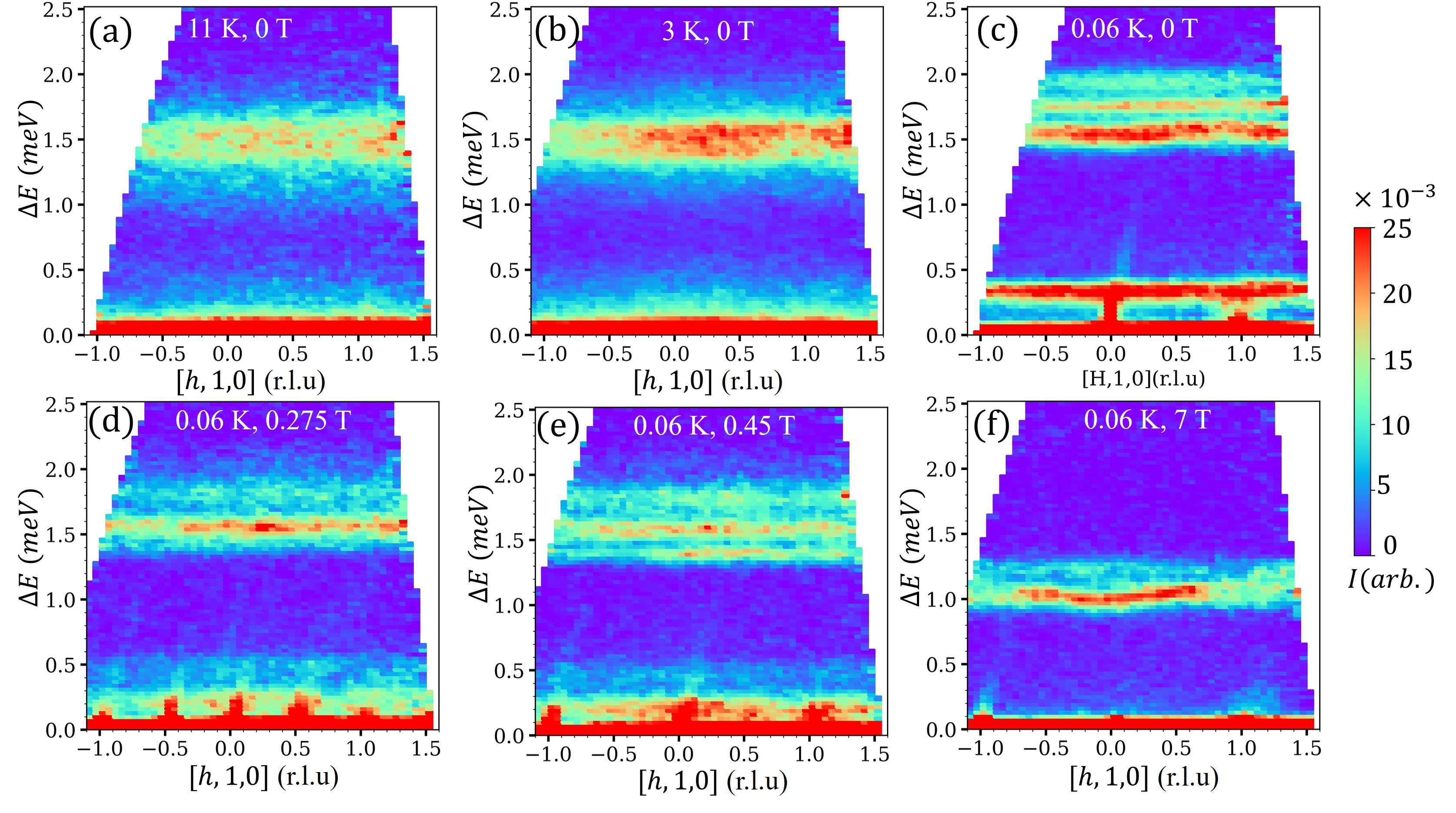} \caption{Scattering intensities as a function of energy transfer for \ErSSL{}. Panels (a–f) present INS data for a single crystal with $H \parallel [001]$. At 11 K and 3 K (panels a and b), the paramagnetic phase has only a 1.5 meV CEF band. At 0.06 K and 0 T (panel c), a new flat-band excitation appears at 0.35 meV, indicating a gapped feature. With increasing field (panels d and e), corresponding to $m=1/4$ and $m=1/2$ phases, this band broadens, redistributing intensity between 0 to 0.5 meV. At 7 T (panel f), the excitation moves to $\approx$1 meV, reflecting contributions from ferromagnetic order and CEF effects. Note that the sharp vertical features extending from the elastic channel originate from the magnetic Bragg peaks.} 
\label{fig:INS_hyspec} 
\end{figure*}
%%%%%%%%%%%%%%%%%%%%%%%%%%%%%%%%%%%%%%%%%%%%%%%%%%%%%%%%%%%%%%%%%

%\subsection{Single Crystal Neutron scattering results -- Inelastic}

The low-energy excitations observed in the inelastic channel are presented in Fig.~\ref{fig:INS_hyspec}. In the paramagnetic phase at 11~K and 3~K, a CEF band centered around \(\approx 1.6\)~meV is observed, consistent with previous CEF analyses. At 3~K, as the temperature approaches the ordering temperature, this 1.6~meV band begins to split, possibly due to the molecular field generated by the onset of long-range spin correlations. At 0.06~K, a new flat-band excitation emerges at 0.35~meV, showing no detectable dispersion within the energy resolution. This energy band appears to be well-gapped. Under applied magnetic fields of 0.275~T and 0.45~T, corresponding to the \(m=1/4\) and \(m=1/2\) phases, the flat energy band broadens, with its intensity redistributing and shifting to 0.5~meV and 0.25~meV, respectively. Notably, magnetic Bragg peaks from the elastic line extend into the inelastic channel, creating continuum-like features at the corresponding \((H, K, L)\) positions. Although low-energy excitation remains confined below 0.5~meV, the CEF Kramers doublet at 1.6~meV evolves across these phases (see Appendix \ref{CEF_app} for line cuts), likely influenced by the development of a local ordered moment. Finally, at 7~T, the excitation shifts upward to approximately 1--1.25~meV, exhibiting weak dispersion. This behavior likely arises from a combination of ferromagnetic spin order and field-dependent modifications to the CEF levels.

%%%%%%%%%%%%%%%%%%%%%%%%%%%%%%% THEORY %%%%%%%%%%%%%%%%%%%%%%%%%%%%%%
\section{Theoretical Results}

The two ground states of the CEF Hamiltonian have their largest components along the fully polarized states  $\ket{\pm15/2}$ (see table II in SI), suggesting that a good starting point to describe the magnetic properties of \ErSSL{} is the Ising model:

%%%%%%%%%%%%%%%%%%%%%%%%%% EQUATION 2%%%%%%%%%%%%%%%%%%%%
\begin{equation}
    H_{\text{Ising}}=\sum_{i,j}J_{ij}^{zz}\sigma_i\sigma_j -h\sum_{i}\sigma_i,\,\,\sigma_i\in\{-1,1\} \label{eq:general_ising_model}
\end{equation}
%%%%%%%%%%%%%%%%%%%%%%%%%%%%%%%%%%%%%%%%%%%%%%%%%%%%%%%%%

This is further supported by the absence of significant intermediate phases between the fractional magnetization plateaus (FIG. \ref{magnetization}) and by the dispersionless mode in inelastic neutron scattering (FIG. \ref{fig:INS_hyspec}). Recent studies on the theory of Rare-Earth Shastry-Sutherland compounds \cite{liu2024theoryrareearthkramersmagnets} have also shown that intra-dimer quantum fluctuations allow the existence of a $m=0$ plateau at small fields only if the zero-field ground state is a singlet dimer product state. Given that both an $m=0$ plateau and long-range magnetic order are observed in \ErSSL{}, the zero-field ground state may not be a dimer product state and instead is more likely to be an Ising order stabilized by farther-range interactions. The possible effect of residual off-diagonal interactions will be discussed later in this section.

The most remarkable property of \ErSSL{} is the existence of 1/4 and 1/2 magnetization plateaus in a field $\mathbf{H}\parallel[001]$. This is in apparent contradiction with the Ising description since the Ising model on the SSL with intra-dimer coupling $J_1$ and inter-dimer coupling $J_2$, is known to have a single fractional magnetization plateau at $m=1/3$ \cite{dublenychGroundStatesIsing2012}. To resolve this discrepancy within the Ising approximation, the model may be modified in two ways: by considering farther-range interactions or by allowing the Ising couplings to not be fully symmetric under the space group \SpaceGroup{}. 

We first consider the effect of long-range dipolar interactions on top of intra and inter-dimer exchange interactions: 

%%%%%%%%%%%%%%%%%%%%EQUATION 3%%%%%%%%%%%%%%%%
\begin{equation}
\begin{aligned}
    H_{\text{Ising,LRI}}= & \sum_{\langle i,j\rangle_1}J_{1}\,\sigma_i\sigma_j
    +\sum_{\langle i,j\rangle_2}J_2\,\sigma_i\sigma_j \\
    & +\sum_{i,j}\frac{J_{\text{d}}}{|\mathbf{r}_i-\mathbf{r}_j|^3}\sigma_i\sigma_j
    -h\sum_{i}\sigma_i\ .
\end{aligned}
\label{eq:long_range_ising_model}
\end{equation}
%%%%%%%%%%%%%%%%%%%%%%%%%%%%%%%%%%%%%%%%%%

The ground state of $H_{\text{Ising,LRI}}$ has been determined by systematic enumeration of magnetic unit cells and constrained energy minimization within each unit cell \cite{koziolSystematicAnalysisCrystalline2023,koziol_quantum_2024}.
For methodological details, see section~\ref{sec:unit_cell_technique}.
A typical result for the magnetization curve of this model is shown in Fig.~\ref{fig:devil_staircase}.
This approach shows that dipolar interactions lead to the stabilization of an infinite sequence of fractional magnetization plateaus (a devil's staircase), and that the 1/3 plateau remains the dominant one, in stark contrast with the properties of \ErSSL{}.
These two aspects persist independently of the specific choice of couplings in $H_{\text{Ising,LRI}}$.
%%%%%%%%%%%%%%%%%%%%%%%%%%%%%%%%%%%%%%%%%%%%%

\begin{figure}
    \centering
    \includegraphics[width=\linewidth]{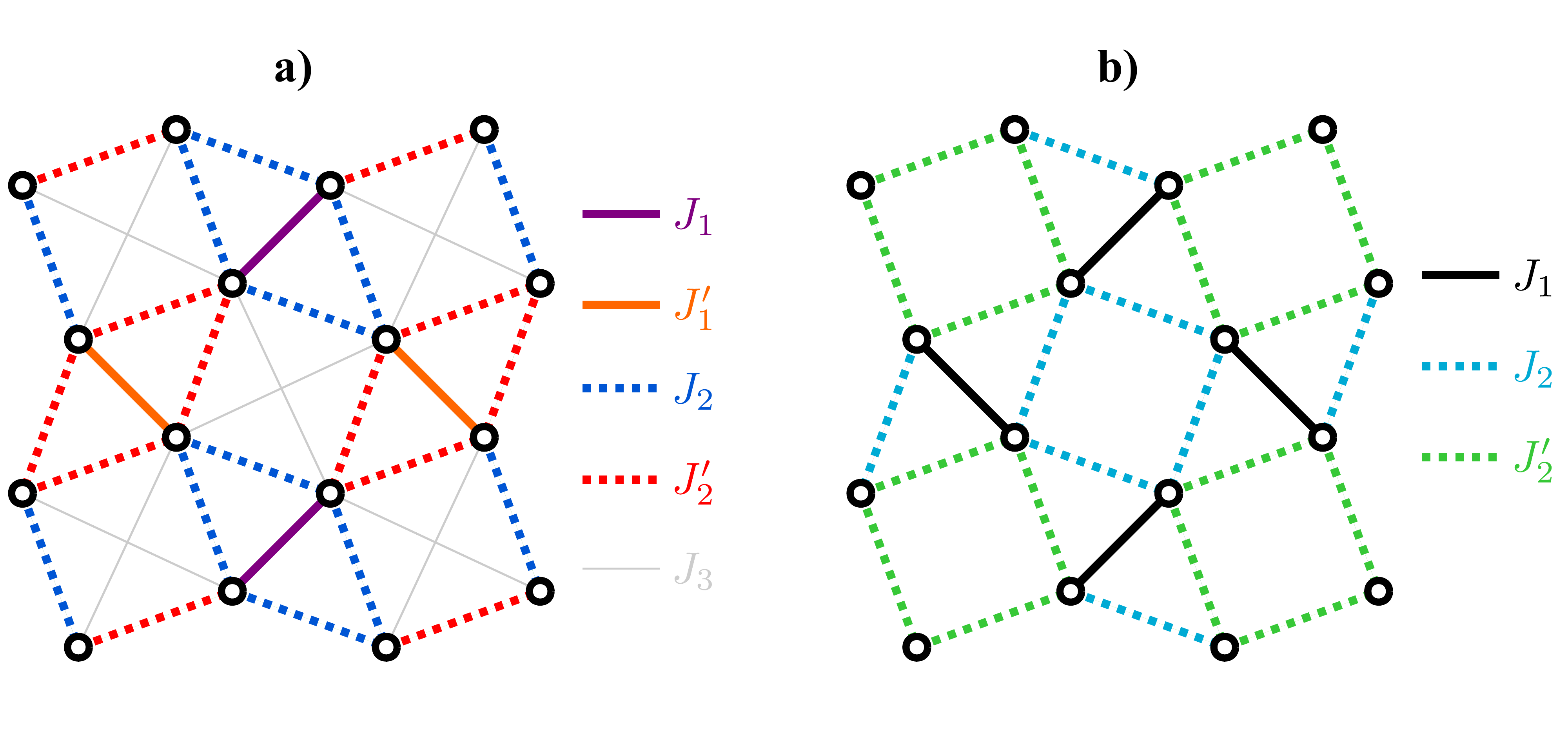}
    \caption{Inequivalent couplings on the anisotropic Shastry-Sutherland lattice Ising model with a) $Cmm2$ and b) $P4$ symmetry groups.}
    \label{fig:subgroups_ising_ssl}
\end{figure}

%%%%%%%%%%%%%%%%%%%%%%%%%%%%%%%%%%%%%%
The other possibility is to allow for the existence of a spatial anisotropy in the Ising couplings $J_{ij}^{zz}$. This hypothesis is supported by the single crystal neutron diffraction (FIG. \ref{sc_diff}-b),  where a weak but noticeable intensity is observed at two reflections - (100) and (010) - which are forbidden in the \SpaceGroup{} space group. These forbidden reflections 
persist at temperatures significantly higher than the energy scale of magnetic interactions and hence are likely of nuclear origin. The observed reflections are compatible with two of the maximal subgroups of \SpaceGroup{}: $P4$ and $Cmm2$. The distortions compatible with each of these subgroups induce a spatial anisotropy in the SSL Ising couplings, as illustrated in FIG. \ref{fig:subgroups_ising_ssl}.\par

%%%%%%%%%%%%%%%%%%%%%FIG 7%%%%%%%%%%%%%%%%
%%%%%%%%%%%%%%%%%%%%%%%%%%%%%%%%%%%%%%%

\begin{figure}
    \centering
    \includegraphics[width=\linewidth]{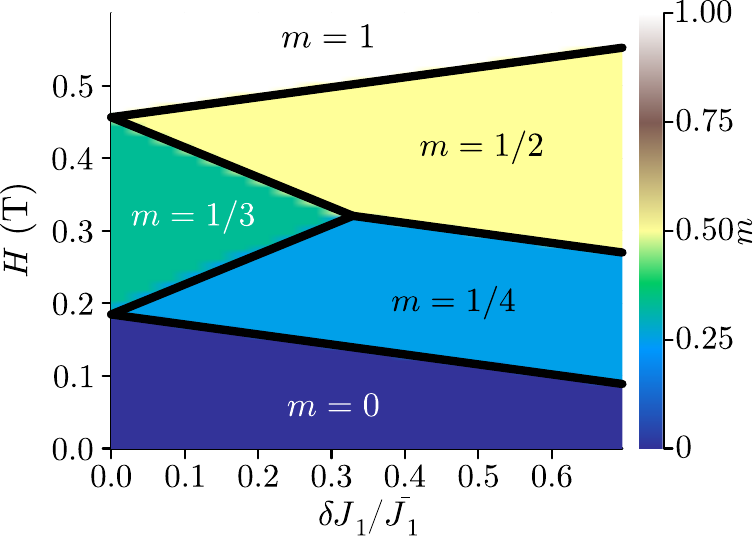}
    \caption{Ground state phase diagram of the ASSLIM as a function of magnetic field and relative anisotropy ($\delta J_1=J_1'-J_1$, $\bar{J_1}=(J_1+J_1')/2$). The isotropic intra-dimer interaction $\bar{J_1}=0.079$ meV and inter-dimer interaction $J_2=0.013$ meV were chosen to be consistent with the parameter estimates written in Fig. \ref{fig:mag_theory_exp}, while the no third-neighbor interaction $J_3$ was considered.}
    \label{fig:phase_diagram_anisotropy}
\end{figure}

\begin{figure*}
    \centering
    \includegraphics[width=\linewidth]{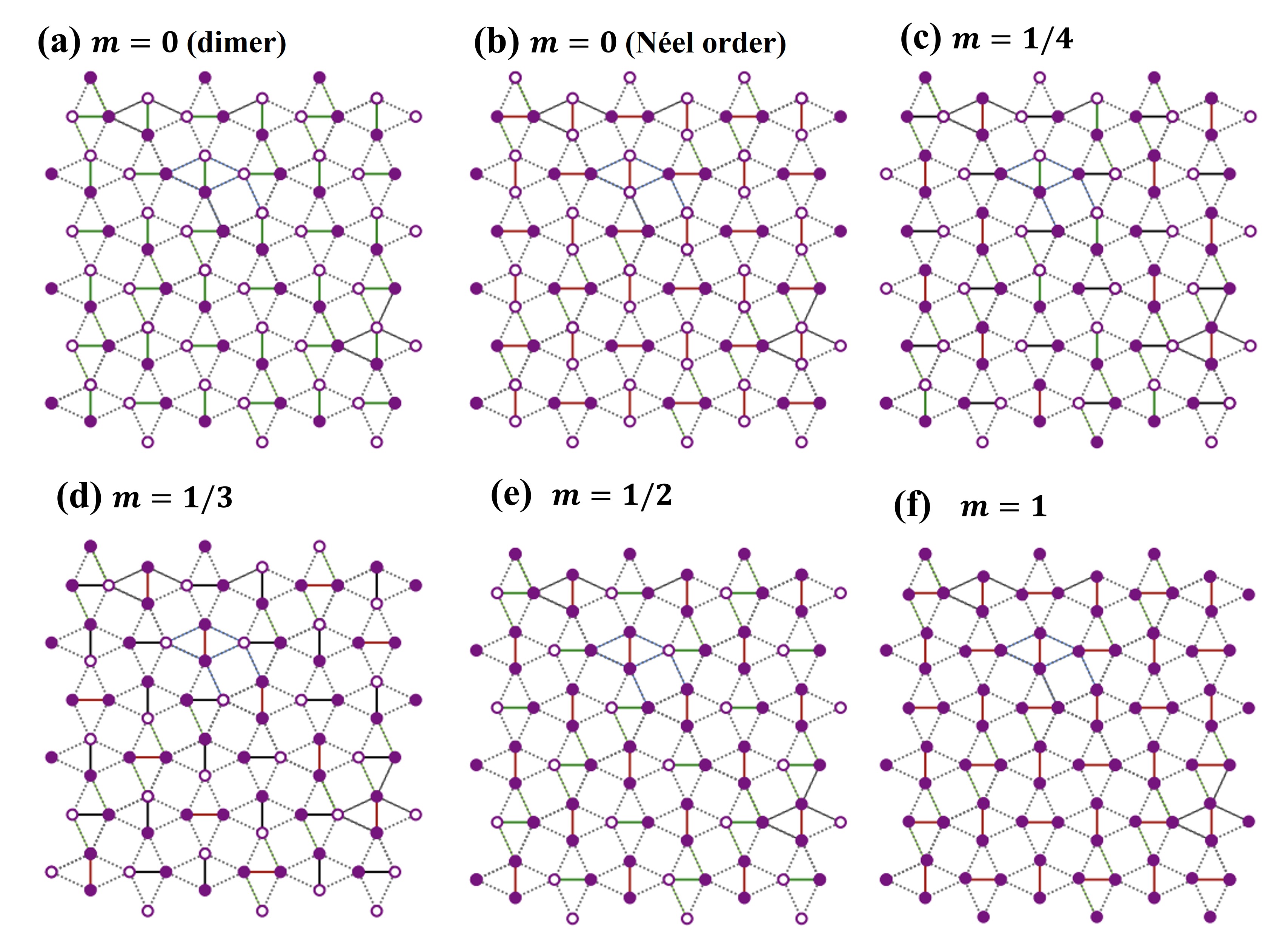}
    \caption{Trial configurations for ground-state magnetic structures across various magnetizations. Filled circles represent positive spins ($\sigma=1$), empty circles represent negative spins ($\sigma=-1$) and ferromagnetic dimer bonds are highlighted in red. The magnetic unit cell is highlighted as a gray rectangle. If $J_3=0$ the bonds highlighted in green can be independently flipped with no energy change, giving rise to an infinitely degenerate family of ground states. The energy of the configurations are as follows: (a) Stripe order, $m=0$, energy $E=-(J_1 + J_1')/4 -J_{3}$. (b) Néel order, $m=0$, energy $E=-2J_2+(J_1 + J_1')/4 + J_{3}$. (c) $m=1/4$, energy $E=-h/4- J_2/2 - J_1/4$. (d) $m=1/3$, energy $E=-h/3 -2J_2/3 - (J_1 + J_1')/12 +J_{3}/3$. (e) $m=1/2$, energy $E=-h/2 + (J_1' - J_1)/4$. (f) $m=1$, energy $E=-h + 2J_2 + (J_1 + J_1')/4 + J_{3}$. $J_3>0$ stabilizes long-range order in the $m=0$ and $m=1/2$ plateaus, while a sub-extensive degeneracy remains in the $m=1/4$ plateau.}
    \label{fig:candidate_states}
\end{figure*}

In order to test these anisotropic Ising models, the ground-state energy and magnetization have been determined by comparing ground-state energy lower bounds (obtained as described in section \ref{section:gs_lower_bounds}) with the energy of the candidate states illustrated in FIG.~\ref{fig:candidate_states}. It turns out that, out of the two possible distortions, only the anisotropic Shastry-Sutherland lattice Ising model (ASSLIM) invariant under the $Cmm2$ has a magnetization curve compatible with that of \ErSSL{} with magnetization plateaus at $m=1/4$ and $m=1/2$. An arbitrarily small anisotropy in the intra-dimer interaction is enough to stabilize narrow $m=1/4$ and $m=1/2$ plateaus on the boundaries of the $m=1/3$ plateau, while a relative anisotropy of 33\% leads to the complete suppression of the $m=1/3$ plateau (see the ground-state phase diagram in Fig. \ref{fig:phase_diagram_anisotropy}).\par
The role of intra-dimer anisotropy in destabilizing the 1/3 plateau can be intuitively understood by the following simple argument. The 1/3 plateau structure is highly favorable in any Shastry--Sutherland system dominated by antiferromagnetic intra- and inter-dimer interactions because, in this structure, all up-down dimers gain inter-dimer exchange energy through their interactions with a nearby polarized dimer. By contrast, in the 1/4 resp. 1/2 plateau, some resp. all up-down dimers can be flipped and do not gain any energy through the inter-dimer exchange. However, the 1/3 plateau requires putting half the polarized dimers in each direction and is penalized by the intra-dimer anisotropy ($J_1-J_1'$), while in the 1/4 and 1/2 plateaus all polarized dimers are in the same direction and can take advantage of the anisotropy. And since the 1/4 and 1/2 plateaus are degenerate with the 1/3 plateau at the lower and upper critical fields, even a small anisotropy immediately stabilizes 1/4 and 1/2 plateaus at the boundaries of the 1/3 plateau.\par

%%%%%%%%%%%%%%%%%%%%%%%%%FIG 8 %%%%%%%%%%%%%%%%%%
%%%%%%%%%%%%%%%%%%%%%%%%%%%%%%%%%%%%%%%%%%%%%%%%%%
%%%%%%%%%%%%%%%%%%%%%%%%%FIG 9 %%%%%%%%%%%%%%%%%%
%%%%%%%%%%%%%%%%%%%%%%%%%%%%%%%%%%%%%%%

%%%%%%%%%%%%%%%%%%%%TABLE 1%%%%%%%%%%%%%%%%%%%%
\begin{table*}
    \centering
    \begin{tabular}{|p{0.15\linewidth}|c|c|c|c|c|c|}
        \hline
        \centering Model& & $m=0$ & $m=1/4$ & $m=1/2$ & $m=1$ \\
        \hline
        \multirow{3}{\linewidth}{\centering $J_3=0$} 
        &$\Omega(N)$& $2^{N/2}$ & $2^{\sqrt{N}/2}\times2^{N/8}$ & $2^{N/4}$ & $1$ \\
        &$s_0$& $\log(2)/2$ & $\log(2)/8$ & $\log(2)/4$ & $0$ \\
        &$G(T\rightarrow\infty)/G(T=0)$& $\mathds{1}$ & $\mathds{1}$ & $\mathds{1}$ & $\mathds{1}$ \\
        \hline
        \multirow{3}{\linewidth}{\centering $J_3>0$} 
        &$\Omega(N)$ & $4$ & $2^{3\sqrt{N}/4}$ & $2$ & $1$ \\
        &$s_0$ & $0$ & $0$ & $0$ & $0$ \\
        &$G(T\rightarrow\infty)/G(T=0)$ & $\mathbb{Z}_2\times\mathbb{Z}_2$ & $\mathds{1}$ & $\mathbb{Z}_2$ & $\mathds{1}$ \\
        \hline
        \multirow{3}{\linewidth}{\centering Dipolar interactions} 
        &$\Omega(N)$ & $4$ & $4$ & $2$ & $1$ \\
        &$s_0$ & $0$ & $0$ & $0$ & $0$ \\
        &$G(T\rightarrow\infty)/G(T=0)$ & $\mathbb{Z}_2\times\mathbb{Z}_2$ & $\mathbb{Z}_2\times\mathbb{Z}_2$ & $\mathbb{Z}_2$ & $\mathds{1}$ \\
        \hline
        \multirow{3}{\linewidth}{\centering Quantum Fluctuations} 
        &$\Omega(N)$ & $1$ & $4$ & $1$ & $1$ \\
        &$s_0$ & $0$ & $0$ & $0$ & $0$ \\
        &$G(T\rightarrow\infty)/G(T=0)$ & $\mathds{1}$ & $\mathbb{Z}_2\times\mathbb{Z}_2$ & $\mathds{1}$ & $\mathds{1}$ \\
        \hline
    \end{tabular}
    \caption{Ground-state degeneracy $(\Omega)$, residual entropy $(s_0)$ and magnetic order (defined as the quotient between the high- and low-temperature symmetry groups, $G(T\rightarrow\infty)/G(T=0)$) of the ASSLIM with and without $J_3$ couplings. The row concerning the effect of quantum fluctuations results from qualitative considerations about the formation of dimer product states and the stabilization of the stripe order shown in FIG. \ref{fig:1_4_plateau_orders} c. }
    \label{tab:residual_entropy_table}
\end{table*}
%%%%%%%%%%%%%%%%%%%%%%%%%%%%%%%%%%%%%%%%

\begin{figure*}
    \centering
    \includegraphics[width=.9\linewidth]{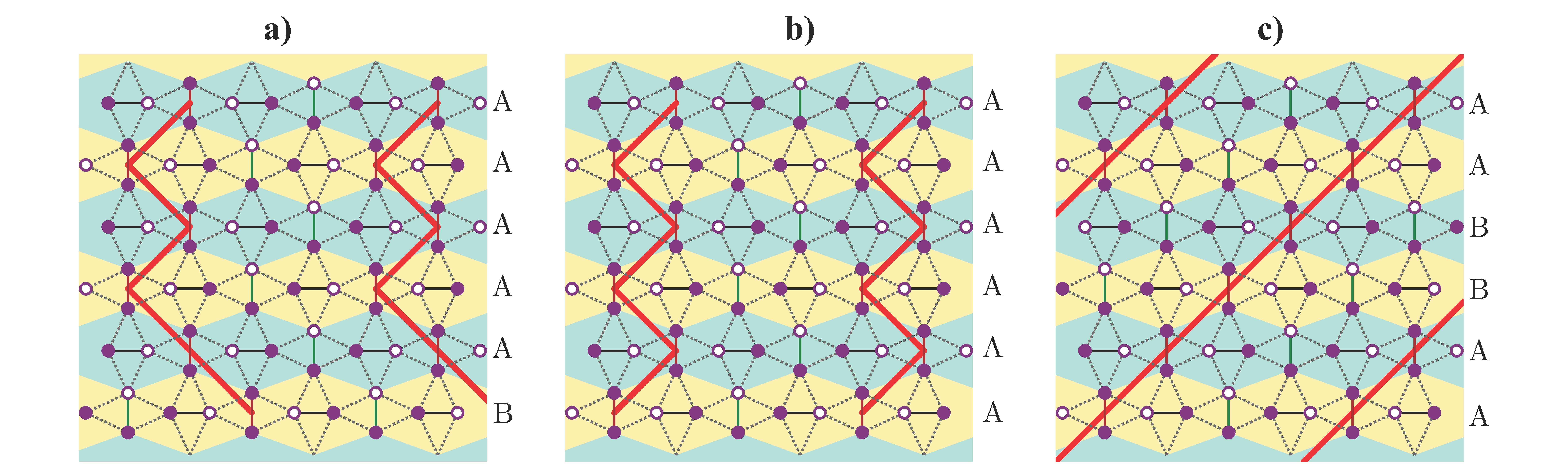}
    \caption{(a): Illustration of the infinite sub-extensive degeneracy of the $m=1/4$ plateau phase corresponding to the independent shifts of rows configurations. (b): Staggered order stabilized by long-range interactions. (c): Stripe order obtained by aligning ferromagnetic dimers in a straight line.}
    \label{fig:1_4_plateau_orders}
\end{figure*}

The identification of the 1/4 and 1/2 plateaus relies on having found one state that saturates the lower bound, but if one only includes $J_1,J_1'$ and $J_2$, the phases with magnetization $m=0$, $m=1/4$ and $m=1/2$ are macroscopically degenerate (their residual entropies are listed in table \ref{tab:residual_entropy_table}): In all plateaus, there are dimers that can flipped independently (shown in green in FIGS. \ref{fig:candidate_states}-a and e). Besides, in the 1/4 plateau some rows can be shifted independently (see  FIG. \ref{fig:1_4_plateau_orders}). However, the macroscopic degeneracy of the plateaus is incompatible with the experimental data: (i) With the specific heat since  no hint of a residual entropy was found (FIG. \ref{hc}-e); (ii) With neutron scattering, which revealed long-range order in all the plateaus. This indicates that residual interactions, either in the form of longer-range interactions or quantum fluctuations, must be considered.

Let us first look at the effect of longer-range interactions. Third-neighbor interactions ($J_3$ shown in FIG. \ref{fig:subgroups_ising_ssl}-a) are enough to lift all \textit{dimer flip} degeneracies, stabilizing the ordered states shown in FIG. \ref{fig:candidate_states} for the $m=0$ and $m=1/2$ plateaus. Note that the stripe ordered phase stabilized by $J_3$ in the $m=0$ plateau  leads to magnetic peaks either in the [100] or [010] reflections while both were observed in \ErSSL{}. This points to the coexistence of magnetic domains with net moments in each of the four allowed directions, which also explains why no net magnetization was observed at zero field in \ErSSL{}. 

By contrast, the \textit{row shift} degeneracy of the $m=1/4$ plateau is remarkably resilient to longer-range interactions and subsists even if couplings up to the sixth nearest neighbor are considered. The \textit{row shift} degeneracies eventually get lifted if full long-range dipolar interactions are taken into account in a distorted version of the SSL with $Cmm2$ symmetry, but in that case the ordered zig-zag state shown in FIG.~\ref{fig:1_4_plateau_orders}-b is selected. This order is incompatible with neutron scattering since it would induce reflections at $(h \pm 1/4,k \pm 1/4,0)$ and $(h\pm 1/4,k\mp 1/4,0)$, whereas magnetic peaks at $(h\pm 1/2,k,0)$ and $(h,k\pm 1/2,0)$ are found in the $m=1/4$ plateau of \ErSSL{}. So another explanation has to be found for the 1/4 plateau. Interestingly enough, in the spin-1/2 Heisenberg Shastry-Sutherland compound SCBO \ the 1/4 plateau has the structure of Fig.~\ref{fig:1_4_plateau_orders}-c \cite{Takigawa2013}, which would be compatible with the peaks observed in single-crystal neutron scattering in \ErSSL{}. So it is plausible that residual quantum fluctuations are responsible for the stabilization of this plateau in \ErSSL{}, but a proof that realistic off-diagonal interactions are sufficient remains to be seen. Note that for the 1/2 plateau the absence of a phase transition in the specific heat also suggests that quantum fluctuations, rather than long-range interactions, lift the degeneracy. Indeed, in both cases quantum fluctuations are expected to stabilize an entangled state (singlet or triplet) on the unpolarized dimers that would not be seen in neutron scattering \cite{liu2024theoryrareearthkramersmagnets}.\par

\begin{figure*}
    \centering
    \includegraphics[width=0.95\linewidth]{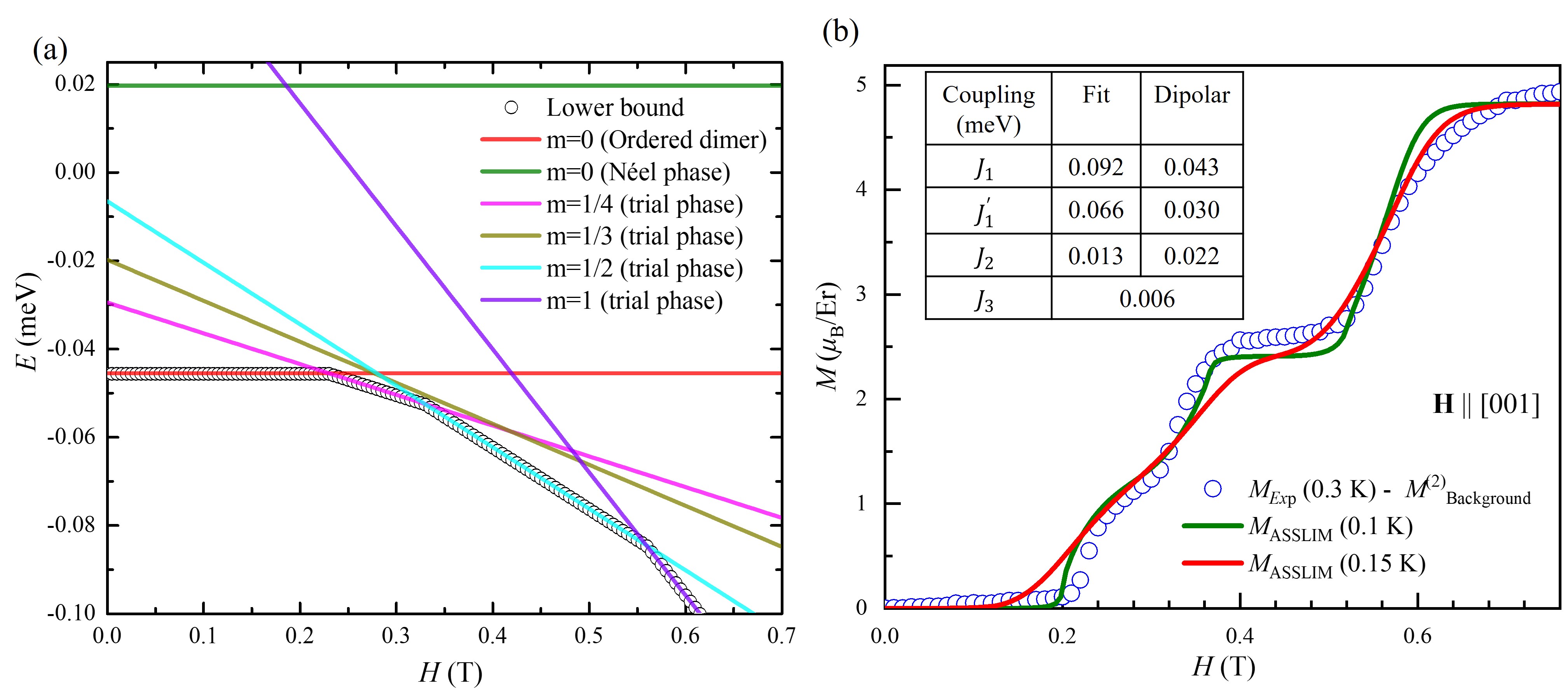}
    \caption{Ground-state energy (a) and magnetization (b) of \ErSSL{} after subtracting the calculated background contribution from the population of excited CEF levels, compared with the theoretical prediction of the ASSLIM model. The theoretical prediction was numerically computed using the Corner Transfer Matrix Renormalization Group method. Inset of (b) shows the values of interaction parameters required for agreement between of the critical fields predicted by the ASSLIM and those measured in \ErSSL{}, in comparison with the magnitude of dipole-dipole interactions. The third-neighbor interaction $J_3$ was assumed to be entirely due to dipole-dipole interaction.}
    \label{fig:mag_theory_exp}
\end{figure*}

Next we try to be more quantitative. Assuming, as usual in the context of rare earths, that only the nearest couplings $J_1$, $J_1'$, and $J_2$ are influenced by exchange, all other couplings are taken to be dipolar, which leads in particular to $J_3=0.006$ meV for the magnetic moments deduced from the CEF. The three critical fields that separate the plateaus at 0, 1/4, 1/2 and saturation can then be used to extract the nearest coupling, leading to $J_1=0.092$ meV, $J_1'=0.066$ meV and $J_2=0.013$ meV (see inset of Fig. \ref{fig:mag_theory_exp}). They have been obtained using the conversion $h = (g^{zz}/2)\mu_BH^{z} = H^z \times 0.287 \text{ meV/T}$, where we make use of the effective $g$-tensor calculated in section \ref{section:zeeman_vv}.

To check the consistency of our theory, we now discuss three other experimental sets of data: the flat mode in inelastic neutron scattering, the temperature dependence of the magnetization curve, and the temperature of the phase transition in zero field. In all cases, it turns out to be important to keep further dipolar couplings. The largest ones are actually the ferromagnetic nearest and next-nearest neighbor inter-layer couplings $J^1_{\perp} = -0.021 \text{ meV}$ and $J^2_{\perp} = -0.006 \text{ meV}$.

The flat mode observed in inelastic neutron scattering has an energy given by $2J_1+4J_3-4J^1_\perp+4J^2_\perp=0.24$ meV, which is  smaller than the experimental value of $0.35$ meV but already much closer than if the inter-layer interactions were neglected.

%%%%%%%%%%%%%%%%%%%%%%%%FIG 10 %%%%%%%%%%%%%%%%%%%%%%%
%%%%%%%%%%%%%%%%%%%%%%%%%%%%%%%%%%%%%%%%%%%%

%%%%%%%%%%%%%%%%%%%%%%%%%%%%%%% FIG 11 %%%%%%%%%%%%%%%%%%%%%%%%
%%%%%%%%%%%%%%%%%%%%%%%%%%%%%%%%%%%
Since $J_1,J_1'\gg J_3$, the melting temperature of the long-range stripe order at zero field in the absence of inter-layer coupling can be estimated by an effective model where the two antiferromagnetic states of each dimer are described by an Ising variable which interacts with its four neighbors due to $J_3$, leading to the critical temperature: 

%%%%%%%%%%%%%%%%%%%% EQUATION 4 %%%%%%%%%%%%%
\begin{equation}
    T_c(J_3)= \frac{2 J_3}{\log\left(1+\sqrt{2}\right)k_B}=0.16\text{ K},
\end{equation}
%%%%%%%%%%%%%%%%%%%%%%%%%%%%%%%%%%%%%

significantly lower than the observed temperature of 0.88 K.
If one includes the dominant inter-layer couplings, the resulting effective model is an anisotropic 3D Ising model with anisotropy parameter $\Delta = J_3/(2J^2_\perp-2J^1_\perp)=0.22$, for which Monte-Carlo simulations \cite{graimMonteCarloStudy1981} determined the critical temperature:

%%%%%%%%%%%%%%%%%%%% EQUATION 5 %%%%%%%%%%%%%
\begin{equation}
    T_c(J_\perp,J_3)=1.88 J_{\perp}/k_B = 0.65 \text{ K}.
\end{equation}
%%%%%%%%%%%%%%%%%%%%%%%%%%%%%%

The transition temperature corrected with $J_{\perp}$ is much closer to the experimentally measured temperature of 0.88 K. The persisting mismatch is probably due to other residual interactions. 

Next, we calculated the temperature dependence of the magnetization for the purely 2D model using the Corner Transfer Matrix Renormalization Group (CTMRG), a numerical method briefly described in section \ref{CTMRG}. To compare with the experimental data, we have subtracted the Van Vleck contribution to the magnetization ($M_{\text{VV}}^{(2)}$ in eq. \ref{eq:m_background}). The overall shape of the curve at low but non-zero temperature is compatible with the experimental data, further supporting the hypothesis that the system is well described by an Ising model, and that the broadening of the jumps between plateaus is thermal. Note also that, after the background is subtracted, the saturation magnetization of \ErSSL{} is compatible with the size of the magnetic moment of the CEF ground state doublets $\ket{\pm}$ given in Eq.\ref{eq:CEF_moment}. There is however a mismatch between the theoretical temperature, 0.15 K, needed to reproduce the experimental curve obtained at 0.3 K. This mismatch is again attributed to further couplings. 

Finally, we comment on the most prominent feature of the specific heat data, a very high and narrow peak at the boundary between the 1/4 and 1/2 plateau. The interpretation of this peak, which  has all the characteristics of a phase transition, is based on the following key observations : (i)
There is no phase transition in the 1/2 plateau; (ii) There is a jump in magnetization at low temperature between 1/4 and 1/2, hence a first-order transition. (iii) The thermal phase transition in the 1/4 plateau takes place at 0.54 K, and it gives rise to a small peak that does not seem to connect to the very pronounced peak at 0.56 K and 0.34 T.
Now, a first order transition can either terminate at a critical point, or be connected to a continuous phase transition through a tricritical point. In the field-temperature phase diagram (Fig. \ref{hc} (e)), the specific heat data are more consistent with a horizontal line at 0.54 K for the thermal transition of the 1/4 plateau phase cut by a first-order transition line terminating at a critical point that gives rise to a much stronger peak. So the most natural interpretation is that this prominent peak in the specific heat is a critical point terminating a first-order transition line starting at the transition between the 1/4 and 1/2 plateaus at zero temperature. 
A similar interpretation has been proposed for the peak observed in \SCBO\  at the transition between the dimer phase and the plaquette phase\cite{julio2021}.

%%%%%%%%%%%%%%%%%%%%%% CONCLUSIONS %%%%%%%%%%%%%%%%%%%%%
\section{Conclusions}

Through a comprehensive combination of cutting-edge experimental techniques and advanced theoretical modeling, we have discovered and elucidated the unprecedented magnetic properties of \ErSSL{}, establishing it as a unique and powerful platform for exploring exotic phases of matter in Shastry-Sutherland lattice (SSL) systems. Our results mark the first observation of 1/4 and 1/2 fractional magnetization plateaus in an Ising SSL compound — an outcome that defies long-standing theoretical expectations of a 1/3 plateau and signals a fundamental shift in our understanding of such systems.

We demonstrated that this remarkable plateau sequence arises from a subtle orthorhombic distortion that introduces spatial anisotropy in dimer interactions, captured effectively by the anisotropic Shastry-Sutherland Ising model (ASSLIM). This model successfully reproduces the observed magnetization behavior and is validated by detailed structural and neutron scattering data, including the detection of symmetry-forbidden reflections and anisotropic bond lengths. 

Notably, our experimental results reveal a complete absence of residual entropy across all phases, despite theoretical expectations of macroscopic degeneracy. This striking result is fully accounted for by incorporating two key factors: longer-range Ising interactions extending beyond second neighbors, and quantum fluctuations arising from off-diagonal terms in the effective Hamiltonian. Impressively, we find that degeneracy is lifted at zero field through third-neighbor interactions, while in the 1/4 and 1/2 magnetization plateaus, it is suppressed by the emergence of entangled dimer states on non-polarized dimers—a direct manifestation of quantum fluctuations stabilizing unique ordered phases.

Our work provides a clear roadmap for the design and investigation of related rare-earth SSL systems, positioning \ErSSL{} and the broader melilite family as exceptional platforms to uncover and understand new phenomena in frustrated magnetism. The accessible energy scales and field ranges in \ErSSL{}, compared to other SSL compounds, open avenues for high-precision measurements and controlled theoretical investigations.

Looking ahead, our findings open several compelling avenues for future research that can deepen the understanding of \ErSSL{} and related systems. In particular, the distinct mechanisms responsible for lifting degeneracy across the various plateau phases—ranging from longer-range Ising interactions to quantum fluctuations—present a rich landscape for further theoretical and experimental investigation. Additionally, our work suggests that incorporating long-range interactions is essential to accurately capture the full energy and temperature scales, pointing to the value of refined modeling approaches. The observation of satellite peaks indicative of incommensurate correlations at the boundary between the 1/4 and 1/2 plateaus raises intriguing questions about emergent spin textures and competing orders in this regime. Moreover, determining the universality class of the phase transitions at zero field and within the 1/4 plateau remains an open challenge that could shed light on the critical behavior in frustrated Ising systems. Addressing these questions will benefit from a deeper understanding of the underlying microscopic interactions, potentially guided by ab-initio calculations and advanced numerical methods, paving the way for new discoveries in rare-earth-based Shastry-Sutherland systems.

Beyond this compound, our study underscores the enduring potential of the SSL geometry to promote unexpected phases and transitions, and it highlights rare-earth-based SSL materials as fertile ground for discovering and controlling exotic quantum states.
%%%%%%%%%%%%%%%%%%%%%%%%%%%%%%%%%%%%%%%%%%%%%%%%%%%%%%%%

%%%%%%%%%%%%%%%%%%%%%%%% ACKNOWLEDGMENTS %%%%%%%%%%%%%%%%%%%%%%%%%%%%%%
\section{Acknowledgments}
The research performed at Duke University was supported by NSF award number DMR-2327555. A portion of this research used resources at the High Flux Isotope Reactor and Spallation Neutron Source, a DOE Office of Science User Facility operated by the Oak Ridge National Laboratory. A portion of this work was performed at the National High Magnetic Field Laboratory, which is supported by National Science Foundation Cooperative Agreement No. DMR-2128556 and the State of Florida. The research performed at EPFL has been supported by the Swiss National Science Foundation Grant No. 212082.
The research performed at FAU was supported by the Munich Quantum Valley, which is supported by the Bavarian state government with funds from the Hightech Agenda Bayern Plus.
For the numerical calculations of the unit-cell based ground-state search, we acknowledge the scientific support and HPC resources provided by the Erlangen National High Performance Computing Center (NHR@FAU) of the Friedrich-Alexander-Universität Erlangen-Nürnberg.
%We thank Pratyay Ghosh and Samuel Nyckees for useful discussions.
%%%%%%%%%%%%%%%%%%%%%%%%%%%%%%%%%%%%%%%%%%%%%%%%%%%%%%%%%%%%%%%%%%%%%

%%%%%%%%%%%%%%%%%%%%%%%%%%%% AUTHOR CONTRIBUTIONS %%%%%%%%%%%%%%%%%%%%%%
\section{Author Contributions}
Research conceived by S.H.; L.Y. , R.B. and S.H. synthesized samples; L.Y. , R.B. and S.H. performed thermodynamics measurements; L.Y. , M.E., C.dela C., A.I.K, O.G., K.M.T and S.H. conducted neutron scattering experiments; L.Y., M.E., and D.G. performed magneto-transport measurements; A.R., J.A.K., K.P.S., and F.M. provided theoretical interpretations; L.Y., A.R., J.A.K., K.P.S., F.M., and S.H. wrote the manuscript with comments from all authors.

\appendix
%%EXPERIMENTAL METHODS %%%%%%%%%%%%%%%%%%%%%%%%%%%%
%TC:ignore
\section{ Methods}
\subsection{Synthesis and thermodynamic measurements}

Polycrystalline \ErSSL{} powder was prepared by a solid-state reaction route using the starting precursors of \ch{Er_2O_3} (99.9~\%, Alfa Aesar) with \ch{BeO} (99.99~\%, 
Alfa Aesar) and \ch{GeO2} (99.99~\%, Alfa Aesar). The starting precursors were weighed in the 1 : 2 : 1 molar ratio and the mixture was sintered at 1285~\degree C for 48 hours with intermediate grindings. Powder X-ray diffraction (PXRD) data were analyzed by performing Rietveld refinement using FullProf Suite. Once the phase purity of the powder sample was confirmed, the single crystals of \ErSSL{} were grown using four mirrors optical floating zone furnace (Model: FZ-T-12000-X-VII-VPO-PC, Crystal System Corporation, Japan). The grown crystals were analyzed and oriented using the Laue diffractometer (MULTIWIRE LABS MWL120) and subsequently cut to the required dimensions using a wire saw. Heat capacity was measured on oriented single crystal samples of \ErSSL{} and powder samples of \LuSSL{} (non-magnetic) using Helium-4 (1.8 K $\leq T \leq$ 300 K) and dilution refrigerator (0.06 K $\leq T \leq$ 2 K) set up attached to the Physical Properties Measurement Systems, Quantum design (PPMS Dynacool, QD, USA) accompanied by 14 T magnets. Magnetic measurements from 300 K to 2 K were performed using the Cryogenic Ltd SQUID (superconducting quantum interference device) magnetometer. Additional measurements from 1.8 K to 0.3 K were performed using the Helium-3 probe in SQUID magnetometer. 
 
\subsection{Neutron scattering}

Neutron powder diffraction (NPD) measurements to study the magnetic structure of \ErSSL{} were performed using the HB-2A \cite{garlea2010high} diffractometer at the High Flux Isotope Reactor (HFIR) in Oak Ridge National Laboratory (ORNL). For the HB-2A experiment, 2.7 grams of the powder sample were loaded in an Al can and sealed under a Helium environment, and subsequently loaded onto a $^3$He cryo-stick in a Cryostat. The measurements were performed using a collimation of open-21$''$-12$''$ and a Ge monochromator to select incident neutrons of wavelength $\lambda = 2.41$ \AA. Diffraction patterns were collected at 0.3 K and 100 K by counting for 120 sec at each point ranging between $2\theta = 5^\circ$ and 130$^\circ$. Single crystal neutron scattering measurements were performed at the HYSPEC spectrometer \cite{zaliznyak2017polarized} at ORNL. Approximately 1 gram of the single crystal sample of \ErSSL{}\ was oriented in the [$h$,$k$,0] scattering plane, with a magnetic field of up to 7 T applied along the crystallographic $c$-axis. The sample environment included a dilution refrigerator capable of reaching a base temperature of 60 mK. An incident neutron energy of 3.8 meV was selected using a Fermi chopper operating at 180 Hz. The paramagnetic phase dataset, collected at 12 K, was employed to subtract the nuclear component from the scattering.

Inelastic neutron scattering experiments were performed on powder samples of \ErSSL{} and \LuSSL{} (for phonon contributions) at SEQUOIA \cite{granroth2010sequoia} spectrometer at ORNL to probe crystal electric field (CEF) levels. Powder samples were loaded in a Al can and sealed under $^4$He atmosphere to ensure good temperature coupling. Measurements were collected at incident energy $E_i~$=~150 meV in high flux mode and at $E_i~$=~60 meV and 30 meV in high-resolution mode at 6 K, 50 K and 250 K. Data analysis was performed using the DAVE MSlice \cite{azuah2009dave}, and the PyCrystalField \cite{scheie2021pycrystalfield} packages, to determine the crystal electric field (CEF) parameters.

\subsection{Crystal Electric Field (CEF) Analysis}
\label{method_cef}
The crystal field effect on the magnetic ion can be parameterized using the single-ion crystal field Hamiltonian:
 \begin{equation}
{H_{CEF}= \sum_{n,m}B^m_nO^m_n}.
\label{eq:crystal_field_hamiltonian}
\end{equation}
Here, \(B^m_n\) are the CEF parameters that parameterize the effects of the ligand environment on the magnetic ion, and \(O^m_n\) are the Stevens operators \cite{hutchings1964point,stevens1952matrix,scheie2021pycrystalfield}.

The fits to these observed CEF levels in the INS spectrum were carried out using  PyCrystalField software\cite{scheie2021pycrystalfield} where $B^m_n$ were optimized during the fitting procedure. The number of CEF parameters in the Hamiltonian was reduced to fifteen by rotating the coordinates so that the quantization axis $z$ coincided with [001], while $x$ and $y$ were assigned to [110] and [1-10], respectively, corresponding to the Er1 site. This coordinate system ensures that the mirror axis [-0.707  0.707  0] lies in the $xy$ plane and is parallel to $y$, ensuring that all imaginary CEF parameters are zero. The initial $B^m_n$ were estimated using an electrostatic point charge model based on the 8 \ch{O^{2-}} ligands surrounding the \ch{Er^{3+}} ion. Starting with these CEF parameters, $2D$ $Q-\Delta E$ slices at all three temperatures (6 K, 50 K and 250 K) for both $E_i=30$ meV and $E_i=60$ meV were fitted. To further constrain the fit, magnetic susceptibilities ($\chi$) along $[001]$ and $[010]$ and isothermal magnetization data at 5 K, 10 K, and 20 K in the same directions were included, ensuring the solutions captured the pronounced anisotropy observed in this system. The initial and the optimized $B^m_n$ parameters are provided in the table in Appendix \ref{CEF_app}. A comprehensive list of eigenvalues and their corresponding eigenvectors can also be found in the Appendix \ref{CEF_app}.

\subsubsection{Effective spin-1/2 Hamiltonian from the Crystal Electric Field}
\label{section:CEF_eff_hamiltonian}
At temperatures much smaller than the smallest CEF gap, $T\ll1.63 \text{ meV}/k_B=18.7\text{ K}$ only the lowest lying CEF levels, which we label as $\ket{+}$ and $\ket{-}$, will be populated and the system can be described by an effective spin-1/2 Hamiltonian. In order to derive this effective Hamiltonian, one must start from a Hamiltonian including the most general two-body interaction \cite{rauFrustratedQuantumRareEarth2019}
\begin{equation}
\begin{aligned}
    H &= H_{\text{int}}+H_{\text{CEF}} \\
      &= \frac{1}{2} \sum_{i j}\sum_{k,q}\sum_{k', q'} 
      \mathcal{J}_{i j}^{k q, k' q'} T_{q}^{(k)}(\mathbf{J}_i) T_{q'}^{(k')}(\mathbf{J}_j)\\
      &+ \sum_i H_{\text{CEF}}(\mathbf{J}_i)\,.
\end{aligned}
\label{eq:your_label}
\end{equation}
where $T_{q}^{(k)}(\mathbf{J}_i)$ are spherical tensors of rank $k$, built as polynomials of order $k$ in the angular momentum $\mathbf{J}_i$. They include the usual bilinear exchanges such as $\mathbf{J}_i\cdot\mathbf{J}_j$ as well as interactions between higher-order multipoles, up to the maximum rank $k_{\text{max}}=2j$. The components of the $J_i^{\alpha}$ are represented in the local reference frame $\hat{x}_i,\hat{y}_i,\hat{z}_i$, related to each-other by the symmetry operations of the space group of the lattice.\par
The effective spin-1/2 Hamiltonian can be obtained to leading order in $\mathcal{J}/\Delta_{\text{CEF}}$ by projecting $H_{\text{int}}$ in the ground-state manifold of $H_{\text{CEF}}$
\begin{equation}
    H_{\text{eff}} =PH_{\text{int}}P  = \frac{1}{2}\sum_{i,j}\sum_{\alpha,\beta}J_{i,j}^{\alpha,\beta}S^{\alpha}_iS^{\beta}_j,\label{eq:eff_hamiltonian_cef}
\end{equation}
where $P$ is the projector onto the ground-state manifold of $H_{\text{CEF}}$ and $S^{\alpha}_i$ are the effective spin-1/2 spin operators, defined as
\begin{equation}
    S^{\alpha}_i=\frac{1}{2}\sum_{\sigma\in\{+,-\}}\sum_{\sigma'\in\{+,-\}} \tau_{\sigma\sigma'}^{\alpha}\ket{\sigma}_i\bra{\sigma'}_i,
\end{equation}
where $\tau_{\sigma\sigma'}^{\alpha}$ are the Pauli matrices and the operators $S^{\alpha}_i$ act non-trivially only on site $i$. The spin-1/2 couplings $J_{i,j}^{\alpha,\beta}$ can be calculated from the physical Hamiltonian $H_{\text{int}}$ as
\begin{align}
J_{i,j}^{\alpha,\beta} &= \sum_{k,q}\sum_{k',q'}\mathcal{J}_{i,j}^{kq,k'q'} \nonumber \\
& \times \left(\sum_{\sigma_i,\sigma_i'\in\{+,-\} }(\tau_{\sigma_i,\sigma_i'}^{\alpha})^*
\bra{\sigma_i}T_q^{(k)}(\mathbf{J}_i)\ket{\sigma_i'}\right) \nonumber \\
& \times \left(\sum_{\sigma_j,\sigma_j'\in\{+,-\}}(\tau_{\sigma_j,\sigma_j'}^{\beta})^*
\bra{\sigma_j}T_{q'}^{(k')}(\mathbf{J}_j)\ket{\sigma_j'}\right).
\end{align}

\indent In some rare-earth compounds, such as the pyrochlore \ch{Dy2Ti2O7} and \ch{Ho2Ti2O7} \cite{rauMagnitudeQuantumEffects2015}), $H_{\text{eff}}$ is dominated by diagonal terms proportional to $J_{ij}^{zz}$, which means that their low-temperature magnetic properties can be well approximated by a classical Ising model. The dominance of the diagonal terms is a consequence of the fact that the strength of multipole-multipole interactions $\mathcal{J}_{i,j}^{kq,k'q'}$ - which result from a combination of physical processes such as magnetic dipole-dipole interactions, direct exchange, superexchange and lattice mediated interactions - are highly suppressed at rank $k=8$ or higher. This fact, combined with the Wigner-Eckart theorem \cite{messiahRepresentationIrreducibleTensor1961} which states that the only non-zero matrix elements in
\begin{equation}
    \bra{m_J}T_q^{(k)}(\mathbf{J})\ket{m_J'}
\end{equation}
have $|m_J-m_J'|\leq k$, implies that interactions can only connect states $\ket{m_J}$ and $\ket{m_J'}$ in first-order perturbation theory if $|m_J-m_J'|\leq7$. In the cases of \ch{Dy2Ti2O7} and \ch{Ho2Ti2O7}, the CEF ground state doublets are almost collinear with the fully polarized states, $\ket{\pm15/2}$ and $\ket{\pm 8}$ respectively \cite{rosenkranzCrystalfieldInteractionPyrochlore2000}. This means that, to first order in $\mathcal{J}/\Delta_{\text{CEF}}$, the effective spin-1/2 Hamiltonian is an Ising model with interaction constants given by $J_{i,j}^{zz}$. Similarly, in \ErSSL{} the ground state doublets of the CEF have their largest component collinear with the fully polarized states $\ket{\pm15/2}$ (see SM Table II), albeit to a lesser extent than in \ch{Dy2Ti2O7} and \ch{Ho2Ti2O7}. This justifies us in using an Ising model as a starting point to understand the magnetic properties of \ErSSL{}, and in invoking quantum fluctuations due to transverse exchange in the discussion of some properties.\par
Due to the Kramers degeneracy, there is some freedom in the choice of the basis of the ground-state manifold of $H_{\text{CEF}}$. In the absence of a detailed model for the physical \ce{Er^3+} moments, we choose the basis of the CEF ground state doublet such that the projection of intra-dimer Heisenberg interaction leads to a diagonal coupling matrix:
\begin{equation}
    J^{\alpha\beta}_{ij}=\delta_{\alpha\beta}J^{\alpha\alpha}_{ij}.
\end{equation}
The CEF states presented in table \ref{eigen} reflect this choice of basis.
\subsubsection{Zeeman splitting and Van-Vleck paramagnetism}\label{section:zeeman_vv}
The Zeeman splitting Hamiltonian due to the application of an external magnetic field is
\begin{equation}
    H_{\text{Zeeman}}=-\mu_Bg_J\sum_i\mathbf{H}\cdot\mathbf{J}_i,
\end{equation}
where $g_J$ is the free-ion Landé $g$-factor. At small fields and temperatures the effect of the magnetic field can be determined by perturbation theory of the total hamiltonian
\begin{equation}
    H = H_{\text{CEF}} + H_{\text{Int.}}+H_{\text{Zeeman}}.
\end{equation}
To first order in $H_{\text{Zeeman}}$ we obtain
\begin{equation}
    H_{\text{Zeeman}}^{\text{eff}}=PH_{\text{Zeeman}}P = \mu_B\sum_i\sum_{\alpha,\mu}g_i^{\alpha \mu}S_i^{\alpha}H^{\mu},\label{eq:effective_zeeman_h}
\end{equation} 
where $g^{\alpha,\mu}$ is the effective $g$-tensor for the pseudo-spin 1/2 variables and can be calculated as
\begin{equation}
    g_i^{\alpha \mu} = \sum_{\sigma,\sigma'}(\tau^{\alpha}_{\sigma\sigma'})^{*}g_J\bra{\sigma}J_i^{\mu}\ket{\sigma'}.
\end{equation}
For the case of \ErSSL{}, from the CEF levels written in the Appendix \ref{CEF_app} we obtain 
\begin{equation}
    g_i^{zz} = 9.920
\end{equation}
(the full $g$-tensor is given in Appendix \ref{CEF_app}). \par
The first order approximation amounts to neglecting the population of higher Crystal Electric Field levels due to the magnetic field and predicts a zero-temperature magnetization
\begin{equation}
    M^{\mu} = \frac{\partial E_{GS}}{\partial H^{\mu}}= \mu_B\sum_{\alpha} \sum_i g_i^{\alpha\mu}\bra{0}S_i^{\alpha}\ket{0}.
\end{equation}
Under the further approximation that $H_{\text{int}}$ solely has Ising couplings,  the ground state of $H_{\text{CEF}}+H_{\text{int}}$ is a tensor-product state
\begin{equation}
    \ket{0} = \bigotimes_{i}\ket{\sigma_i},\sigma_i\in\{+,-\}.
\end{equation}
and we obtain
\begin{equation}
    M^{\mu} = \frac{\mu_B}{2} \sum_i g_i^{z\mu}\sigma_i.\label{eq:mag_first_order}
\end{equation}
Under the assumptions for which Eq. \ref{eq:mag_first_order} is valid, the magnetization predicted at zero temperature must be piecewise constant. On the contrary, the magnetization curve of \ErSSL{} shows a significant slope, even at $0.3$ K and away from any critical fields. This background slope to the magnetization can be calculated by taking the perturbative calculation of the ground-state energy to second order in $H^z$
\begin{equation}
    E_{GS}^{(2)} =  (H^z)^2\mu_B^2g_J^2\sum_{i=1}^N\sum_{E_k\in\{\text{Exc. states}\}}\frac{|\bra{k}J_i^z\ket{0}|^2}{E_0-E_k}.
\end{equation}
In the previous equation, for the sake of simplicity, we have assumed the magnetic field to point along the $z$ direction. If $H_{\text{int}}$ only contains Ising-like terms,
the only excited states for which the matrix element $\bra{k}J_i^z\ket{0}$ is non-zero are those which differ from $\ket{0}$ in a single site. Taking into account that the energy scale of the excited CEF levels ($\sim10$K) is much larger than the energy scale of interactions in \ErSSL{} ($\sim1$ K), we can neglect the effect of interactions in the energy difference $E_k-E_0$ and obtain
\begin{equation}
    E_{GS}^{(2)} = -H_z^2\mu_B^2g_J^2\sum_{c_i=3}^{16}\frac{|\bra{c_i}J_i^z\ket{\sigma_i}|^2}{E_{\text{CEF}}(c_i)},
\end{equation}
where $E_{\text{CEF}}(c_i)$ are the energies of the excited CEF levels.  $E_{GS}^{(2)}$ remarkably doesn't depend on the Ising configuration $\{\sigma_i\}$. The second-order correction to $E_{GS}^{(2)}$ leads to a Van-Vleck paramagnetic contribution to the magnetization
\begin{equation}
    M^{z} =  \frac{\mu_B}{2} \sum_i g_i^{zz}\sigma_i + M_{\text{VV}}^{(2)},\label{eq:magnetization_corrected}
\end{equation}
where
\begin{equation}
    M_{\text{VV}}^{(2)}=2H_z\mu_B^2g_J^2\sum_{c=3}^{13}\frac{|\bra{c}J_i^z\ket{+}|}{E_{\text{CEF}}(c_i)}=1.40 ~\mu_B/T \times H^z.\label{eq:m_background}
\end{equation}
\subsection{Ising ground-state energy lower bounds}
\label{section:gs_lower_bounds}
Lower bounds to the ground-state energy of an Ising model (defined by the Hamiltonian in Eq. \ref{eq:general_ising_model}) can be obtained by independently minimizing the energy of finite clusters of spins \cite{vanheckeSolvingFrustratedIsing2021} \cite{huangFindingProvingExact2016}.\par
It is useful to first define the lattice translations
\begin{equation}
\mathcal{L}(L)=\{\mathbf{r}_{k}+x\mathbf{a_1}+y\mathbf{a_2}: x,y\in \mathbb{Z}/L\mathbb{Z},k\in \{1,\cdots,p\}\},    
\end{equation}
where $\mathbf{r}_k, k\in \{1,\cdots,p\}$ are the positions of the $p$ sites in a unit cell, while $x,y$ index the unit cell in a system with periodic boundary conditions where positions translated by $L$ in any of the crystal directions are identified: $\mathbf{r}
\equiv \mathbf{r}+L\mathbf{a}_1\equiv\mathbf{r}+L\mathbf{a}_2$. The lattice $\mathcal{L}(L)$ can be expressed as a union of translated copies of a generating cluster $t_{0,0}\subset\mathcal{L}(L)$
\begin{align}
    t_{x,y}&=\{i+x\,\mathbf{u_1}+y\,\mathbf{u_2}:i\in t_{0,0}\}\\\mathcal{L}(L)&=\bigcup_{(x,y)\in(\mathbb{Z}/\Tilde{L}\mathbb{Z})^2}t_{x,y}.
\end{align}
The integer $\Tilde{L}$ determines the number of translated  copies of $t_{0,0}$ necessary to cover the whole lattice $\mathcal{L}(L)$ and satisfies the formula:
\begin{equation}
    \frac{\Tilde{L}^2}{L^2}=\frac{
    \det(\mathbf{a}_1,\mathbf{a}_2)
    }{\det(\mathbf{u}_1,\mathbf{u}_2)}
\end{equation}
If the generating tile and the cluster translation vectors $(\mathbf{u_1},\mathbf{u_2})$ are chosen such that every pair of interacting spins is contained in at least one cluster, the Hamiltonian can be decomposed as a sum over clusters. If different clusters overlap, some spins or bonds will be contained in several clusters and the decomposition is not unique. The possible decompositions are parameterized by the weights  $0\leq\alpha_{i,j}\leq 1$ and $0\leq\beta_{i}\leq 1$, responsible for splitting the energy of every site and interacting bond among the clusters that share it:
\begin{align}
    H(L,\{\sigma\})&=\sum_{x,y\in (\mathbb{Z}/\Tilde{L}\mathbb{Z})^2}H_{\rm loc}(\sigma|_{t_{x,y}},\mathbf{\alpha},\mathbf{\beta}) \label{eq:hamiltonian_tesselation}\\
    H_{\rm loc}(\sigma|_{t_{x,y}},\mathbf{\alpha},\mathbf{\beta})&=\sum_{i \in t_{x,y}}\sum_{j \in t_{x,y}} J_{i,j}\alpha_{i,j} \sigma_i\sigma_j\nonumber\\
    &+h \sum_{i \in t_{x,y}}\beta_i\sigma_i.
\end{align}
For the decomposition to be consistent, the weights must satisfy the constraints:
\begin{align}
    &\sum_{x,y}\sum_{i,i' \in t_{0,0}}\delta_{i+x\mathbf{a}_1+y\mathbf{a}_2,i'} \beta_{i}=1 \label{eq:weight_constraint1}\\
    &\sum_{x,y}\sum_{i,i'\in t_{0,0}}\sum_{j,j' \in t_{0,0}}\delta_{i+x\mathbf{a}_1+y\mathbf{a}_2,i'}\delta_{j+x\mathbf{a}_1+y\mathbf{a}_2,j'} \alpha_{i,j}=1.\label{eq:weight_constraint2}
\end{align}

%%%%%%%%%%%%%%%%%%%%%%% FIG 12 %%%%%%%%%%%%%%%%%%%
\begin{figure}
    \centering
    \includegraphics[width=0.75\linewidth]{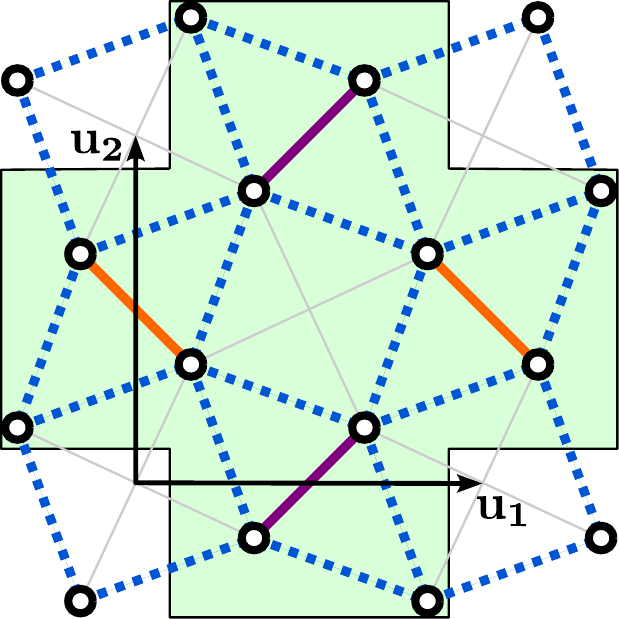}
    \caption{Illustration of the 12 site cluster $t_{x,y}$ and the cluster translation vectors $\mathbf{u_1}$,$\mathbf{u_2}$.}
    \label{fig:gs_lower_bound_tile}
\end{figure}
%%%%%%%%%%%%%%%%%%%%%%%%%%%%%%%%%%%%%%%%%%%%%%%

In the case of the Shastry-Sutherland Ising model, we take as generating cluster the set of 12 spins highlighted in fig. \ref{fig:gs_lower_bound_tile}. Independent minimisation of  the energy of each cluster leads to a lower bound on the ground state energy
\begin{align}
    &E_0=\lim_{L\rightarrow\infty}\min_{\{\sigma\}}\frac{H(L,\{\sigma\})}{p L^2}\\
    &H(L,\{\sigma\})\geq \Tilde{L}^2\min_{\sigma|_{t_{0,0}}}H_{\rm loc}(\sigma|_{t_{0,0}},\mathbf{\alpha},\mathbf{\beta})\implies\\
    &E_0\geq \frac{\Tilde{L}^2}{pL^2}\, E_{0,{\rm loc}}(\alpha,\beta)=\frac{\Tilde{L}^2}{pL^2}\min_{\sigma|_{t_{0,0}}}H_{\rm loc}(\sigma|_{t_{0,0}},\mathbf{\alpha},\mathbf{\beta}),
\end{align}
Each value of  $(\alpha,\beta)$ compatible with the constraints  \ref{eq:weight_constraint1} and \ref{eq:weight_constraint2} defines a lower bound on the ground state energy. Therefore, the most restrictive lower bound can be obtained by optimising over the weights as
\begin{equation}
    E_{\text{Lower Bound}}^{\text{opt}} = \frac{\Tilde{L}^2}{pL^2} \max_{\alpha,\beta}\left\{\min_{\sigma|_{t_{0,0}}}H_{\rm loc}(\sigma|_{t_{0,0}},\mathbf{\alpha},\mathbf{\beta})\right\}.
\end{equation}

If the ground state energy lower bound obtained in this way matches the energy of any of the candidate states, the bound is said to be \textit{saturated} and we can be sure to have found the correct ground-state energy.

\subsection{Unit-cell based ground-state search for long-range interactions}
\label{sec:unit_cell_technique}

%%%%%%%%%%%%%%%%%%%%% FIG 13 %%%%%%%%%%%%%%%%%%%%
\begin{figure}
    \centering
    \includegraphics[width=\linewidth]{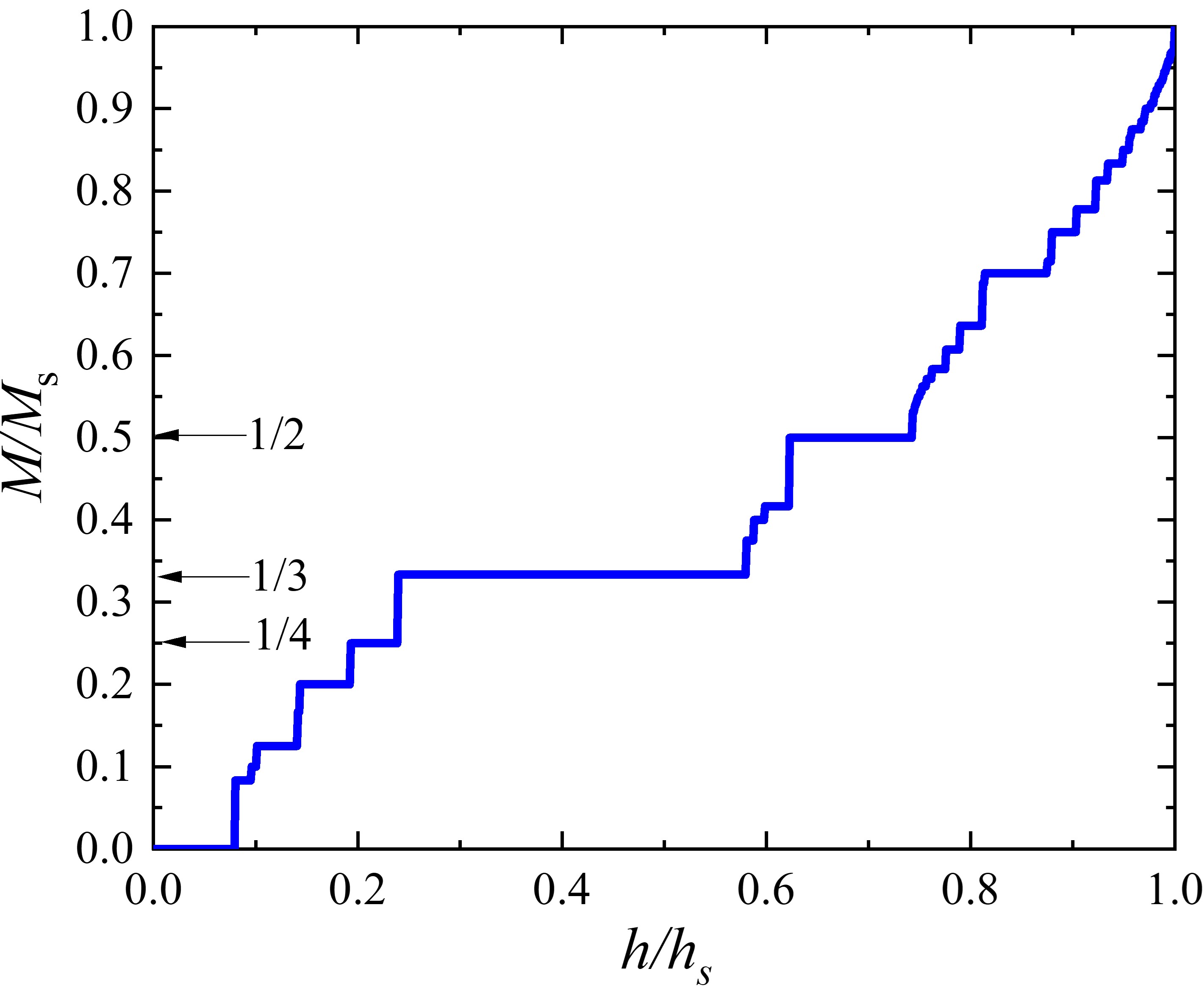}
    \caption{
    Magnetization of the ground state of the SSL Ising model Eq.~\eqref{eq:long_range_ising_model} with only dipolar interactions.
    }
    \label{fig:devil_staircase}
\end{figure}
%%%%%%%%%%%%%%%%%%%%%%%%%%%%%%%%%%%%%%%%%%%%%%%
{
\renewcommand{\vec}[1]{\mathbf{#1}}
A recently proposed way to determine ground states for Hamiltonians of the form of Eq.~\eqref{eq:long_range_ising_model} is to consider all possible magnetic unit cells up to a certain limitation and optimize an effective Hamiltonian on each of the unit cells with appropriately resummed couplings \cite{koziolSystematicAnalysisCrystalline2023, koziol_quantum_2024}. These couplings are chosen so that the energy per site of the pattern on the unit cell corresponds to the thermodynamic limit being tiled by this unit cell. The treatment of finite-range interactions in this framework is well-established \cite{Dorier2008}. 
The key insight that makes this approach possible for long-range interactions is that one can rewrite a diagonal long-range interaction for a periodic pattern with a $K$-site unit cell and translational vectors $\vec a_1$ and $\vec a_2$
\begin{align}
        \label{eq:hamiltonian_on_unit_cell}
		\sum_{i\neq j}\frac{J}{|\vec r_i -\vec r_j|^\alpha}S_i^z S_j^z=\sum_{i=1}^K \bar J^{\alpha}_{i,i}S_i^z S_i^z + \sum_{i\neq j}^K \bar J^{\alpha}_{i,j}S_i^z S_j^z
\end{align}
as sums over the unit cell of the pattern. 
The appropriately resummed couplings
\begin{align}
\bar J^{\alpha}_{i,j} 
&= J \sum_{k=-\infty}^{\infty}\sum_{l=-\infty}^{\infty} 
    \frac{1}{|\vec r_i -\vec r_j + l\vec a_1 - k \vec a_2|^\alpha} \nonumber \\
&= J \sum_{\vec u \in \mathcal{L}(\vec a_1, \vec a_2)} 
    \frac{1}{|\vec r_i -\vec r_j + \vec u|^\alpha} \nonumber \\
&= J \ \zeta_{\mathcal{L}(\vec a_1, \vec a_2), \alpha}(\vec r_i - \vec r_j, \vec 0)
\end{align}

\begin{align}
\bar J^{\alpha}_{i,i} 
&= J \sum_{k=-\infty}^{\infty}\sum_{l=-\infty}^{\infty} 
    \frac{(1-\delta_{l,0}\delta_{k,0})}{| l\vec a_1 - k \vec a_2|^\alpha} \nonumber \\
&= J \sum_{\vec u \in \mathcal{L}(\vec a_1, \vec a_2)\setminus\vec 0} 
    \frac{1}{|\vec u|^\alpha} \nonumber \\
&= J \ \zeta_{\mathcal{L}(\vec a_1, \vec a_2), \alpha}(\vec 0, \vec 0)
\end{align}

with $\delta_{i,j}$ being the Kronecker delta, $\mathcal{L}(\vec a_1, \vec a_2)$ the lattice spanned by $\vec a_1$ and $\vec a_2$ and $\zeta_{\mathcal{L}, \alpha}(\vec x, \vec y)$ being the Epstein $\zeta$-function.
Inverting the logic, this means that finding the ground state of \eqref{eq:hamiltonian_on_unit_cell} on a unit cell gives the lowest energy state for the thermodynamic limit that fits the considered cell.

Therefore, the workflow is as follows: First, we determine the unit cells for the respective (non-)distorted SSL lattice with atomic positions provided by experimental measurements following the procedure described in \cite{koziolSystematicAnalysisCrystalline2023}.
Then we evaluate the resummed couplings for each unit cell using efficient implementations of the Epstein $\zeta$-function \cite{Buchheit2024CodePaper}.
In the end, a parameter study is conducted for varying the $h$ values. Here, for each $h$ value, an optimization of the spin state is performed on each unit cell and the configuration with the overall lowest energy in the thermodynamic limit is selected.
From the resulting state, observables such as the magnetization or static structure factors can be computed.

This method is limited by the number and shape of the unit cells considered, as well as the optimization algorithm on the respective unit cells.
For this study, we choose unit cells with no more than 16 elementary four site unit cells of the respective SSL lattice, with no more than six elementary unit cells in one linear direction.
}

\subsection{CTMRG}
\label{CTMRG}
\begin{figure}
    \centering
    \includegraphics[width=0.95\linewidth]{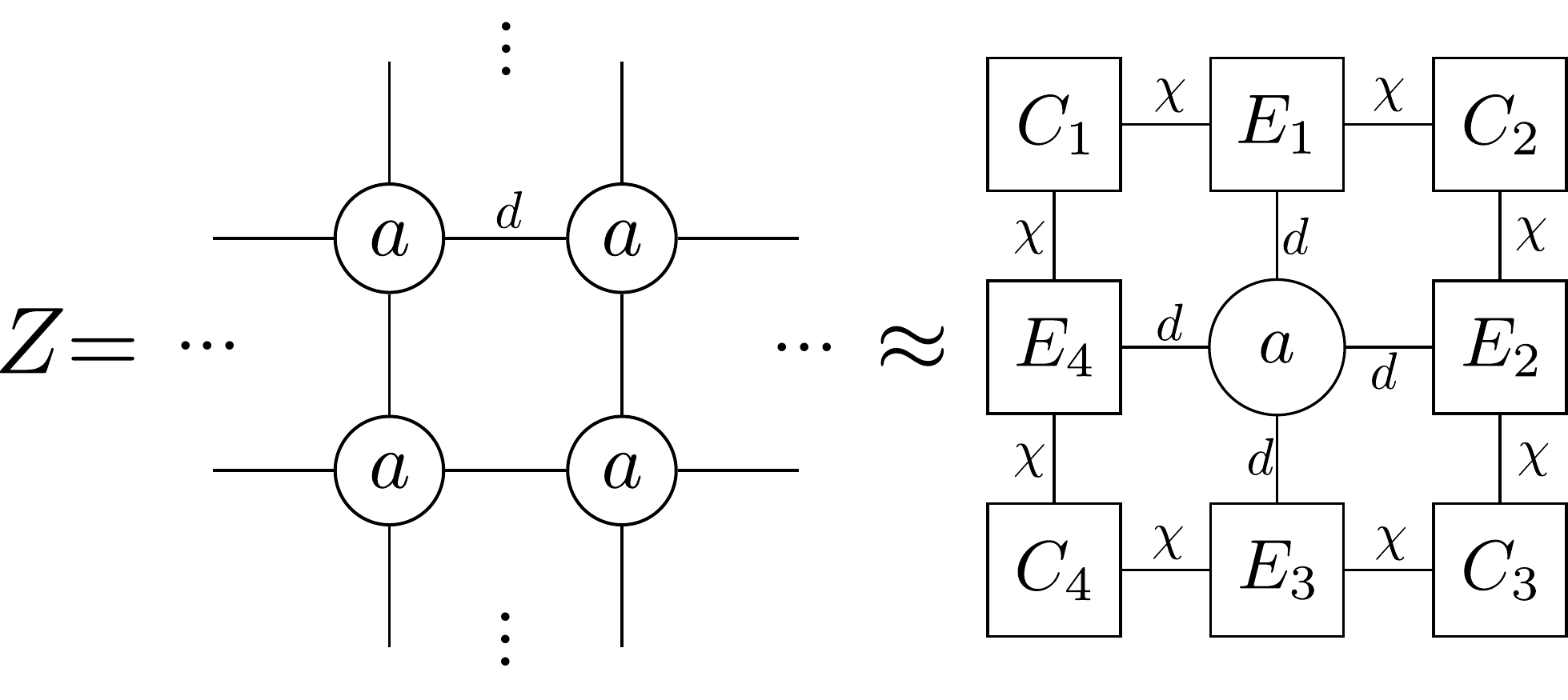}
    \caption{Illustration of the Corner Transfer Matrix Renormalisation Group (CTMRG) algorithm.}
    \label{fig:ctmrg_main}
\end{figure}

The Corner Transfer Matrix Renormalization Group (CTMRG) algorithm \cite{nishinoCornerTransferMatrix1996,orusSimulationTwodimensionalQuantum2009,vanheckeSolvingFrustratedIsing2021} is a numerical method for 2D classical lattice models which relies on the approximate contraction of a tensor network formulation of the partition function. Since the cost of exact contraction of a 2D tensor network grows exponentially with system size, the CTMRG algorithm probes the thermodynamic limit by approximating the environment around a tensor by finite-dimensional corner ($C_i$) and edge ($E_i$) tensors accounting respectively for a quadrant or half-column of the infinite two-dimensional tensor network, as illustrated in Fig.~\ref{fig:ctmrg_main}. After the environment tensors are initialized with the desired boundary conditions, sites are iteratively added to the lattice until convergence has been reached. Once the environment tensors $C_i$ and $E_i$ have converged, the expectation value of a local operator can be calculated by performing a contraction similar to that of FIG.~\ref{fig:ctmrg_main} where the tensor $a$ in the center is replaced by a tensor representation of the operator.\par
%%%%%%%%%%%%%%%%%%% FIG 14 %%%%%%%%%%%%%%%%
%%%%%%%%%%%%%%%%%%%%%%%%%%%%%%%%%%%%%%%
The cut-off bond dimension $\chi$ to which the corner and edge tensors are truncated after each CTMRG step controls the precision of the algorithm.

%%%%%%%%%%%%%%%%%%%%%%%%%%%%% FIG 15 %%%%%%%%%%%%%%%%
\section{Crystal Structure and Phase Purity of \ErSSL{}}
\label{App:Crystal Structure and Phase Purity}
\begin{figure}[]
\centering
\includegraphics[width=0.48\textwidth]{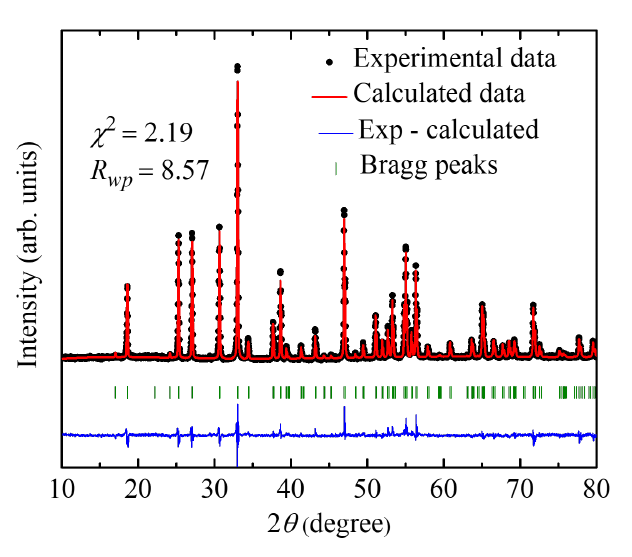}
\caption{Powder X-ray diffraction pattern of \ErSSL{} at room temperature. The observed (black markers) and calculated (red line) patterns are shown along with their difference (blue line) and positions of Bragg peaks (green ticks). The calculated pattern is obtained through Rietveld refinement using the tetragonal structure with space group $P\bar 42_1m$. The good agreement between observed and calculated pattern is evidenced by a weighted profile $R$ factor ($R_{wp}$) of 8.57.}
\label{XRD}
\end{figure}
%%%%%%%%%%%%%%%%%%%%%%%%%%%%%%%%%%%%%%%%%%

\ErSSL{} belongs to the melilite family and crystallizes in the tetragonal structure with space group $P\bar 42_1m$ (no. 113). The crystal structure and phase purity were analyzed using powder X-ray diffraction. FIG. \ref{XRD} displays the diffractogram collected at room temperature, alongside the calculated pattern via Rietveld refinement for the proposed space group $P\bar 42_1m$. The fitted profile exhibits an $R_{wp}$ of 8.57, indicating a good agreement with the experimental data. The phase purity exceeds 99.9\%, with no detectable impurity phase. \ch{Er^{3+}} ions form a 2D network of SSL, with the nearest neighbor Er-Er bond length of 3.31~\text{\AA} and the next nearest neighbor bond length of 3.92~\text{\AA}. The 2D Er layers are separated by non-magnetic [\ch{GeBe2O7}] layers, with an interlayer separation of 4.72~\text{\AA}. After successfully synthesizing the polycrystalline sample, centimeter-sized large single crystals were grown using a floating zone furnace. The phase purity of the grown crystals was confirmed using a powder x-ray diffraction pattern on crushed \ErSSL{} single crystal samples.

%%%%%%%%%%%%%%%%%%% FIG 16 %%%%%%%%%%%%%%%%%%%%%%
\begin{figure}[]
\centering
\includegraphics[width=0.48\textwidth]{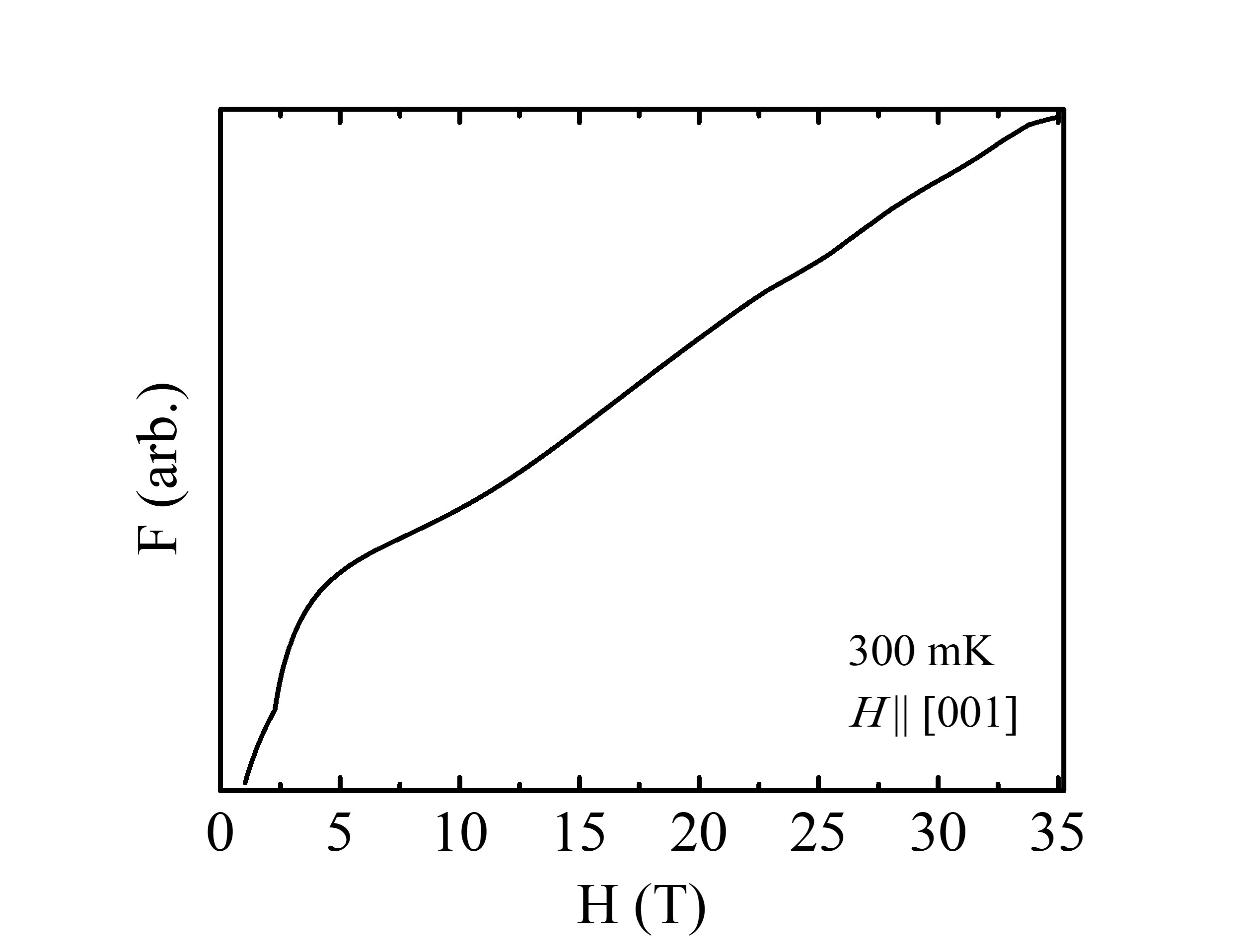}
\caption{Tunnel diode oscillator measurement showing the frequency response of the sample as a field up to 35 T is applied using resistive magnets at 300 mK. In the higher field range, specifically from 7 T to 35 T, no additional peaks are observed, indicating that no further magnetic transitions occur at higher fields.}
\label{TDO}
\end{figure}
%%%%%%%%%%%%%%%%%%%%%%%%%%%%%%%%%%%%%%%%

\section {Tunnel Diode Oscillator measurement}
\label{App_TDO}

To investigate further magnetic transitions at fields higher than 7 T in \ErSSL{}, tunnel diode oscillator (TDO) measurements were conducted at the DC Field Facility of the National High Magnetic Field Laboratory in Tallahassee. A bar-shaped single-crystal sample, approximately 1.8 mm in length and 0.9 mm in width, was placed inside a detection coil to align the [001] direction along the coil axis and the direction of the applied field. This setup formed the inductive component of an LC circuit connected to a tunnel diode, adjusted to resonate within a frequency range of 10 to 50 MHz. The shift in resonance frequency \(f\), reflecting changes in magnetization \(M\) (\(f \propto dM/dH\)), was observed as the magnetic field was increased to 35 T using resistive magnets, with the sample's temperature maintained at 300 mK. The measurements, shown in FIG.~\ref{TDO}, show a smooth linear frequency response up to 35 T, without any indication of peaks or transitions beyond 7 T.

\section{Additional Single Crystal Neutron Diffraction} \label{Wand}
\begin{figure}[]
\centering
\includegraphics[width=0.48\textwidth]{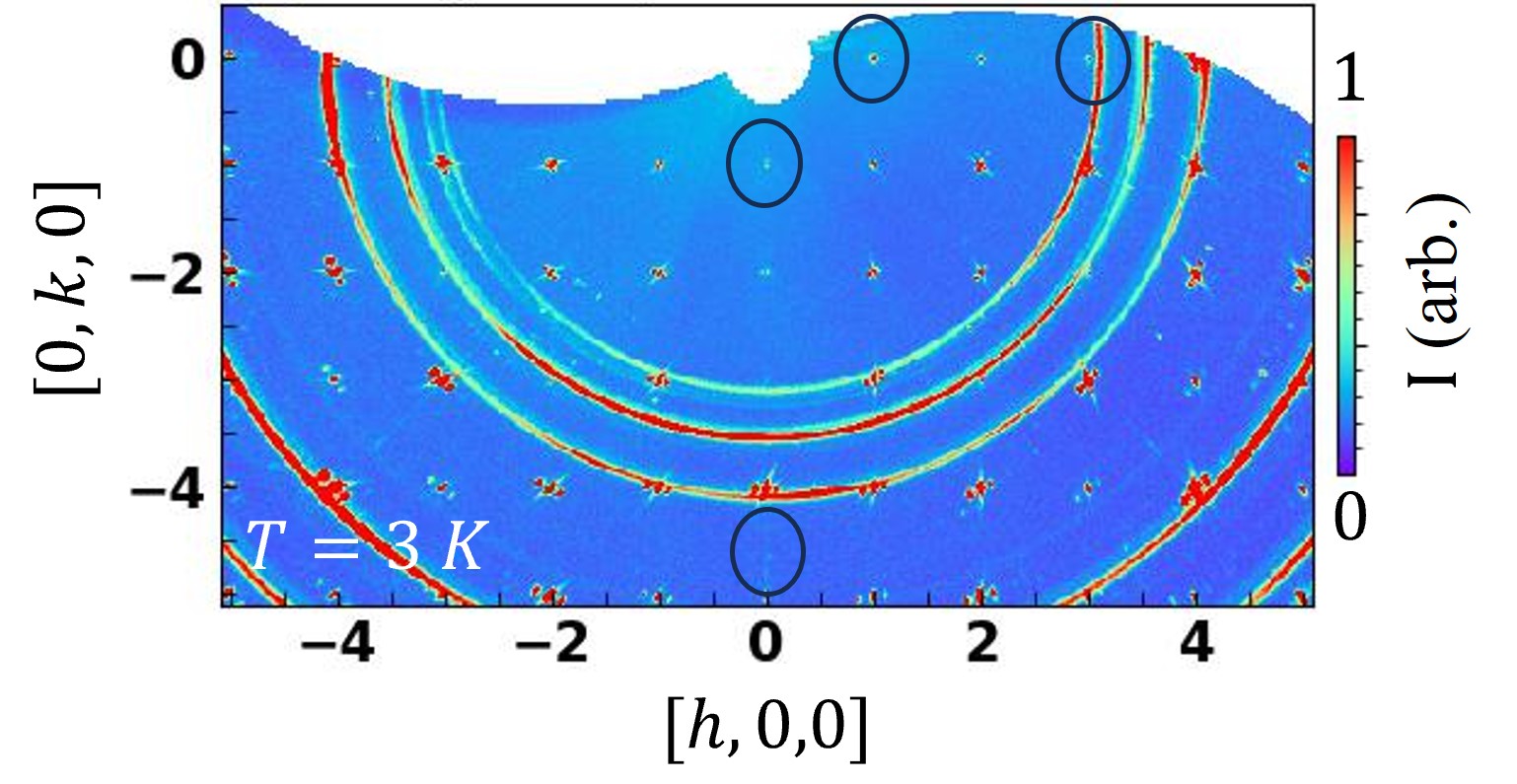}
\caption{ Neutron diffraction data from the WAND$^2$ diffractometer at ORNL, showing peaks corresponding to the $(2h+1,0,0)$ and $(0,2k+1,0)$ families, indicated by black circles. The data was collected in the paramagnetic phase at 3 K, where only structural peaks are expected. The visible powder rings are due to the Cu mount and aluminum foil used to cover the sample, in accordance with Beryllium-containing sample handling guidelines at ORNL.}
\label{wand}
\end{figure}

Neutron diffraction measurements were performed on a single crystal using the WAND$^2$ diffractometer at ORNL to examine the structural properties of the sample in the paramagnetic state. As shown in Fig.~\ref{wand}, the diffraction pattern collected at $T = 3$~K reveals peaks corresponding to the $(2h+1,0,0)$ and $(0,2k+1,0)$ families of reflections, which are marked by black circles. These reflections are consistent with the expected crystallographic symmetry and arise solely from the nuclear structure, as no magnetic ordering is present at this temperature. The absence of magnetic Bragg peaks confirms that the measurement was conducted well above any magnetic transition temperature. In addition to the structural peaks, powder rings are visible in the diffraction image. These features originate from the copper sample mount and the aluminum foil used to enclose the crystal, both of which are required for safe handling of beryllium-containing materials at ORNL. These background features have been taken into account in the analysis and do not affect the identification of the intrinsic structural reflections.

%%%%%%%%%%%%%%%%%%%%%%%%%%%%%% TABLE II %%%%%%%%%%%%%%%%%%%%%%%%
%\clearpage
\begin{table}
\centering
\caption{Crystal field parameters \( B_n^m \) calculated using the point charge model and after fitting the experimental data. The values are reported in units of meV. }
\label{Blm_table}
\begin{tabular}{|c|c|c|}
\hline
\( B_n^m \) & Point Charge Model &  Fit \\
\hline
\( B_2^0 \times 10^{3} \) & -229 & -2.611 \\
\( B_2^1 \times 10^{1} \) & -6.56 & -4.736 \\
\( B_2^2 \times 10^{1} \) & 2.19 & 1.325 \\
\( B_4^0 \times 10^{4} \) & -5.47 & -0.670 \\
\( B_4^1 \times 10^{3} \) & -1.57 & 2.273 \\
\( B_4^2 \times 10^{3} \) & 0.489 & -2.988 \\
\( B_4^3 \times 10^{4} \) & 41.05 & -13.240 \\
\( B_4^4 \times 10^{5} \) & -96.54 & -57.425 \\
\( B_6^0 \times 10^{6} \) & 2.88 & -0.889 \\
\( B_6^1 \times 10^{4} \) & 0.214 & 1.101 \\
\( B_6^2 \times 10^{5} \) & 0.070 & -6.066 \\
\( B_6^3 \times 10^{5} \) & -1.21 & 7.818 \\
\( B_6^4 \times 10^{5} \) & -0.289 & 0.557 \\
\( B_6^5 \times 10^{5} \) & -1.42 & 7.547 \\
\( B_6^6 \times 10^{6} \) & -4.08 & 3.262 \\
\hline
\end{tabular}
\end{table}
%%%%%%%%%%%%%%%%%%%%%%%%%%%%%%%%%%%%%%%%%%%%%%%%%%%%%%%%%

%%%%%%%%%%%%%%%%%%%%%%%% FIG 17 %%%%%%%%%%%%%%%%%
%%%%%%%%%%%%%%%%%%%%%%%%%%%%%%%%%%%%%%%%%%%%%

%%%%%%%%%%%%%%%%%%%%%% FIG 18 %%%%%%%%%%%%%%%%%%%
\begin{figure}[]
\centering
\includegraphics[width=0.48\textwidth]{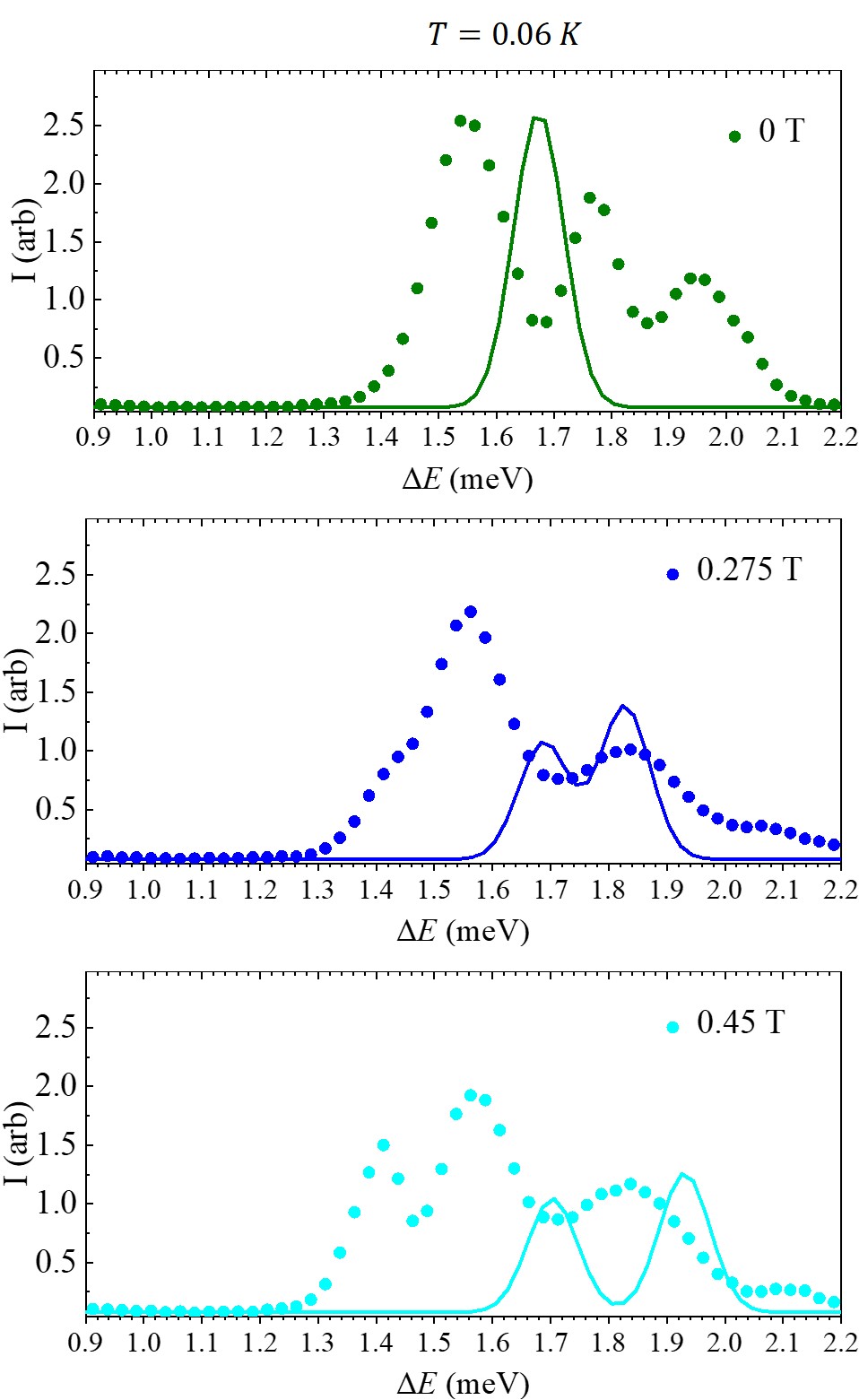}
\caption{Field dependence of the first crystal electric field (CEF) level at \(\approx 1.6\) meV at 0.06 K. The scatter points represent the experimental data obtained by integrating over the full momentum coverage accessible in the HYSPEC experiment. The solid lines correspond to the calculated field dependence of the CEF level using a single-ion CEF Hamiltonian for \(\mathbf{H} \parallel [001]\). As expected, the single-ion model predicts the splitting of the Kramers doublet into two levels with increasing magnetic field. However, experimentally, the CEF doublet shows additional splitting into multiple levels, likely due to molecular fields caused by the ordered spins. }
\label{fig:hyspeclinecuts}
\end{figure}
%%%%%%%%%%%%%%%%%%%%%%%%%%%%%%%%%%%%%%%%%%%%%%%%%%%%%%%%%%%%%%%%%%%%%%%

\section{Experimental determination of the Crystal Electric Field Hamiltonian}
\label{CEF_app}
The inelastic neutron scattering data of \ErSSL{} were analyzed using the single-ion CEF Hamiltonian discussed in the main text. The fitting was constrained to match the six observed energy levels and their corresponding intensities, along with magnetization and susceptibility data for the $[001]$ and $[010]$ directions. The position of the seventh level was not constrained; however, its intensity was restricted to remain within the background level, as it could not be discerned and was expected to be of the same order as the background. The full list of energy levels and the corresponding eigenvectors derived from the fitting is listed in Table~\ref{eigen}. The initial and the optimized $B^m_n$ parameters are provided in Table~\ref{Blm_table}. Note that the the crystal-field parameters presented in table \ref{eigen} were calculated in a coordinate system which has $x$ parallel to [110] and $z$ parallel to [001], as discussed in the main text.
The basis of the CEF ground state was chosen such that the projection of intra-dimer Heisenberg interactions leads to a diagonal effective coupling matrix.
The \( g \)-tensor calculated from the ground state eigenvectors is as follows :
\begin{equation}
\label{eq_gtensor}
g = \begin{pmatrix}
-3.527 & 0 & 0.580 \\
0 & -5.755 & 0 \\
-1.531 & 0 & 9.920
\end{pmatrix}.
\end{equation}
In the HYSPEC experiment, we resolved the first excited crystal electric field state around 1.6 meV and tracked its evolution as a function of applied magnetic field along the [001] direction. The results, shown in Fig.~\ref{fig:hyspeclinecuts}, reveal a splitting pattern that is only partially captured by the single-ion CEF model. While the calculated spectrum correctly predicts the linear Zeeman splitting of the Kramers doublet, the experimental data show additional field-induced features that are likely due to internal molecular fields originating from the long-range magnetic order.

%%%%%%%%%%%%%%%%%%%%%%%%%%%%%%%%% TABLE III %%%%%%%%%%%%%%%
\begin{table*}[!htb]
\centering
\caption{Eigenvalues and their corresponding eigenvectors for the single-ion CEF Hamiltonian, derived from fitting the \ErSSL{} INS data as discussed in the main text. Wavefunctions are presented in the \(\ket{m_J}\) basis.}
\label{eigen}
\renewcommand{\arraystretch}{1.5}
\begin{tabular}{|c|*{16}{c|}}
\hline
E (meV) & \(\ket{\frac{15}{2}}\) & \(\ket{\frac{13}{2}}\) & \(\ket{\frac{11}{2}}\) & \(\ket{\frac{9}{2}}\) & \(\ket{\frac{7}{2}}\) & \(\ket{\frac{5}{2}}\) & \(\ket{\frac{3}{2}}\) & \(\ket{\frac{1}{2}}\) & \(\ket{\frac{-1}{2}}\) & \(\ket{\frac{-3}{2}}\) & \(\ket{\frac{-5}{2}}\) & \(\ket{\frac{-7}{2}}\) & \(\ket{\frac{-9}{2}}\) & \(\ket{\frac{-11}{2}}\) & \(\ket{\frac{-13}{2}}\) & \(\ket{\frac{-15}{2}}\) \\
\hline
0.000 & -0.094 & -0.109 & 0.08 & -0.206 & 0.128 & 0.165 & -0.039 & -0.257 & -0.117 & 0.251 & -0.198 & -0.1 & 0.283 & 0.513 & 0.061 & 0.589 \\
0.000 & 0.589 & -0.061 & 0.513 & -0.283 & -0.1 & 0.198 & 0.251 & 0.117 & -0.257 & 0.039 & 0.165 & -0.128 & -0.206 & -0.08 & -0.109 & 0.094  \\
1.674 & -0.650 & -0.334 & -0.022 & -0.322 & 0.047 & 0.189 & 0.149 & -0.023 & -0.254 & 0.081 & 0.175 & -0.173 & -0.305 & -0.228 & -0.155 & -0.034 \\
1.674 & 0.034 & -0.155 & 0.228 & -0.305 & 0.173 & 0.175 & -0.081 & -0.254 & 0.023 & 0.149 & -0.189 & 0.047 & 0.322 & -0.022 & 0.334 & -0.650 \\
6.240 & 0.208 & 0.246 & -0.145 & 0.093 & 0.291 & 0.075 & -0.320 & -0.481 & -0.009 & 0.342 & -0.117 & -0.272 & -0.298 & -0.293 & -0.252 & 0.009 \\
6.240 & 0.009 & 0.252 & -0.293 & 0.298 & -0.272 & 0.117 & 0.342 & 0.009 & -0.481 & 0.320 & 0.075 & -0.291 & 0.093 & 0.145 & 0.246 & -0.208 \\
15.604 & 0.001 & 0.169 & -0.156 & -0.005 & -0.189 & 0.538 & 0.440 & -0.369 & 0.199 & -0.212 & -0.132 & 0.348 & 0.080 & -0.130 & -0.209 & 0.050 \\
15.604 & -0.050 & -0.209 & 0.130 & 0.080 & -0.348 & -0.132 & 0.212 & 0.199 & 0.369 & 0.440 & -0.538 & -0.189 & 0.005 & -0.156 & -0.169 & 0.001 \\
23.191 & -0.034 & 0.450 & -0.137 & -0.329 & 0.565 & -0.070 & 0.396 & 0.374 & 0.155 & 0.130 & -0.083 & -0.025 & 0.025 & -0.002 & -0.008 & 0.000 \\
23.191 & 0.000 & 0.008 & -0.002 & -0.025 & -0.025 & 0.083 & 0.130 & -0.155 & 0.374 & -0.396 & -0.070 & -0.565 & -0.329 & 0.137 & 0.450 & 0.034 \\
28.002 & 0.001 & -0.071 & 0.108 & -0.011 & 0.076 & -0.691 & 0.446 & -0.497 & -0.167 & -0.131 & -0.043 & 0.043 & 0.029 & -0.030 & -0.066 & -0.006 \\
28.002 & 0.006 & -0.066 & 0.030 & 0.029 & -0.043 & -0.043 & 0.131 & -0.167 & 0.497 & 0.446 & 0.691 & 0.076 & 0.011 & 0.108 & 0.071 & 0.001 \\
45.056 & -0.265 & 0.631 & 0.265 & -0.384 & -0.475 & -0.157 & -0.229 & -0.097 & 0.012 & 0.017 & 0.028 & -0.005 & -0.005 & 0.001 & 0.003 & 0.000 \\
45.056 & 0.000 & -0.003 & 0.001 & 0.005 & -0.005 & -0.028 & 0.017 & -0.012 & -0.097 & 0.229 & -0.157 & 0.475 & -0.384 & -0.265 & 0.631 & 0.265 \\
62.195 & 0.000 & -0.000 & 0.000 & 0.000 & -0.004 & -0.004 & 0.009 & -0.011 & -0.003 & 0.083 & -0.157 & 0.269 & -0.565 & 0.655 & -0.206 & -0.323 \\
62.195 & 0.323 & -0.206 & -0.655 & -0.565 & -0.269 & -0.157 & -0.083 & -0.003 & 0.011 & 0.009 & 0.004 & -0.004 & -0.000 & 0.000 & 0.000 & 0.000 \\
\hline
\end{tabular}
\end{table*}
%%%%%%%%%%%%%%%%%%%%%%%%%%%%%%%%%%%%%%%%%%%%

To verify that the forbidden (1,0,0) peak observed in the HYSPEC experiment is not due to multiple scattering or higher-order wavelength contamination, we conducted a neutron diffraction experiment at the WAND$^2$ diffractometer at ORNL at 3 K in the paramagnetic phase of the \ErSSL{}. This experiment utilized a Ge(113) monochromator with an incident wavelength of 1.49 \AA{} and employed a different single crystal sample aligned in the $[h,k,0]$ plane. Artifacts such as multiple scattering or higher-order contamination can produce spurious forbidden peaks in single crystal diffraction. However, in this WAND experiment, we observed peaks corresponding to the $(2h+1,0,0)$ and $(0,2k+1,0)$ families (Fig. \ref{wand}), extending to higher $h$ and $k$, effectively ruling out accidental multiple reflections. Since forbidden peaks were observed in both experiments, which used different incident wavelengths and different single crystal samples, we conclude that the forbidden peak arises from intrinsic structural changes, not experimental artifacts.

%\clearpage

\bibliography{references}

\begin{thebibliography}{46}%
\makeatletter
\providecommand \@ifxundefined [1]{%
 \@ifx{#1\undefined}
}%
\providecommand \@ifnum [1]{%
 \ifnum #1\expandafter \@firstoftwo
 \else \expandafter \@secondoftwo
 \fi
}%
\providecommand \@ifx [1]{%
 \ifx #1\expandafter \@firstoftwo
 \else \expandafter \@secondoftwo
 \fi
}%
\providecommand \natexlab [1]{#1}%
\providecommand \enquote  [1]{``#1''}%
\providecommand \bibnamefont  [1]{#1}%
\providecommand \bibfnamefont [1]{#1}%
\providecommand \citenamefont [1]{#1}%
\providecommand \href@noop [0]{\@secondoftwo}%
\providecommand \href [0]{\begingroup \@sanitize@url \@href}%
\providecommand \@href[1]{\@@startlink{#1}\@@href}%
\providecommand \@@href[1]{\endgroup#1\@@endlink}%
\providecommand \@sanitize@url [0]{\catcode `\\12\catcode `\$12\catcode
  `\&12\catcode `\#12\catcode `\^12\catcode `\_12\catcode `\%12\relax}%
\providecommand \@@startlink[1]{}%
\providecommand \@@endlink[0]{}%
\providecommand \url  [0]{\begingroup\@sanitize@url \@url }%
\providecommand \@url [1]{\endgroup\@href {#1}{\urlprefix }}%
\providecommand \urlprefix  [0]{URL }%
\providecommand \Eprint [0]{\href }%
\providecommand \doibase [0]{https://doi.org/}%
\providecommand \selectlanguage [0]{\@gobble}%
\providecommand \bibinfo  [0]{\@secondoftwo}%
\providecommand \bibfield  [0]{\@secondoftwo}%
\providecommand \translation [1]{[#1]}%
\providecommand \BibitemOpen [0]{}%
\providecommand \bibitemStop [0]{}%
\providecommand \bibitemNoStop [0]{.\EOS\space}%
\providecommand \EOS [0]{\spacefactor3000\relax}%
\providecommand \BibitemShut  [1]{\csname bibitem#1\endcsname}%
\let\auto@bib@innerbib\@empty
%</preamble>
\bibitem [{\citenamefont {Balents}(2010)}]{Balents2010SpinMagnets}%
  \BibitemOpen
  \bibfield  {author} {\bibinfo {author} {\bibfnamefont {L.}~\bibnamefont
  {Balents}},\ }\bibfield  {title} {\bibinfo {title} {Spin liquids in
  frustrated magnets},\ }\href@noop {} {\bibfield  {journal} {\bibinfo
  {journal} {nature}\ }\textbf {\bibinfo {volume} {464}},\ \bibinfo {pages}
  {199} (\bibinfo {year} {2010})}\BibitemShut {NoStop}%
\bibitem [{\citenamefont {Knolle}\ and\ \citenamefont
  {Moessner}(2019)}]{Knolle2019ALiquids}%
  \BibitemOpen
  \bibfield  {author} {\bibinfo {author} {\bibfnamefont {J.}~\bibnamefont
  {Knolle}}\ and\ \bibinfo {author} {\bibfnamefont {R.}~\bibnamefont
  {Moessner}},\ }\bibfield  {title} {\bibinfo {title} {{A field guide to spin
  liquids}},\ }\href {https://doi.org/10.1146/annurev-conmatphys-031218-013401}
  {\bibfield  {journal} {\bibinfo  {journal} {Annual Review of Condensed Matter
  Physics}\ }\textbf {\bibinfo {volume} {10}},\ \bibinfo {pages} {451}
  (\bibinfo {year} {2019})}\BibitemShut {NoStop}%
\bibitem [{\citenamefont {Takigawa}\ and\ \citenamefont
  {Mila}(2011)}]{TakigawaMila2011}%
  \BibitemOpen
  \bibfield  {author} {\bibinfo {author} {\bibfnamefont {M.}~\bibnamefont
  {Takigawa}}\ and\ \bibinfo {author} {\bibfnamefont {F.}~\bibnamefont
  {Mila}},\ }\bibinfo {title} {Magnetization plateaus},\ in\ \href
  {https://doi.org/10.1007/978-3-642-10589-0_10} {\emph {\bibinfo {booktitle}
  {Introduction to Frustrated Magnetism: Materials, Experiments, Theory}}},\
  \bibinfo {editor} {edited by\ \bibinfo {editor} {\bibfnamefont
  {C.}~\bibnamefont {Lacroix}}, \bibinfo {editor} {\bibfnamefont
  {P.}~\bibnamefont {Mendels}},\ and\ \bibinfo {editor} {\bibfnamefont
  {F.}~\bibnamefont {Mila}}}\ (\bibinfo  {publisher} {Springer},\ \bibinfo
  {address} {Berlin, Heidelberg},\ \bibinfo {year} {2011})\ p.\ \bibinfo
  {pages} {241}\BibitemShut {NoStop}%
\bibitem [{\citenamefont {Kageyama}\ \emph {et~al.}(1999)\citenamefont
  {Kageyama}, \citenamefont {Onizuka}, \citenamefont {Yamauchi}, \citenamefont
  {Ueda}, \citenamefont {Hane}, \citenamefont {Mitamura}, \citenamefont {Goto},
  \citenamefont {Yoshimura},\ and\ \citenamefont {Kosuge}}]{Kageyama1999}%
  \BibitemOpen
  \bibfield  {author} {\bibinfo {author} {\bibfnamefont {H.}~\bibnamefont
  {Kageyama}}, \bibinfo {author} {\bibfnamefont {K.}~\bibnamefont {Onizuka}},
  \bibinfo {author} {\bibfnamefont {T.}~\bibnamefont {Yamauchi}}, \bibinfo
  {author} {\bibfnamefont {Y.}~\bibnamefont {Ueda}}, \bibinfo {author}
  {\bibfnamefont {S.}~\bibnamefont {Hane}}, \bibinfo {author} {\bibfnamefont
  {H.}~\bibnamefont {Mitamura}}, \bibinfo {author} {\bibfnamefont
  {T.}~\bibnamefont {Goto}}, \bibinfo {author} {\bibfnamefont {K.}~\bibnamefont
  {Yoshimura}},\ and\ \bibinfo {author} {\bibfnamefont {K.}~\bibnamefont
  {Kosuge}},\ }\href {https://doi.org/10.1143/JPSJ.68.1821} {\emph {\bibinfo
  {title} {Journal of the Physical Society of Japan}}},\ \bibinfo {type} {Tech.
  Rep.}\ \bibinfo {number} {6}\ (\bibinfo {year} {1999})\BibitemShut {NoStop}%
\bibitem [{\citenamefont {Onizuka}\ \emph {et~al.}(2000)\citenamefont
  {Onizuka}, \citenamefont {Kageyama}, \citenamefont {Narumi}, \citenamefont
  {Kindo}, \citenamefont {Ueda},\ and\ \citenamefont {Goto}}]{Onizuka2000}%
  \BibitemOpen
  \bibfield  {author} {\bibinfo {author} {\bibfnamefont {K.}~\bibnamefont
  {Onizuka}}, \bibinfo {author} {\bibfnamefont {H.}~\bibnamefont {Kageyama}},
  \bibinfo {author} {\bibfnamefont {Y.}~\bibnamefont {Narumi}}, \bibinfo
  {author} {\bibfnamefont {K.}~\bibnamefont {Kindo}}, \bibinfo {author}
  {\bibfnamefont {Y.}~\bibnamefont {Ueda}},\ and\ \bibinfo {author}
  {\bibfnamefont {T.}~\bibnamefont {Goto}},\ }\bibfield  {title} {\bibinfo
  {title} {1/3 magnetization plateau in {SrCu$_{2}$(BO$_{3}$)$_{2}$} - stripe
  order of excited triplets -},\ }\href {https://doi.org/10.1143/JPSJ.69.1016}
  {\bibfield  {journal} {\bibinfo  {journal} {J. Phys. Soc. Jpn.}\ }\textbf
  {\bibinfo {volume} {69}},\ \bibinfo {pages} {1016} (\bibinfo {year}
  {2000})}\BibitemShut {NoStop}%
\bibitem [{\citenamefont {Kodama}\ \emph {et~al.}(2002)\citenamefont {Kodama},
  \citenamefont {Takigawa}, \citenamefont {Horvati{\'{c}}}, \citenamefont
  {Berthier}, \citenamefont {Kageyama}, \citenamefont {Ueda}, \citenamefont
  {Miyahara}, \citenamefont {Becca},\ and\ \citenamefont {Mila}}]{Kodama2002}%
  \BibitemOpen
  \bibfield  {author} {\bibinfo {author} {\bibfnamefont {K.}~\bibnamefont
  {Kodama}}, \bibinfo {author} {\bibfnamefont {M.}~\bibnamefont {Takigawa}},
  \bibinfo {author} {\bibfnamefont {M.}~\bibnamefont {Horvati{\'{c}}}},
  \bibinfo {author} {\bibfnamefont {C.}~\bibnamefont {Berthier}}, \bibinfo
  {author} {\bibfnamefont {H.}~\bibnamefont {Kageyama}}, \bibinfo {author}
  {\bibfnamefont {Y.}~\bibnamefont {Ueda}}, \bibinfo {author} {\bibfnamefont
  {S.}~\bibnamefont {Miyahara}}, \bibinfo {author} {\bibfnamefont
  {F.}~\bibnamefont {Becca}},\ and\ \bibinfo {author} {\bibfnamefont
  {F.}~\bibnamefont {Mila}},\ }\bibfield  {title} {\bibinfo {title} {{Magnetic
  superstructure in the two-dimensional quantum antiferromagnet
  \ch{SrCu2(BO3)2}}},\ }\href {https://doi.org/10.1126/science.1075045}
  {\bibfield  {journal} {\bibinfo  {journal} {Science}\ }\textbf {\bibinfo
  {volume} {298}},\ \bibinfo {pages} {395} (\bibinfo {year}
  {2002})}\BibitemShut {NoStop}%
\bibitem [{\citenamefont {Jaime}\ \emph {et~al.}(2012)\citenamefont {Jaime},
  \citenamefont {Daou}, \citenamefont {Crooker}, \citenamefont {Weickert},
  \citenamefont {Uchida}, \citenamefont {Feiguin}, \citenamefont {Batista},
  \citenamefont {Dabkowska},\ and\ \citenamefont {Gaulin}}]{Jaime2012}%
  \BibitemOpen
  \bibfield  {author} {\bibinfo {author} {\bibfnamefont {M.}~\bibnamefont
  {Jaime}}, \bibinfo {author} {\bibfnamefont {R.}~\bibnamefont {Daou}},
  \bibinfo {author} {\bibfnamefont {S.~A.}\ \bibnamefont {Crooker}}, \bibinfo
  {author} {\bibfnamefont {F.}~\bibnamefont {Weickert}}, \bibinfo {author}
  {\bibfnamefont {A.}~\bibnamefont {Uchida}}, \bibinfo {author} {\bibfnamefont
  {A.~E.}\ \bibnamefont {Feiguin}}, \bibinfo {author} {\bibfnamefont {C.~D.}\
  \bibnamefont {Batista}}, \bibinfo {author} {\bibfnamefont {H.~A.}\
  \bibnamefont {Dabkowska}},\ and\ \bibinfo {author} {\bibfnamefont {B.~D.}\
  \bibnamefont {Gaulin}},\ }\bibfield  {title} {\bibinfo {title}
  {Magnetostriction and magnetic texture to 100.75 tesla in frustrated srcu2
  (bo3) 2},\ }\href@noop {} {\bibfield  {journal} {\bibinfo  {journal}
  {Proceedings of the National Academy of Sciences}\ }\textbf {\bibinfo
  {volume} {109}},\ \bibinfo {pages} {12404} (\bibinfo {year}
  {2012})}\BibitemShut {NoStop}%
\bibitem [{\citenamefont {Takigawa}\ \emph {et~al.}(2013)\citenamefont
  {Takigawa}, \citenamefont {Horvati\'{c}}, \citenamefont {Waki}, \citenamefont
  {Kr\"amer}, \citenamefont {Berthier}, \citenamefont {L\'evy-Bertrand},
  \citenamefont {Sheikin}, \citenamefont {Kageyama}, \citenamefont {Ueda},\
  and\ \citenamefont {Mila}}]{Takigawa2013}%
  \BibitemOpen
  \bibfield  {author} {\bibinfo {author} {\bibfnamefont {M.}~\bibnamefont
  {Takigawa}}, \bibinfo {author} {\bibfnamefont {M.}~\bibnamefont
  {Horvati\'{c}}}, \bibinfo {author} {\bibfnamefont {T.}~\bibnamefont {Waki}},
  \bibinfo {author} {\bibfnamefont {S.}~\bibnamefont {Kr\"amer}}, \bibinfo
  {author} {\bibfnamefont {C.}~\bibnamefont {Berthier}}, \bibinfo {author}
  {\bibfnamefont {F.}~\bibnamefont {L\'evy-Bertrand}}, \bibinfo {author}
  {\bibfnamefont {I.}~\bibnamefont {Sheikin}}, \bibinfo {author} {\bibfnamefont
  {H.}~\bibnamefont {Kageyama}}, \bibinfo {author} {\bibfnamefont
  {Y.}~\bibnamefont {Ueda}},\ and\ \bibinfo {author} {\bibfnamefont
  {F.}~\bibnamefont {Mila}},\ }\bibfield  {title} {\bibinfo {title} {Incomplete
  devil's staircase in the magnetization curve of
  {SrCu$_{2}$(BO$_{3}$)$_{2}$}},\ }\href
  {https://doi.org/10.1103/PhysRevLett.110.067210} {\bibfield  {journal}
  {\bibinfo  {journal} {Phys. Rev. Lett.}\ }\textbf {\bibinfo {volume} {110}},\
  \bibinfo {pages} {067210} (\bibinfo {year} {2013})}\BibitemShut {NoStop}%
\bibitem [{\citenamefont {Matsuda}\ \emph {et~al.}(2013)\citenamefont
  {Matsuda}, \citenamefont {Abe}, \citenamefont {Takeyama}, \citenamefont
  {Kageyama}, \citenamefont {Corboz}, \citenamefont {Honecker}, \citenamefont
  {Manmana}, \citenamefont {Foltin}, \citenamefont {Schmidt},\ and\
  \citenamefont {Mila}}]{Matsuda2013}%
  \BibitemOpen
  \bibfield  {author} {\bibinfo {author} {\bibfnamefont {Y.~H.}\ \bibnamefont
  {Matsuda}}, \bibinfo {author} {\bibfnamefont {N.}~\bibnamefont {Abe}},
  \bibinfo {author} {\bibfnamefont {S.}~\bibnamefont {Takeyama}}, \bibinfo
  {author} {\bibfnamefont {H.}~\bibnamefont {Kageyama}}, \bibinfo {author}
  {\bibfnamefont {P.}~\bibnamefont {Corboz}}, \bibinfo {author} {\bibfnamefont
  {A.}~\bibnamefont {Honecker}}, \bibinfo {author} {\bibfnamefont {S.~R.}\
  \bibnamefont {Manmana}}, \bibinfo {author} {\bibfnamefont {G.~R.}\
  \bibnamefont {Foltin}}, \bibinfo {author} {\bibfnamefont {K.~P.}\
  \bibnamefont {Schmidt}},\ and\ \bibinfo {author} {\bibfnamefont
  {F.}~\bibnamefont {Mila}},\ }\bibfield  {title} {\bibinfo {title}
  {{Magnetization of \ch{SrCu2(BO3)2} in ultrahigh magnetic fields up to 118
  T}},\ }\bibfield  {journal} {\bibinfo  {journal} {Physical Review Letters}\
  }\textbf {\bibinfo {volume} {111}},\ \href
  {https://doi.org/10.1103/PhysRevLett.111.137204}
  {10.1103/PhysRevLett.111.137204} (\bibinfo {year} {2013})\BibitemShut
  {NoStop}%
\bibitem [{\citenamefont {Corboz}\ and\ \citenamefont
  {Mila}(2014)}]{Corboz2014}%
  \BibitemOpen
  \bibfield  {author} {\bibinfo {author} {\bibfnamefont {P.}~\bibnamefont
  {Corboz}}\ and\ \bibinfo {author} {\bibfnamefont {F.}~\bibnamefont {Mila}},\
  }\bibfield  {title} {\bibinfo {title} {{Crystals of bound states in the
  magnetization plateaus of the shastry-sutherland model}},\ }\href
  {https://doi.org/10.1103/PhysRevLett.112.147203} {\bibfield  {journal}
  {\bibinfo  {journal} {Physical Review Letters}\ }\textbf {\bibinfo {volume}
  {112}},\ \bibinfo {pages} {1} (\bibinfo {year} {2014})}\BibitemShut {NoStop}%
\bibitem [{\citenamefont {{Sriram Shastry}}\ and\ \citenamefont
  {Sutherland}(1981)}]{SriramShastry1981}%
  \BibitemOpen
  \bibfield  {author} {\bibinfo {author} {\bibfnamefont {B.}~\bibnamefont
  {{Sriram Shastry}}}\ and\ \bibinfo {author} {\bibfnamefont {B.}~\bibnamefont
  {Sutherland}},\ }\bibfield  {title} {\bibinfo {title} {{Exact ground state of
  a quantum mechanical antiferromagnet}},\ }\href
  {https://doi.org/10.1016/0378-4363(81)90838-X} {\bibfield  {journal}
  {\bibinfo  {journal} {Physica B+C}\ }\textbf {\bibinfo {volume} {108}},\
  \bibinfo {pages} {1069} (\bibinfo {year} {1981})}\BibitemShut {NoStop}%
\bibitem [{\citenamefont {Meng}\ and\ \citenamefont
  {Wessel}(2008)}]{mengPhasesMagnetizationProcess2008}%
  \BibitemOpen
  \bibfield  {author} {\bibinfo {author} {\bibfnamefont {Z.~Y.}\ \bibnamefont
  {Meng}}\ and\ \bibinfo {author} {\bibfnamefont {S.}~\bibnamefont {Wessel}},\
  }\bibfield  {title} {\bibinfo {title} {Phases and magnetization process of an
  anisotropic {{Shastry-Sutherland}} model},\ }\href
  {https://doi.org/10.1103/PhysRevB.78.224416} {\bibfield  {journal} {\bibinfo
  {journal} {Physical Review B}\ }\textbf {\bibinfo {volume} {78}},\ \bibinfo
  {pages} {224416} (\bibinfo {year} {2008})}\BibitemShut {NoStop}%
\bibitem [{\citenamefont {Dublenych}(2012)}]{dublenychGroundStatesIsing2012}%
  \BibitemOpen
  \bibfield  {author} {\bibinfo {author} {\bibfnamefont {{\relax Yu}.~I.}\
  \bibnamefont {Dublenych}},\ }\bibfield  {title} {\bibinfo {title} {{Ground
  {{States}} of the {{Ising Model}} on the {{Shastry-Sutherland Lattice}} and
  the {{Origin}} of the {{Fractional Magnetization Plateaus}} in
  {{Rare-Earth-Metal Tetraborides}}}},\ }\href
  {https://doi.org/10.1103/PhysRevLett.109.167202} {\bibfield  {journal}
  {\bibinfo  {journal} {Physical Review Letters}\ }\textbf {\bibinfo {volume}
  {109}},\ \bibinfo {pages} {167202} (\bibinfo {year} {2012})}\BibitemShut
  {NoStop}%
\bibitem [{\citenamefont {Brunt}\ \emph {et~al.}(2018)\citenamefont {Brunt},
  \citenamefont {Balakrishnan}, \citenamefont {Mayoh}, \citenamefont {Lees},
  \citenamefont {Gorbunov}, \citenamefont {Qureshi},\ and\ \citenamefont
  {Petrenko}}]{bruntMagnetisationProcessRare2018}%
  \BibitemOpen
  \bibfield  {author} {\bibinfo {author} {\bibfnamefont {D.}~\bibnamefont
  {Brunt}}, \bibinfo {author} {\bibfnamefont {G.}~\bibnamefont {Balakrishnan}},
  \bibinfo {author} {\bibfnamefont {D.~A.}\ \bibnamefont {Mayoh}}, \bibinfo
  {author} {\bibfnamefont {M.~R.}\ \bibnamefont {Lees}}, \bibinfo {author}
  {\bibfnamefont {D.}~\bibnamefont {Gorbunov}}, \bibinfo {author}
  {\bibfnamefont {N.}~\bibnamefont {Qureshi}},\ and\ \bibinfo {author}
  {\bibfnamefont {O.~A.}\ \bibnamefont {Petrenko}},\ }\bibfield  {title}
  {\bibinfo {title} {{Magnetisation Process in the Rare Earth Tetraborides,
  \ch{NdB4} and \ch{HoB4}}},\ }\href
  {https://doi.org/10.1038/s41598-017-18301-1} {\bibfield  {journal} {\bibinfo
  {journal} {Scientific Reports}\ }\textbf {\bibinfo {volume} {8}},\ \bibinfo
  {pages} {232} (\bibinfo {year} {2018})}\BibitemShut {NoStop}%
\bibitem [{\citenamefont {Suzuki}\ \emph {et~al.}(2010)\citenamefont {Suzuki},
  \citenamefont {Tomita}, \citenamefont {Kawashima},\ and\ \citenamefont
  {Sengupta}}]{suzukiFinitetemperaturePhaseTransition2010}%
  \BibitemOpen
  \bibfield  {author} {\bibinfo {author} {\bibfnamefont {T.}~\bibnamefont
  {Suzuki}}, \bibinfo {author} {\bibfnamefont {Y.}~\bibnamefont {Tomita}},
  \bibinfo {author} {\bibfnamefont {N.}~\bibnamefont {Kawashima}},\ and\
  \bibinfo {author} {\bibfnamefont {P.}~\bibnamefont {Sengupta}},\ }\bibfield
  {title} {\bibinfo {title} {Finite-temperature phase transition to the m = 1 2
  plateau phase in the spin- 1 2 {{X X Z}} model on the {{Shastry-Sutherland}}
  lattices},\ }\href {https://doi.org/10.1103/PhysRevB.82.214404} {\bibfield
  {journal} {\bibinfo  {journal} {Physical Review B}\ }\textbf {\bibinfo
  {volume} {82}},\ \bibinfo {pages} {214404} (\bibinfo {year}
  {2010})}\BibitemShut {NoStop}%
\bibitem [{\citenamefont {Orend{\'a}{\v c}}\ \emph {et~al.}(2021)\citenamefont
  {Orend{\'a}{\v c}}, \citenamefont {Gab{\'a}ni}, \citenamefont {Farka{\v
  s}ovsk{\'y}}, \citenamefont {Ga{\v z}o}, \citenamefont {Ka{\v c}mar{\v
  c}{\'i}k}, \citenamefont {Marcin}, \citenamefont {Prist{\'a}{\v s}},
  \citenamefont {Siemensmeyer}, \citenamefont {Shitsevalova},\ and\
  \citenamefont {Flachbart}}]{orendacGroundStateStability2021}%
  \BibitemOpen
  \bibfield  {author} {\bibinfo {author} {\bibfnamefont {M.}~\bibnamefont
  {Orend{\'a}{\v c}}}, \bibinfo {author} {\bibfnamefont {S.}~\bibnamefont
  {Gab{\'a}ni}}, \bibinfo {author} {\bibfnamefont {P.}~\bibnamefont {Farka{\v
  s}ovsk{\'y}}}, \bibinfo {author} {\bibfnamefont {E.}~\bibnamefont {Ga{\v
  z}o}}, \bibinfo {author} {\bibfnamefont {J.}~\bibnamefont {Ka{\v c}mar{\v
  c}{\'i}k}}, \bibinfo {author} {\bibfnamefont {M.}~\bibnamefont {Marcin}},
  \bibinfo {author} {\bibfnamefont {G.}~\bibnamefont {Prist{\'a}{\v s}}},
  \bibinfo {author} {\bibfnamefont {K.}~\bibnamefont {Siemensmeyer}}, \bibinfo
  {author} {\bibfnamefont {N.}~\bibnamefont {Shitsevalova}},\ and\ \bibinfo
  {author} {\bibfnamefont {K.}~\bibnamefont {Flachbart}},\ }\bibfield  {title}
  {\bibinfo {title} {Ground state and stability of the fractional plateau phase
  in metallic {{Shastry}}--{{Sutherland}} system \ch{TmB4}},\ }\href
  {https://doi.org/10.1038/s41598-021-86353-5} {\bibfield  {journal} {\bibinfo
  {journal} {Scientific Reports}\ }\textbf {\bibinfo {volume} {11}},\ \bibinfo
  {pages} {6835} (\bibinfo {year} {2021})}\BibitemShut {NoStop}%
\bibitem [{\citenamefont {Ishii}\ \emph {et~al.}(2021)\citenamefont {Ishii},
  \citenamefont {Sala}, \citenamefont {Stone}, \citenamefont {Garlea},
  \citenamefont {Calder}, \citenamefont {Chen}, \citenamefont {Yoshida},
  \citenamefont {Fukuoka}, \citenamefont {Yan}, \citenamefont {{Dela Cruz}},
  \citenamefont {Du}, \citenamefont {Parker}, \citenamefont {Zhang},
  \citenamefont {Batista}, \citenamefont {Yamaura},\ and\ \citenamefont
  {Christianson}}]{Ishii2021}%
  \BibitemOpen
  \bibfield  {author} {\bibinfo {author} {\bibfnamefont {Y.}~\bibnamefont
  {Ishii}}, \bibinfo {author} {\bibfnamefont {G.}~\bibnamefont {Sala}},
  \bibinfo {author} {\bibfnamefont {M.~B.}\ \bibnamefont {Stone}}, \bibinfo
  {author} {\bibfnamefont {V.~O.}\ \bibnamefont {Garlea}}, \bibinfo {author}
  {\bibfnamefont {S.}~\bibnamefont {Calder}}, \bibinfo {author} {\bibfnamefont
  {J.}~\bibnamefont {Chen}}, \bibinfo {author} {\bibfnamefont {H.~K.}\
  \bibnamefont {Yoshida}}, \bibinfo {author} {\bibfnamefont {S.}~\bibnamefont
  {Fukuoka}}, \bibinfo {author} {\bibfnamefont {J.}~\bibnamefont {Yan}},
  \bibinfo {author} {\bibfnamefont {C.}~\bibnamefont {{Dela Cruz}}}, \bibinfo
  {author} {\bibfnamefont {M.~H.}\ \bibnamefont {Du}}, \bibinfo {author}
  {\bibfnamefont {D.~S.}\ \bibnamefont {Parker}}, \bibinfo {author}
  {\bibfnamefont {H.}~\bibnamefont {Zhang}}, \bibinfo {author} {\bibfnamefont
  {C.~D.}\ \bibnamefont {Batista}}, \bibinfo {author} {\bibfnamefont
  {K.}~\bibnamefont {Yamaura}},\ and\ \bibinfo {author} {\bibfnamefont {A.~D.}\
  \bibnamefont {Christianson}},\ }\bibfield  {title} {\bibinfo {title}
  {{Magnetic properties of the Shastry-Sutherland lattice material
  \ch{BaNd2ZnO5}}},\ }\bibfield  {journal} {\bibinfo  {journal} {Physical
  Review Materials}\ }\textbf {\bibinfo {volume} {5}},\ \href
  {https://doi.org/10.1103/PhysRevMaterials.5.064418}
  {10.1103/PhysRevMaterials.5.064418} (\bibinfo {year} {2021})\BibitemShut
  {NoStop}%
\bibitem [{\citenamefont {Marshall}\ \emph {et~al.}(2023)\citenamefont
  {Marshall}, \citenamefont {Billingsley}, \citenamefont {Bai}, \citenamefont
  {Ma}, \citenamefont {Kong},\ and\ \citenamefont {Cao}}]{Marshall2023}%
  \BibitemOpen
  \bibfield  {author} {\bibinfo {author} {\bibfnamefont {M.}~\bibnamefont
  {Marshall}}, \bibinfo {author} {\bibfnamefont {B.~R.}\ \bibnamefont
  {Billingsley}}, \bibinfo {author} {\bibfnamefont {X.}~\bibnamefont {Bai}},
  \bibinfo {author} {\bibfnamefont {Q.}~\bibnamefont {Ma}}, \bibinfo {author}
  {\bibfnamefont {T.}~\bibnamefont {Kong}},\ and\ \bibinfo {author}
  {\bibfnamefont {H.}~\bibnamefont {Cao}},\ }\bibfield  {title} {\bibinfo
  {title} {{Field-induced partial disorder in a Shastry-Sutherland lattice}},\
  }\href {https://doi.org/10.1038/s41467-023-39409-1} {\bibfield  {journal}
  {\bibinfo  {journal} {Nature Communications}\ }\textbf {\bibinfo {volume}
  {14}},\ \bibinfo {pages} {1} (\bibinfo {year} {2023})}\BibitemShut {NoStop}%
\bibitem [{\citenamefont {Ashtar}\ \emph {et~al.}(2021)\citenamefont {Ashtar},
  \citenamefont {Bai}, \citenamefont {Xu}, \citenamefont {Wan}, \citenamefont
  {Wei}, \citenamefont {Liu}, \citenamefont {Marwat},\ and\ \citenamefont
  {Tian}}]{Ashtar2021}%
  \BibitemOpen
  \bibfield  {author} {\bibinfo {author} {\bibfnamefont {M.}~\bibnamefont
  {Ashtar}}, \bibinfo {author} {\bibfnamefont {Y.}~\bibnamefont {Bai}},
  \bibinfo {author} {\bibfnamefont {L.}~\bibnamefont {Xu}}, \bibinfo {author}
  {\bibfnamefont {Z.}~\bibnamefont {Wan}}, \bibinfo {author} {\bibfnamefont
  {Z.}~\bibnamefont {Wei}}, \bibinfo {author} {\bibfnamefont {Y.}~\bibnamefont
  {Liu}}, \bibinfo {author} {\bibfnamefont {M.~A.}\ \bibnamefont {Marwat}},\
  and\ \bibinfo {author} {\bibfnamefont {Z.}~\bibnamefont {Tian}},\ }\bibfield
  {title} {\bibinfo {title} {{Structure and Magnetic Properties of Melilite-
  Type Compounds \(\text{RE}_{2}\text{Be}_{2}\text{GeO}_{7}\) (\text{RE}=
  \ch{Pr}, \ch{Nd}, \ch{Gd}--\ch{Yb}) with Rare-Earth Ions on
  Shastry--Sutherland Lattice}},\ }\href
  {https://doi.org/https://doi.org/10.1021/acs.inorgchem.0c03131} {\bibfield
  {journal} {\bibinfo  {journal} {Inorganic Chemistry}\ }\textbf {\bibinfo
  {volume} {60}},\ \bibinfo {pages} {3626} (\bibinfo {year}
  {2021})}\BibitemShut {NoStop}%
\bibitem [{\citenamefont {Liu}\ \emph {et~al.}(2024{\natexlab{a}})\citenamefont
  {Liu}, \citenamefont {Song}, \citenamefont {Cao}, \citenamefont {Ge},
  \citenamefont {Bu}, \citenamefont {Zhou}, \citenamefont {Qin}, \citenamefont
  {Zeng}, \citenamefont {Li}, \citenamefont {Ling}, \citenamefont {Tong},
  \citenamefont {Sheng}, \citenamefont {Yang}, \citenamefont {Wu},
  \citenamefont {Guo},\ and\ \citenamefont {Tian}}]{NdSSL_2024}%
  \BibitemOpen
  \bibfield  {author} {\bibinfo {author} {\bibfnamefont {A.}~\bibnamefont
  {Liu}}, \bibinfo {author} {\bibfnamefont {F.}~\bibnamefont {Song}}, \bibinfo
  {author} {\bibfnamefont {Y.}~\bibnamefont {Cao}}, \bibinfo {author}
  {\bibfnamefont {H.}~\bibnamefont {Ge}}, \bibinfo {author} {\bibfnamefont
  {H.}~\bibnamefont {Bu}}, \bibinfo {author} {\bibfnamefont {J.}~\bibnamefont
  {Zhou}}, \bibinfo {author} {\bibfnamefont {Y.}~\bibnamefont {Qin}}, \bibinfo
  {author} {\bibfnamefont {Q.}~\bibnamefont {Zeng}}, \bibinfo {author}
  {\bibfnamefont {J.}~\bibnamefont {Li}}, \bibinfo {author} {\bibfnamefont
  {L.}~\bibnamefont {Ling}}, \bibinfo {author} {\bibfnamefont {W.}~\bibnamefont
  {Tong}}, \bibinfo {author} {\bibfnamefont {J.}~\bibnamefont {Sheng}},
  \bibinfo {author} {\bibfnamefont {M.}~\bibnamefont {Yang}}, \bibinfo {author}
  {\bibfnamefont {L.}~\bibnamefont {Wu}}, \bibinfo {author} {\bibfnamefont
  {H.}~\bibnamefont {Guo}},\ and\ \bibinfo {author} {\bibfnamefont
  {Z.}~\bibnamefont {Tian}},\ }\bibfield  {title} {\bibinfo {title} {{Distinct
  magnetic ground states in Shastry-Sutherland lattice materials:
  \ch{Pr2Be2GeO7} versus \ch{Nd2Be2GeO7}}},\ }\href
  {https://doi.org/10.1103/PhysRevB.109.184413} {\bibfield  {journal} {\bibinfo
   {journal} {Phys. Rev. B}\ }\textbf {\bibinfo {volume} {109}},\ \bibinfo
  {pages} {184413} (\bibinfo {year} {2024}{\natexlab{a}})}\BibitemShut
  {NoStop}%
\bibitem [{\citenamefont {Pula}\ \emph {et~al.}(2024)\citenamefont {Pula},
  \citenamefont {Sharma}, \citenamefont {Gautreau}, \citenamefont {K.~P.},
  \citenamefont {Kanigel}, \citenamefont {Frontzek}, \citenamefont {Dolling},
  \citenamefont {Clark}, \citenamefont {Dunsiger}, \citenamefont {Ghara},\ and\
  \citenamefont {Luke}}]{pula2024}%
  \BibitemOpen
  \bibfield  {author} {\bibinfo {author} {\bibfnamefont {M.}~\bibnamefont
  {Pula}}, \bibinfo {author} {\bibfnamefont {S.}~\bibnamefont {Sharma}},
  \bibinfo {author} {\bibfnamefont {J.}~\bibnamefont {Gautreau}}, \bibinfo
  {author} {\bibfnamefont {S.}~\bibnamefont {K.~P.}}, \bibinfo {author}
  {\bibfnamefont {A.}~\bibnamefont {Kanigel}}, \bibinfo {author} {\bibfnamefont
  {M.~D.}\ \bibnamefont {Frontzek}}, \bibinfo {author} {\bibfnamefont {T.~N.}\
  \bibnamefont {Dolling}}, \bibinfo {author} {\bibfnamefont {L.}~\bibnamefont
  {Clark}}, \bibinfo {author} {\bibfnamefont {S.}~\bibnamefont {Dunsiger}},
  \bibinfo {author} {\bibfnamefont {A.}~\bibnamefont {Ghara}},\ and\ \bibinfo
  {author} {\bibfnamefont {G.~M.}\ \bibnamefont {Luke}},\ }\bibfield  {title}
  {\bibinfo {title} {Candidate for a quantum spin liquid ground state in the
  shastry-sutherland lattice material
  ${\mathrm{yb}}_{2}{\mathrm{be}}_{2}{\mathrm{geo}}_{7}$},\ }\href
  {https://doi.org/10.1103/PhysRevB.110.014412} {\bibfield  {journal} {\bibinfo
   {journal} {Phys. Rev. B}\ }\textbf {\bibinfo {volume} {110}},\ \bibinfo
  {pages} {014412} (\bibinfo {year} {2024})}\BibitemShut {NoStop}%
\bibitem [{\citenamefont {Perez-Mato}\ \emph {et~al.}(2015)\citenamefont
  {Perez-Mato}, \citenamefont {Gallego}, \citenamefont {Tasci}, \citenamefont
  {Elcoro}, \citenamefont {de~la Flor},\ and\ \citenamefont
  {Aroyo}}]{perez2015symmetry}%
  \BibitemOpen
  \bibfield  {author} {\bibinfo {author} {\bibfnamefont {J.}~\bibnamefont
  {Perez-Mato}}, \bibinfo {author} {\bibfnamefont {S.}~\bibnamefont {Gallego}},
  \bibinfo {author} {\bibfnamefont {E.}~\bibnamefont {Tasci}}, \bibinfo
  {author} {\bibfnamefont {L.}~\bibnamefont {Elcoro}}, \bibinfo {author}
  {\bibfnamefont {G.}~\bibnamefont {de~la Flor}},\ and\ \bibinfo {author}
  {\bibfnamefont {M.}~\bibnamefont {Aroyo}},\ }\bibfield  {title} {\bibinfo
  {title} {Symmetry-based computational tools for magnetic crystallography},\
  }\href@noop {} {\bibfield  {journal} {\bibinfo  {journal} {Annual Review of
  Materials Research}\ }\textbf {\bibinfo {volume} {45}},\ \bibinfo {pages}
  {217} (\bibinfo {year} {2015})}\BibitemShut {NoStop}%
\bibitem [{\citenamefont {Liu}\ \emph {et~al.}(2024{\natexlab{b}})\citenamefont
  {Liu}, \citenamefont {Duan},\ and\ \citenamefont
  {Yu}}]{liu2024theoryrareearthkramersmagnets}%
  \BibitemOpen
  \bibfield  {author} {\bibinfo {author} {\bibfnamefont {C.}~\bibnamefont
  {Liu}}, \bibinfo {author} {\bibfnamefont {G.}~\bibnamefont {Duan}},\ and\
  \bibinfo {author} {\bibfnamefont {R.}~\bibnamefont {Yu}},\ }\href
  {https://arxiv.org/abs/2412.00757} {\bibinfo {title} {Theory of rare-earth
  kramers magnets on a shastry-sutherland lattice: dimer phases in presence of
  strong spin-orbit coupling}} (\bibinfo {year} {2024}{\natexlab{b}}),\ \Eprint
  {https://arxiv.org/abs/2412.00757} {arXiv:2412.00757 [cond-mat.str-el]}
  \BibitemShut {NoStop}%
\bibitem [{\citenamefont {Mulders}\ \emph {et~al.}(1997)\citenamefont
  {Mulders}, \citenamefont {Yaouanc}, \citenamefont {Dalmas~de R\'eotier},
  \citenamefont {Gubbens}, \citenamefont {Moolenaar}, \citenamefont {F\aa{}k},
  \citenamefont {Ressouche}, \citenamefont {Proke\ifmmode~\check{s}\else
  \v{s}\fi{}}, \citenamefont {Menovsky},\ and\ \citenamefont
  {Buschow}}]{mulders1997}%
  \BibitemOpen
  \bibfield  {author} {\bibinfo {author} {\bibfnamefont {A.~M.}\ \bibnamefont
  {Mulders}}, \bibinfo {author} {\bibfnamefont {A.}~\bibnamefont {Yaouanc}},
  \bibinfo {author} {\bibfnamefont {P.}~\bibnamefont {Dalmas~de R\'eotier}},
  \bibinfo {author} {\bibfnamefont {P.~C.~M.}\ \bibnamefont {Gubbens}},
  \bibinfo {author} {\bibfnamefont {A.~A.}\ \bibnamefont {Moolenaar}}, \bibinfo
  {author} {\bibfnamefont {B.}~\bibnamefont {F\aa{}k}}, \bibinfo {author}
  {\bibfnamefont {E.}~\bibnamefont {Ressouche}}, \bibinfo {author}
  {\bibfnamefont {K.}~\bibnamefont {Proke\ifmmode~\check{s}\else \v{s}\fi{}}},
  \bibinfo {author} {\bibfnamefont {A.~A.}\ \bibnamefont {Menovsky}},\ and\
  \bibinfo {author} {\bibfnamefont {K.~H.~J.}\ \bibnamefont {Buschow}},\
  }\bibfield  {title} {\bibinfo {title}
  {${\mathrm{prru}}_{2}{\mathrm{si}}_{2}$: A giant anisotropic induced magnet
  with a singlet crystal-field ground state},\ }\href
  {https://doi.org/10.1103/PhysRevB.56.8752} {\bibfield  {journal} {\bibinfo
  {journal} {Phys. Rev. B}\ }\textbf {\bibinfo {volume} {56}},\ \bibinfo
  {pages} {8752} (\bibinfo {year} {1997})}\BibitemShut {NoStop}%
\bibitem [{\citenamefont {Zaliznyak}\ \emph
  {et~al.}(2017{\natexlab{a}})\citenamefont {Zaliznyak}, \citenamefont
  {Savici}, \citenamefont {Ovidiu~Garlea}, \citenamefont {Winn}, \citenamefont
  {Filges}, \citenamefont {Schneeloch}, \citenamefont {Tranquada},
  \citenamefont {Gu}, \citenamefont {Wang},\ and\ \citenamefont
  {Petrovic}}]{Zaliznyak_2017}%
  \BibitemOpen
  \bibfield  {author} {\bibinfo {author} {\bibfnamefont {I.~A.}\ \bibnamefont
  {Zaliznyak}}, \bibinfo {author} {\bibfnamefont {A.~T.}\ \bibnamefont
  {Savici}}, \bibinfo {author} {\bibfnamefont {V.}~\bibnamefont
  {Ovidiu~Garlea}}, \bibinfo {author} {\bibfnamefont {B.}~\bibnamefont {Winn}},
  \bibinfo {author} {\bibfnamefont {U.}~\bibnamefont {Filges}}, \bibinfo
  {author} {\bibfnamefont {J.}~\bibnamefont {Schneeloch}}, \bibinfo {author}
  {\bibfnamefont {J.~M.}\ \bibnamefont {Tranquada}}, \bibinfo {author}
  {\bibfnamefont {G.}~\bibnamefont {Gu}}, \bibinfo {author} {\bibfnamefont
  {A.}~\bibnamefont {Wang}},\ and\ \bibinfo {author} {\bibfnamefont
  {C.}~\bibnamefont {Petrovic}},\ }\bibfield  {title} {\bibinfo {title}
  {Polarized neutron scattering on hyspec: the hybrid spectrometer at sns},\
  }\href {https://doi.org/10.1088/1742-6596/862/1/012030} {\bibfield  {journal}
  {\bibinfo  {journal} {Journal of Physics: Conference Series}\ }\textbf
  {\bibinfo {volume} {862}},\ \bibinfo {pages} {012030} (\bibinfo {year}
  {2017}{\natexlab{a}})}\BibitemShut {NoStop}%
\bibitem [{\citenamefont {Koziol}\ \emph {et~al.}(2023)\citenamefont {Koziol},
  \citenamefont {Duft}, \citenamefont {Morigi},\ and\ \citenamefont
  {Schmidt}}]{koziolSystematicAnalysisCrystalline2023}%
  \BibitemOpen
  \bibfield  {author} {\bibinfo {author} {\bibfnamefont {J.~A.}\ \bibnamefont
  {Koziol}}, \bibinfo {author} {\bibfnamefont {A.}~\bibnamefont {Duft}},
  \bibinfo {author} {\bibfnamefont {G.}~\bibnamefont {Morigi}},\ and\ \bibinfo
  {author} {\bibfnamefont {K.~P.}\ \bibnamefont {Schmidt}},\ }\bibfield
  {title} {\bibinfo {title} {Systematic analysis of crystalline phases in
  bosonic lattice models with algebraically decaying density-density
  interactions},\ }\href {https://doi.org/10.21468/SciPostPhys.14.5.136}
  {\bibfield  {journal} {\bibinfo  {journal} {SciPost Physics}\ }\textbf
  {\bibinfo {volume} {14}},\ \bibinfo {pages} {136} (\bibinfo {year}
  {2023})}\BibitemShut {NoStop}%
\bibitem [{\citenamefont {Koziol}\ \emph {et~al.}(2024)\citenamefont {Koziol},
  \citenamefont {Morigi},\ and\ \citenamefont {Schmidt}}]{koziol_quantum_2024}%
  \BibitemOpen
  \bibfield  {author} {\bibinfo {author} {\bibfnamefont {J.~A.}\ \bibnamefont
  {Koziol}}, \bibinfo {author} {\bibfnamefont {G.}~\bibnamefont {Morigi}},\
  and\ \bibinfo {author} {\bibfnamefont {K.~P.}\ \bibnamefont {Schmidt}},\
  }\bibfield  {title} {\bibinfo {title} {Quantum phases of hardcore bosons with
  repulsive dipolar density-density interactions on two-dimensional lattices},\
  }\href {https://doi.org/10.21468/SciPostPhys.17.4.111} {\bibfield  {journal}
  {\bibinfo  {journal} {SciPost Physics}\ }\textbf {\bibinfo {volume} {17}},\
  \bibinfo {pages} {111} (\bibinfo {year} {2024})}\BibitemShut {NoStop}%
\bibitem [{\citenamefont {Graim}\ and\ \citenamefont
  {Landau}(1981)}]{graimMonteCarloStudy1981}%
  \BibitemOpen
  \bibfield  {author} {\bibinfo {author} {\bibfnamefont {T.}~\bibnamefont
  {Graim}}\ and\ \bibinfo {author} {\bibfnamefont {D.~P.}\ \bibnamefont
  {Landau}},\ }\bibfield  {title} {\bibinfo {title} {Monte {{Carlo}} study of
  three-dimensional---to---one-dimensional crossover in the {{Ising}} model},\
  }\href {https://doi.org/10.1103/PhysRevB.24.5156} {\bibfield  {journal}
  {\bibinfo  {journal} {Physical Review B}\ }\textbf {\bibinfo {volume} {24}},\
  \bibinfo {pages} {5156} (\bibinfo {year} {1981})}\BibitemShut {NoStop}%
\bibitem [{\citenamefont {Jiménez}\ \emph {et~al.}(2021)\citenamefont
  {Jiménez}, \citenamefont {Crone}, \citenamefont {Fogh}, \citenamefont
  {Zayed}, \citenamefont {Lortz}, \citenamefont {Pomjakushina}, \citenamefont
  {Conder}, \citenamefont {Läuchli}, \citenamefont {Weber}, \citenamefont
  {Wessel}, \citenamefont {Honecker}, \citenamefont {Normand}, \citenamefont
  {Corboz}, \citenamefont {Rønnow},\ and\ \citenamefont {Mila}}]{julio2021}%
  \BibitemOpen
  \bibfield  {author} {\bibinfo {author} {\bibfnamefont {J.~L.}\ \bibnamefont
  {Jiménez}}, \bibinfo {author} {\bibfnamefont {S.~P.~G.}\ \bibnamefont
  {Crone}}, \bibinfo {author} {\bibfnamefont {E.}~\bibnamefont {Fogh}},
  \bibinfo {author} {\bibfnamefont {M.~E.}\ \bibnamefont {Zayed}}, \bibinfo
  {author} {\bibfnamefont {R.}~\bibnamefont {Lortz}}, \bibinfo {author}
  {\bibfnamefont {E.}~\bibnamefont {Pomjakushina}}, \bibinfo {author}
  {\bibfnamefont {K.}~\bibnamefont {Conder}}, \bibinfo {author} {\bibfnamefont
  {A.~M.}\ \bibnamefont {Läuchli}}, \bibinfo {author} {\bibfnamefont
  {L.}~\bibnamefont {Weber}}, \bibinfo {author} {\bibfnamefont
  {S.}~\bibnamefont {Wessel}}, \bibinfo {author} {\bibfnamefont
  {A.}~\bibnamefont {Honecker}}, \bibinfo {author} {\bibfnamefont
  {B.}~\bibnamefont {Normand}}, \bibinfo {author} {\bibfnamefont {C.~R.~P.}\
  \bibnamefont {Corboz}}, \bibinfo {author} {\bibfnamefont {H.~M.}\
  \bibnamefont {Rønnow}},\ and\ \bibinfo {author} {\bibfnamefont
  {F.}~\bibnamefont {Mila}},\ }\bibfield  {title} {\bibinfo {title} {A quantum
  magnetic analogue to the critical point of water},\ }\href
  {https://doi.org/10.1038 s41586-021-03411-8} {\bibfield  {journal} {\bibinfo
  {journal} {Nature}\ }\textbf {\bibinfo {volume} {592}},\ \bibinfo {pages}
  {370} (\bibinfo {year} {2021})}\BibitemShut {NoStop}%
\bibitem [{\citenamefont {Garlea}\ \emph {et~al.}(2010)\citenamefont {Garlea},
  \citenamefont {Chakoumakos}, \citenamefont {Moore}, \citenamefont {Taylor},
  \citenamefont {Chae}, \citenamefont {Maples}, \citenamefont {Riedel},
  \citenamefont {Lynn},\ and\ \citenamefont {Selby}}]{garlea2010high}%
  \BibitemOpen
  \bibfield  {author} {\bibinfo {author} {\bibfnamefont {V.~O.}\ \bibnamefont
  {Garlea}}, \bibinfo {author} {\bibfnamefont {B.~C.}\ \bibnamefont
  {Chakoumakos}}, \bibinfo {author} {\bibfnamefont {S.~A.}\ \bibnamefont
  {Moore}}, \bibinfo {author} {\bibfnamefont {G.~B.}\ \bibnamefont {Taylor}},
  \bibinfo {author} {\bibfnamefont {T.}~\bibnamefont {Chae}}, \bibinfo {author}
  {\bibfnamefont {R.~G.}\ \bibnamefont {Maples}}, \bibinfo {author}
  {\bibfnamefont {R.~A.}\ \bibnamefont {Riedel}}, \bibinfo {author}
  {\bibfnamefont {G.~W.}\ \bibnamefont {Lynn}},\ and\ \bibinfo {author}
  {\bibfnamefont {D.~L.}\ \bibnamefont {Selby}},\ }\bibfield  {title} {\bibinfo
  {title} {{The high-resolution powder diffractometer at the high flux isotope
  reactor}},\ }\href {https://doi.org/10.1007/s00339-010-5603-6} {\bibfield
  {journal} {\bibinfo  {journal} {Applied Physics A}\ }\textbf {\bibinfo
  {volume} {99}},\ \bibinfo {pages} {531} (\bibinfo {year} {2010})}\BibitemShut
  {NoStop}%
\bibitem [{\citenamefont {Zaliznyak}\ \emph
  {et~al.}(2017{\natexlab{b}})\citenamefont {Zaliznyak}, \citenamefont
  {Savici}, \citenamefont {Garlea}, \citenamefont {Winn}, \citenamefont
  {Filges}, \citenamefont {Schneeloch}, \citenamefont {Tranquada},
  \citenamefont {Gu}, \citenamefont {Wang},\ and\ \citenamefont
  {Petrovic}}]{zaliznyak2017polarized}%
  \BibitemOpen
  \bibfield  {author} {\bibinfo {author} {\bibfnamefont {I.~A.}\ \bibnamefont
  {Zaliznyak}}, \bibinfo {author} {\bibfnamefont {A.~T.}\ \bibnamefont
  {Savici}}, \bibinfo {author} {\bibfnamefont {V.~O.}\ \bibnamefont {Garlea}},
  \bibinfo {author} {\bibfnamefont {B.}~\bibnamefont {Winn}}, \bibinfo {author}
  {\bibfnamefont {U.}~\bibnamefont {Filges}}, \bibinfo {author} {\bibfnamefont
  {J.}~\bibnamefont {Schneeloch}}, \bibinfo {author} {\bibfnamefont {J.~M.}\
  \bibnamefont {Tranquada}}, \bibinfo {author} {\bibfnamefont {G.}~\bibnamefont
  {Gu}}, \bibinfo {author} {\bibfnamefont {A.}~\bibnamefont {Wang}},\ and\
  \bibinfo {author} {\bibfnamefont {C.}~\bibnamefont {Petrovic}},\ }\bibfield
  {title} {\bibinfo {title} {Polarized neutron scattering on hyspec: the hybrid
  spectrometer at sns},\ }in\ \href@noop {} {\emph {\bibinfo {booktitle}
  {Journal of Physics: Conference Series}}},\ Vol.\ \bibinfo {volume} {862}\
  (\bibinfo {organization} {IOP Publishing},\ \bibinfo {year} {2017})\ p.\
  \bibinfo {pages} {012030}\BibitemShut {NoStop}%
\bibitem [{\citenamefont {Granroth}\ \emph {et~al.}(2010)\citenamefont
  {Granroth}, \citenamefont {Kolesnikov}, \citenamefont {Sherline},
  \citenamefont {Clancy}, \citenamefont {Ross}, \citenamefont {Ruff},
  \citenamefont {Gaulin},\ and\ \citenamefont {Nagler}}]{granroth2010sequoia}%
  \BibitemOpen
  \bibfield  {author} {\bibinfo {author} {\bibfnamefont {G.~E.}\ \bibnamefont
  {Granroth}}, \bibinfo {author} {\bibfnamefont {A.~I.}\ \bibnamefont
  {Kolesnikov}}, \bibinfo {author} {\bibfnamefont {T.~E.}\ \bibnamefont
  {Sherline}}, \bibinfo {author} {\bibfnamefont {J.~P.}\ \bibnamefont
  {Clancy}}, \bibinfo {author} {\bibfnamefont {K.~A.}\ \bibnamefont {Ross}},
  \bibinfo {author} {\bibfnamefont {J.~P.~C.}\ \bibnamefont {Ruff}}, \bibinfo
  {author} {\bibfnamefont {B.~D.}\ \bibnamefont {Gaulin}},\ and\ \bibinfo
  {author} {\bibfnamefont {S.~E.}\ \bibnamefont {Nagler}},\ }\bibfield  {title}
  {\bibinfo {title} {{SEQUOIA: A Newly Operating Chopper Spectrometer at the
  SNS}},\ }\href {https://doi.org/10.1088/1742-6596/251/1/012058} {\bibfield
  {journal} {\bibinfo  {journal} {Journal of Physics: Conference Series}\
  }\textbf {\bibinfo {volume} {251}},\ \bibinfo {pages} {012058} (\bibinfo
  {year} {2010})}\BibitemShut {NoStop}%
\bibitem [{\citenamefont {Azuah}\ \emph {et~al.}(2009)\citenamefont {Azuah},
  \citenamefont {Kneller}, \citenamefont {Qiu}, \citenamefont
  {Tregenna-Piggott}, \citenamefont {Brown}, \citenamefont {Copley},\ and\
  \citenamefont {Dimeo}}]{azuah2009dave}%
  \BibitemOpen
  \bibfield  {author} {\bibinfo {author} {\bibfnamefont {R.~T.}\ \bibnamefont
  {Azuah}}, \bibinfo {author} {\bibfnamefont {L.~R.}\ \bibnamefont {Kneller}},
  \bibinfo {author} {\bibfnamefont {Y.}~\bibnamefont {Qiu}}, \bibinfo {author}
  {\bibfnamefont {P.~L.}\ \bibnamefont {Tregenna-Piggott}}, \bibinfo {author}
  {\bibfnamefont {C.~M.}\ \bibnamefont {Brown}}, \bibinfo {author}
  {\bibfnamefont {J.~R.}\ \bibnamefont {Copley}},\ and\ \bibinfo {author}
  {\bibfnamefont {R.~M.}\ \bibnamefont {Dimeo}},\ }\bibfield  {title} {\bibinfo
  {title} {{DAVE: a comprehensive software suite for the reduction,
  visualization, and analysis of low energy neutron spectroscopic data}},\
  }\href {https://doi.org/10.6028/jres.114.025} {\bibfield  {journal} {\bibinfo
   {journal} {Journal of research of the National Institute of Standards and
  Technology}\ }\textbf {\bibinfo {volume} {114}},\ \bibinfo {pages} {341}
  (\bibinfo {year} {2009})}\BibitemShut {NoStop}%
\bibitem [{\citenamefont {Scheie}(2021)}]{scheie2021pycrystalfield}%
  \BibitemOpen
  \bibfield  {author} {\bibinfo {author} {\bibfnamefont {A.}~\bibnamefont
  {Scheie}},\ }\bibfield  {title} {\bibinfo {title} {Pycrystalfield: software
  for calculation, analysis and fitting of crystal electric field
  hamiltonians},\ }\href@noop {} {\bibfield  {journal} {\bibinfo  {journal}
  {Applied Crystallography}\ }\textbf {\bibinfo {volume} {54}},\ \bibinfo
  {pages} {356} (\bibinfo {year} {2021})}\BibitemShut {NoStop}%
\bibitem [{\citenamefont {Hutchings}(1964)}]{hutchings1964point}%
  \BibitemOpen
  \bibfield  {author} {\bibinfo {author} {\bibfnamefont {M.~T.}\ \bibnamefont
  {Hutchings}},\ }\bibfield  {title} {\bibinfo {title} {Point-charge
  calculations of energy levels of magnetic ions in crystalline electric
  fields},\ }in\ \href@noop {} {\emph {\bibinfo {booktitle} {Solid state
  physics}}},\ Vol.~\bibinfo {volume} {16}\ (\bibinfo  {publisher} {Elsevier},\
  \bibinfo {year} {1964})\ pp.\ \bibinfo {pages} {227--273}\BibitemShut
  {NoStop}%
\bibitem [{\citenamefont {Stevens}(1952)}]{stevens1952matrix}%
  \BibitemOpen
  \bibfield  {author} {\bibinfo {author} {\bibfnamefont {K.}~\bibnamefont
  {Stevens}},\ }\bibfield  {title} {\bibinfo {title} {Matrix elements and
  operator equivalents connected with the magnetic properties of rare earth
  ions},\ }\href@noop {} {\bibfield  {journal} {\bibinfo  {journal}
  {Proceedings of the Physical Society. Section A}\ }\textbf {\bibinfo {volume}
  {65}},\ \bibinfo {pages} {209} (\bibinfo {year} {1952})}\BibitemShut
  {NoStop}%
\bibitem [{\citenamefont {Rau}\ and\ \citenamefont
  {Gingras}(2019)}]{rauFrustratedQuantumRareEarth2019}%
  \BibitemOpen
  \bibfield  {author} {\bibinfo {author} {\bibfnamefont {J.~G.}\ \bibnamefont
  {Rau}}\ and\ \bibinfo {author} {\bibfnamefont {M.~J.~P.}\ \bibnamefont
  {Gingras}},\ }\bibfield  {title} {\bibinfo {title} {Frustrated {{Quantum
  Rare-Earth Pyrochlores}}},\ }\href
  {https://doi.org/10.1146/annurev-conmatphys-022317-110520} {\bibfield
  {journal} {\bibinfo  {journal} {Annual Review of Condensed Matter Physics}\
  }\textbf {\bibinfo {volume} {10}},\ \bibinfo {pages} {357} (\bibinfo {year}
  {2019})}\BibitemShut {NoStop}%
\bibitem [{\citenamefont {Rau}\ and\ \citenamefont
  {Gingras}(2015)}]{rauMagnitudeQuantumEffects2015}%
  \BibitemOpen
  \bibfield  {author} {\bibinfo {author} {\bibfnamefont {J.~G.}\ \bibnamefont
  {Rau}}\ and\ \bibinfo {author} {\bibfnamefont {M.~J.~P.}\ \bibnamefont
  {Gingras}},\ }\bibfield  {title} {\bibinfo {title} {Magnitude of quantum
  effects in classical spin ices},\ }\href
  {https://doi.org/10.1103/PhysRevB.92.144417} {\bibfield  {journal} {\bibinfo
  {journal} {Physical Review B}\ }\textbf {\bibinfo {volume} {92}},\ \bibinfo
  {pages} {144417} (\bibinfo {year} {2015})}\BibitemShut {NoStop}%
\bibitem [{\citenamefont
  {Messiah}(1961)}]{messiahRepresentationIrreducibleTensor1961}%
  \BibitemOpen
  \bibfield  {author} {\bibinfo {author} {\bibfnamefont {A.}~\bibnamefont
  {Messiah}},\ }\bibfield  {title} {\bibinfo {title} {Representation of
  {{Irreducible Tensor Operators}}: {{Wigner-Eckart Theorem}}},\ }in\
  \href@noop {} {\emph {\bibinfo {booktitle} {Quantum {{Mechanics}}}}}\
  (\bibinfo  {publisher} {Elsevier},\ \bibinfo {year} {1961})\ pp.\ \bibinfo
  {pages} {573--575}\BibitemShut {NoStop}%
\bibitem [{\citenamefont {Rosenkranz}\ \emph {et~al.}(2000)\citenamefont
  {Rosenkranz}, \citenamefont {Ramirez}, \citenamefont {Hayashi}, \citenamefont
  {Cava}, \citenamefont {Siddharthan},\ and\ \citenamefont
  {Shastry}}]{rosenkranzCrystalfieldInteractionPyrochlore2000}%
  \BibitemOpen
  \bibfield  {author} {\bibinfo {author} {\bibfnamefont {S.}~\bibnamefont
  {Rosenkranz}}, \bibinfo {author} {\bibfnamefont {A.~P.}\ \bibnamefont
  {Ramirez}}, \bibinfo {author} {\bibfnamefont {A.}~\bibnamefont {Hayashi}},
  \bibinfo {author} {\bibfnamefont {R.~J.}\ \bibnamefont {Cava}}, \bibinfo
  {author} {\bibfnamefont {R.}~\bibnamefont {Siddharthan}},\ and\ \bibinfo
  {author} {\bibfnamefont {B.~S.}\ \bibnamefont {Shastry}},\ }\bibfield
  {title} {\bibinfo {title} {Crystal-field interaction in the pyrochlore magnet
  {{Ho2Ti2O7}}},\ }\href {https://doi.org/10.1063/1.372565} {\bibfield
  {journal} {\bibinfo  {journal} {Journal of Applied Physics}\ }\textbf
  {\bibinfo {volume} {87}},\ \bibinfo {pages} {5914} (\bibinfo {year}
  {2000})}\BibitemShut {NoStop}%
\bibitem [{\citenamefont {Vanhecke}\ \emph {et~al.}(2021)\citenamefont
  {Vanhecke}, \citenamefont {Colbois}, \citenamefont {Vanderstraeten},
  \citenamefont {Verstraete},\ and\ \citenamefont
  {Mila}}]{vanheckeSolvingFrustratedIsing2021}%
  \BibitemOpen
  \bibfield  {author} {\bibinfo {author} {\bibfnamefont {B.}~\bibnamefont
  {Vanhecke}}, \bibinfo {author} {\bibfnamefont {J.}~\bibnamefont {Colbois}},
  \bibinfo {author} {\bibfnamefont {L.}~\bibnamefont {Vanderstraeten}},
  \bibinfo {author} {\bibfnamefont {F.}~\bibnamefont {Verstraete}},\ and\
  \bibinfo {author} {\bibfnamefont {F.}~\bibnamefont {Mila}},\ }\bibfield
  {title} {\bibinfo {title} {{Solving Frustrated {{Ising}} Models Using Tensor
  Networks}},\ }\href {https://doi.org/10.1103/PhysRevResearch.3.013041}
  {\bibfield  {journal} {\bibinfo  {journal} {Physical Review Research}\
  }\textbf {\bibinfo {volume} {3}},\ \bibinfo {pages} {013041} (\bibinfo {year}
  {2021})}\BibitemShut {NoStop}%
\bibitem [{\citenamefont {Huang}\ \emph {et~al.}(2016)\citenamefont {Huang},
  \citenamefont {Kitchaev}, \citenamefont {Dacek}, \citenamefont {Rong},
  \citenamefont {Urban}, \citenamefont {Cao}, \citenamefont {Luo},\ and\
  \citenamefont {Ceder}}]{huangFindingProvingExact2016}%
  \BibitemOpen
  \bibfield  {author} {\bibinfo {author} {\bibfnamefont {W.}~\bibnamefont
  {Huang}}, \bibinfo {author} {\bibfnamefont {D.~A.}\ \bibnamefont {Kitchaev}},
  \bibinfo {author} {\bibfnamefont {S.~T.}\ \bibnamefont {Dacek}}, \bibinfo
  {author} {\bibfnamefont {Z.}~\bibnamefont {Rong}}, \bibinfo {author}
  {\bibfnamefont {A.}~\bibnamefont {Urban}}, \bibinfo {author} {\bibfnamefont
  {S.}~\bibnamefont {Cao}}, \bibinfo {author} {\bibfnamefont {C.}~\bibnamefont
  {Luo}},\ and\ \bibinfo {author} {\bibfnamefont {G.}~\bibnamefont {Ceder}},\
  }\bibfield  {title} {\bibinfo {title} {{Finding and Proving the Exact Ground
  State of a Generalized {{Ising}} Model by Convex Optimization and
  {{MAX-SAT}}}},\ }\href {https://doi.org/10.1103/PhysRevB.94.134424}
  {\bibfield  {journal} {\bibinfo  {journal} {Physical Review B}\ }\textbf
  {\bibinfo {volume} {94}},\ \bibinfo {pages} {134424} (\bibinfo {year}
  {2016})}\BibitemShut {NoStop}%
\bibitem [{\citenamefont {Dorier}\ \emph {et~al.}(2008)\citenamefont {Dorier},
  \citenamefont {Schmidt},\ and\ \citenamefont {Mila}}]{Dorier2008}%
  \BibitemOpen
  \bibfield  {author} {\bibinfo {author} {\bibfnamefont {J.}~\bibnamefont
  {Dorier}}, \bibinfo {author} {\bibfnamefont {K.~P.}\ \bibnamefont
  {Schmidt}},\ and\ \bibinfo {author} {\bibfnamefont {F.}~\bibnamefont
  {Mila}},\ }\bibfield  {title} {\bibinfo {title} {Theory of magnetization
  plateaux in the shastry-sutherland model},\ }\href
  {https://doi.org/10.1103/PhysRevLett.101.250402} {\bibfield  {journal}
  {\bibinfo  {journal} {Phys. Rev. Lett.}\ }\textbf {\bibinfo {volume} {101}},\
  \bibinfo {pages} {250402} (\bibinfo {year} {2008})}\BibitemShut {NoStop}%
\bibitem [{\citenamefont {Buchheit}\ \emph {et~al.}(2024)\citenamefont
  {Buchheit}, \citenamefont {Busse},\ and\ \citenamefont
  {Gutendorf}}]{Buchheit2024CodePaper}%
  \BibitemOpen
  \bibfield  {author} {\bibinfo {author} {\bibfnamefont {A.~A.}\ \bibnamefont
  {Buchheit}}, \bibinfo {author} {\bibfnamefont {J.}~\bibnamefont {Busse}},\
  and\ \bibinfo {author} {\bibfnamefont {R.}~\bibnamefont {Gutendorf}},\ }\href
  {https://arxiv.org/abs/2412.16317} {\bibinfo {title} {Computation and
  properties of the epstein zeta function with high-performance implementation
  in epsteinlib}} (\bibinfo {year} {2024}),\ \Eprint
  {https://arxiv.org/abs/2412.16317} {arXiv:2412.16317 [math.NA]} \BibitemShut
  {NoStop}%
\bibitem [{\citenamefont {Nishino}\ and\ \citenamefont
  {Okunishi}(1996)}]{nishinoCornerTransferMatrix1996}%
  \BibitemOpen
  \bibfield  {author} {\bibinfo {author} {\bibfnamefont {T.}~\bibnamefont
  {Nishino}}\ and\ \bibinfo {author} {\bibfnamefont {K.}~\bibnamefont
  {Okunishi}},\ }\bibfield  {title} {\bibinfo {title} {{Corner {{Transfer
  Matrix Renormalization Group Method}}}},\ }\href
  {https://doi.org/10.1143/JPSJ.65.891} {\bibfield  {journal} {\bibinfo
  {journal} {Journal of the Physical Society of Japan}\ }\textbf {\bibinfo
  {volume} {65}},\ \bibinfo {pages} {891} (\bibinfo {year} {1996})}\BibitemShut
  {NoStop}%
\bibitem [{\citenamefont {Or{\'u}s}\ and\ \citenamefont
  {Vidal}(2009)}]{orusSimulationTwodimensionalQuantum2009}%
  \BibitemOpen
  \bibfield  {author} {\bibinfo {author} {\bibfnamefont {R.}~\bibnamefont
  {Or{\'u}s}}\ and\ \bibinfo {author} {\bibfnamefont {G.}~\bibnamefont
  {Vidal}},\ }\bibfield  {title} {\bibinfo {title} {Simulation of
  two-dimensional quantum systems on an infinite lattice revisited: {{Corner}}
  transfer matrix for tensor contraction},\ }\href
  {https://doi.org/10.1103/PhysRevB.80.094403} {\bibfield  {journal} {\bibinfo
  {journal} {Physical Review B}\ }\textbf {\bibinfo {volume} {80}},\ \bibinfo
  {pages} {094403} (\bibinfo {year} {2009})}\BibitemShut {NoStop}%
\end{thebibliography}%

%TC:endignore
\end{document}